%% file: WWZ.tex
\documentclass[final,5p,times,twocolumn,authoryear]{elsarticle}

\usepackage{amsmath}
\usepackage{amssymb}
\usepackage{longtable}
\usepackage{xcolor}
\usepackage[colorlinks=true, citecolor=blue, linkcolor=blue, urlcolor=blue]{hyperref}

\newcommand{\tablefoot}[1]{\par\smallskip{\footnotesize\raggedright #1\par}}

\journal{High Energy Astrophysics}

\begin{document}

\begin{frontmatter}

\title{A morphological search for stable year-scale modulation in gamma-ray blazars}

\author[ucm]{Adithiya Dinesh\corref{cor1}}
\ead{adinesh@ucm.es}
\author[ucm]{Alberto Dom\'inguez}
\ead{alberto.d@ucm.es}
\author[clemson]{Alba Rico}
\author[clemson]{Pablo Pe\~nil}
\author[clemson]{Marco Ajello}
\author[wuerzburg,desy]{Sara Buson}

\cortext[cor1]{Corresponding author}

\affiliation[ucm]{organization={IPARCOS and Department of EMFTEL, Universidad Complutense de Madrid},
            city={Madrid},
            postcode={E-28040},
            country={Spain}}
\affiliation[clemson]{organization={Department of Physics and Astronomy, Clemson University},
            addressline={Kinard Lab of Physics},
            city={Clemson},
            postcode={29634-0978},
            state={SC},
            country={USA}}
\affiliation[wuerzburg]{organization={Julius-Maximilians-Universität Würzburg, Fakultät für Physik und Astronomie},
            addressline={Emil-Fischer-Str. 31},
            city={Würzburg},
            postcode={D-97074},
            country={Germany}}
\affiliation[desy]{organization={Deutsches Elektronen-Synchrotron DESY},
            addressline={Platanenallee 6},
            city={Zeuthen},
            postcode={D-15738},
            country={Germany}}

\begin{abstract}
Year-scale quasi-periodic oscillations (QPOs) in blazar gamma-ray light curves have been proposed as signatures of supermassive binary black holes and precessing jets. Red noise can mimic periodicity, and detection alone does not establish persistence or stability. We develop and calibrate a morphological framework to characterize temporal behavior of year-scale modulations and constrain the incidence of stable modulation in blazars. We analyze 120 \textit{Fermi}-LAT light curves spanning 18.1~yr. Singular spectrum analysis selects the oscillatory mode and period; weighted wavelet Z-transform ridge tracking on the detrended light curve quantifies strength, persistence, period drift and cycle-to-cycle volatility. Flux randomization, $10\,000$ red-noise simulations and about $20\,000$ injections calibrate the classification. Of 97 quality-selected sources, two show persistent, coherent modulation consistent with a stationary period (Rank~I: PG~1553+113 and B2~1215+30), six show coherent modulation whose stationarity cannot be established (Rank~II), and 89 form Rank~III. Neither Rank~I nor Rank~II significantly exceeds red-noise expectations. Stationary 1--3~yr injections reach Rank~I in 63--97\% of trials at signal-to-noise ratio (SNR) $=1$, depending on period, and 100\% at SNR $=2$; 92\% of drifting injections reach Rank~II. These results limit stable, approximately sinusoidal 1--3~yr modulation to at most 7\% of the sample at SNR $\geq 1$ and 5\% at SNR $\geq 2$ (95\% confidence). PG~1553+113 has the lowest false-alarm probability, 4.7\% after trials. Stable year-scale modulation is uncommon, and no source is individually established as periodic. Calibrated morphology distinguishes persistent, stationary modulation from coherent variability whose period may evolve or fluctuate, but it does not identify the physical mechanism. PG~1553+113 and B2~1215+30 are candidates for continued monitoring and continuous gravitational-wave searches, although their stability does not establish a binary origin. Six Rank~II sources may show evolving modulation, while six coherent longer-period Rank~III sources require a longer baseline to test stability.
\end{abstract}


\end{frontmatter}

\section{Introduction}
\label{sec:intro}

Blazars, a subclass of active galactic nuclei (AGN) hosting relativistic jets oriented close to the observer's line of sight, are among the most variable persistent sources in the Universe. Their emission displays strong variability across the entire electromagnetic spectrum, from radio to gamma rays, spanning timescales from minutes to decades \citep[e.g.,][]{Urry1995}. In the gamma-ray band, long-term monitoring with the \textit{Fermi} Large Area Telescope (\textit{Fermi}-LAT) has revealed that this variability is predominantly stochastic in nature, commonly described by red-noise or broken power-law power spectral densities \citep[e.g.,][]{Abdo2010, Chatterjee2012, Vaughan2016}. Against this noisy background, the possible presence of quasi-periodic oscillations (QPOs) has attracted sustained observational and theoretical interest \citep[e.g.,][]{Ackermann2015, Sandrinelli2016, Penil2020, Penil2025Search, Penil2025Extensive, Rico2025}.

The identification of QPOs in blazar gamma-ray light curves is of particular physical importance because periodic or quasi-periodic variability provides a direct probe of the dynamics of the central engine and the jet-launching region. Establishing the origin of such modulation has far-reaching implications, from understanding disk--jet coupling in AGN \citep[e.g.,][]{Marscher2002, McHardy2006} to identifying electromagnetic counterparts of future low-frequency gravitational-wave sources \citep[e.g.,][]{Sesana2013, BurkeSpolaor2019}.

Nearly two decades of uninterrupted gamma-ray observations with \textit{Fermi}-LAT have revealed that reported QPOs in blazars on year-long timescales are phenomenologically diverse. Some sources show remarkably stable modulations persisting over many cycles \citep[e.g.,][]{Sandrinelli2018, Covino2019, Rueda2022, Penil2025Trends}, whereas others show measurable changes in period, intermittent behavior, or features that appear only over a fraction of the observational baseline \citep[e.g.,][]{Chen2024, Penil2025Tran}. This diversity suggests that QPOs in blazars do not constitute a single homogeneous class, but instead reflect a range of underlying physical processes with different degrees of temporal stability and evolution.

Despite this, much of the existing literature has focused primarily on the statistical detectability of periodic behavior in light curves, typically framed in terms of global significance against a red-noise null hypothesis \citep[e.g.,][]{Vaughan2005, Ackermann2015, Penil2020}. While such tests are essential to guard against false positives, they provide limited information on the physical nature of the variability once a QPO is identified. In particular, a two-way distinction between ``periodic'' and ``stochastic'' variability overlooks the fact that different emission mechanisms predict distinct behaviors in coherence, longevity, and temporal evolution. These characteristics are encoded in the detailed time evolution of the light curve and can only be accessed by explicitly tracking how the modulation changes over long timescales.

The proposed interpretations fall into two broad families. Geometric scenarios, such as orbital motion in systems hosting supermassive binary black holes (SMBBHs) or relativistic precession of the jet or the disk, attribute quasi-periodicity to changes in viewing geometry and Doppler boosting. They therefore tend to generally predict smooth, coherent and long-lived modulation \citep[e.g.,][]{Begelman1980, Camenzind1992, Valtonen2008, Dotti2012, Caproni2017, Liska2018}. Plasma-driven scenarios involve internal changes in the physical state of the emitting region, such as magnetohydrodynamic (MHD) instabilities, quasi-periodic shock formation, magnetic reconnection or turbulence in the jet flow, which naturally lead to less stable and more transient behavior \citep[e.g.,][]{Rieger2004, Dong2020, Shukla2020}. While both classes of models can produce apparent periodicity, they differ fundamentally in how rigidly the characteristic timescale is preserved over years to decades.

However, these scenarios are not mutually exclusive, and the morphological parameters defined below are not in one-to-one correspondence with them. A single system may couple a geometric driver to plasma-driven fluctuations; a rigid driver may be intermittently obscured; and a finite observational baseline limits how tightly stationarity can be bounded. The framework therefore classifies candidates according to the morphology supported by the data, while keeping the remaining physical degeneracies explicit (Sect.~\ref{subsec:degeneracies}). The mappings from scenarios to parameter values given below are phenomenological expectations that motivate the choice of parameters. They depend on model assumptions, such as the relation between warp radius and precession rate or between field fluctuations and period jitter. The data test these expectations through the measured drift and volatility, whose responses are calibrated with injections in Sect.~\ref{subsec:calibration}.

These considerations motivate an analysis framework that goes beyond detection and characterizes the temporal stability and evolution of candidate QPOs over the full \textit{Fermi}-LAT baseline. In this work we introduce a morphology-based approach to year-scale QPOs in blazar gamma-ray light curves that quantifies how coherent, persistent and stable a modulation is, and we calibrate every step of it against red noise and injected signals. Three questions guide the analysis: which of these properties the data can actually measure on an 18-yr baseline, whether the population of coherent, long-lived modulation exceeds what red noise produces, and how common stable year-scale modulation is among gamma-ray blazars. The first question is answered by calibrating each parameter on injected signals. The other two follow from the same calibration, which gives the red-noise false-positive rates and the completeness of the classification.

The paper is organized as follows. Section~\ref{sec:background} outlines the theoretical context and the main physical scenarios considered. In Section~\ref{sec:data} we describe the data set and sample selection, including the construction of the \textit{Fermi}-LAT light curves used in this work. Section~\ref{sec:methods} presents the analysis pipeline used to isolate and track quasi-periodic modulation in the presence of red noise, defines the morphological parameters, and specifies the quality selection. Section~\ref{sec:results} presents the results: the diagnostic planes, the calibration of the classification against red noise and injected signals, and the resulting limit on the incidence of stable modulation. In Section~\ref{sec:discussion} we discuss the physical interpretation of these regimes, define the candidate ranking, and summarize the implications for geometric and plasma-driven scenarios, including their relevance for SMBBH searches. Broader applications of the framework are described in Section~\ref{sec:applications}. Our main conclusions are summarized in Section~\ref{sec:summary}.


\section{Theoretical background}
\label{sec:background}

\subsection{Geometric modulation}

Geometric models attribute quasi-periodicity to the rotation or precession of the jet beam or emission region, governed by conservation of angular momentum. Such mechanisms are expected to produce comparatively smooth and long-lived modulation, with the characteristic timescale evolving slowly over many cycles. We use $\varepsilon$ to describe the mean fractional evolution of this timescale, while $\sigma_{\ln P}$ characterizes cycle-to-cycle fluctuations of the period around its smooth secular evolution. The expected behavior of these quantities depends on whether the underlying geometric driver is rigid or evolving.

\textit{Scenario A.} In an SMBBH system, the fundamental periodicity may be set by the Keplerian orbital period $P_{\rm orb}$. For binaries in the gravitational-wave-dominated regime, and neglecting environmental torques, the secular orbital evolution is governed primarily by gravitational-wave (GW) emission. Following \citet{Peters1964}, the decay timescale $\tau_{\mathrm{GW}}$ for a binary with total mass $M_{\mathrm{tot}}$ and mass ratio $q$ is
\begin{equation}
    \tau_{\mathrm{GW}} = \left| \frac{P}{\dot{P}} \right| \approx \frac{5}{96} \frac{c^5}{G^3} \frac{a^4}{\mu M_{\mathrm{tot}}^2},
\label{eq:tau_gw}
\end{equation}
where $a$ is the orbital separation and $\mu$ is the reduced mass. For a typical blazar binary (assuming a mass ratio $q \approx 1$, $M_{\mathrm{tot}} \sim 2 \times 10^9\,M_{\odot}$, and $P \sim 2$ years), Eq.~(\ref{eq:tau_gw}) gives $\tau_{\mathrm{GW}}\simeq7\times10^{3}$~yr. The scaling is steep, $\tau_{\rm GW}\propto P^{8/3}M_{\rm tot}^{-5/3}$: for plausible combinations of SMBBH masses and year-scale orbital periods, the GW-driven decay timescales span roughly $10^{3}$--$10^{6}$~yr. In terms of the drift rate $\varepsilon$, defined as the mean fractional change of the characteristic period per unit time, the GW-driven regime therefore implies
\begin{equation}
    \varepsilon_{\mathrm{GW}} \sim \frac{1}{\tau_{\mathrm{GW}}} \sim 10^{-3}\text{--}10^{-6}~\mathrm{yr}^{-1}.
\end{equation}
Over decade-long baselines, such GW-driven drift is effectively indistinguishable from zero. We therefore treat binaries in the GW-dominated regime as stationary ($\varepsilon \approx 0$) and adopt the operational stationarity criterion $|\varepsilon| < 0.02~\mathrm{yr}^{-1}$ throughout this work. This threshold is set by the frequency resolution of the data, not by the physics of the binary. It lies about two orders of magnitude above $\varepsilon_{\mathrm{GW}}$ at the fiducial values. GW-driven orbital decay is therefore far below our sensitivity, so such a binary would appear stationary. The adopted threshold should be interpreted as an upper limit on drift rather than as a measured drift.

Two further points complete the picture. First, the estimate does not extend to all SMBBHs in AGN: at larger separations, torques from the surrounding gas may dominate the orbital evolution and produce a different, possibly faster, secular drift. Second, within the GW-dominated picture, stochastic cycle-to-cycle jitter is negligible ($\sigma_{\ln P} \approx 0$). The expected signatures are thus $\varepsilon \approx 0$ and $\sigma_{\ln P} \approx 0$. We refer to this case as the rigid-driver scenario.

\textit{Scenario B.} Geometric modulation may also arise from mechanisms in which the characteristic timescale is set by an evolving geometric configuration instead of a rigid orbital clock. A prominent example is Lense--Thirring (LT) precession, in which a spinning black hole exerts a torque on a misaligned accretion disk. The associated precession period $P_{\mathrm{LT}}$ depends on the warping radius $R_{\mathrm{warp}}$ \citep{Fragile2007},
\begin{equation}
    P_{\mathrm{LT}} \propto M_{\mathrm{BH}} a_{*}^{-1} \left( \frac{R_{\mathrm{warp}}}{R_g} \right)^3,
\end{equation}
where $a_{*}$ is the dimensionless spin and $R_g$ is the gravitational radius. Unlike a binary orbit, $R_{\mathrm{warp}}$ is not fixed. It is determined by the balance between Lense--Thirring torques and internal disk stresses and evolves on the disk viscous timescale $t_{\mathrm{visc}} \sim \alpha^{-1} (H/R)^{-2} t_{\mathrm{dyn}}$, which for typical parameters spans decades to centuries.

Secular changes in the accretion rate $\dot{M}$ can therefore shift $R_{\mathrm{warp}}$ and induce an evolution of the precession period. A small fractional change in radius translates into a threefold fractional change in period,
\begin{equation}
    \varepsilon_{\mathrm{LT}} \approx \frac{1}{P} \frac{dP}{dt} \approx \frac{3}{R_{\mathrm{warp}}} \frac{dR_{\mathrm{warp}}}{dt} \sim \frac{3}{t_{\mathrm{visc}}}.
\end{equation}
For sufficiently rapid viscous evolution, the resulting drift can become measurable on observational timescales. We note that for $t_{\mathrm{visc}}$ at the century end of this range, $\varepsilon_{\mathrm{LT}} \sim 0.03\,\mathrm{yr}^{-1}$ is comparable to our stationarity threshold, whereas for $t_{\mathrm{visc}}$ of several centuries the predicted drift becomes observationally indistinguishable from Scenario~A. LT precession is therefore detectable as an evolving driver only over part of its plausible parameter space, and a subset of genuine LT systems will unavoidably be classified as rigid.

A related class of evolving geometric modulation arises when the emission originates from a compact feature moving helically along the jet magnetic field \citep{Camenzind1992}. In this case, the observed period scales with the transverse jet size $R$ and the azimuthal velocity $v_{\phi}$ as $P \approx 2\pi R / v_{\phi}$. If the emitting knot approximately conserves relativistic angular momentum as it propagates downstream ($\gamma R v_{\phi} \approx \mathrm{const}$), then $v_{\phi} \propto R^{-1}$ and the observed period scales as
\begin{equation}
    P(t) \propto R(t)^2.
\end{equation}
Since relativistic jets are approximately conical on these scales ($R \propto z$), the gradual expansion of the trajectory naturally induces a positive secular drift,
\begin{equation}
    \varepsilon_{\mathrm{helix}} = \frac{1}{P} \frac{dP}{dt} \approx 2 \frac{\dot{R}}{R} > 0.
\end{equation}
This scenario makes a distinctive, testable prediction: the drift should be positive-definite. Evolving candidates whose drifts are equally likely to be positive or negative would therefore disfavor expanding helical trajectories as the dominant cause of period evolution, whereas an excess of positive drifts would support them. We return to this test in Sect.~\ref{subsec:mechanism_map}.

In both cases, the defining signature of Scenario~B is a modulation that remains coherent but shows systematic secular evolution. Unlike orbital drivers confined by gravity (Scenario~A), these evolving geometric configurations are inherently subject to changes in disk structure or jet geometry and are therefore expected to show $|\varepsilon| > 0$, with otherwise smooth modulation and negligible cycle-to-cycle jitter.

\subsection{Plasma-driven modulation}
\label{subsec:plasma}

This corresponds to Scenario C. Quasi-periodicity driven by internal jet physics, such as current-driven kink instabilities or recurrent magnetic reconnection, is regulated by the Alfv\'en crossing time $t_A \sim R_{\mathrm{jet}}/v_A$ and the instability growth rate. In contrast to the rigid rotation characterizing Scenarios A and B, the restoring forces in this case are magnetic tension and pressure, which depend on local plasma conditions.

These plasma parameters, including the magnetic field strength $B(t)$ and density $\rho(t)$, are intrinsically turbulent. This turbulence is expected to manifest as enhanced cycle-to-cycle variability of the characteristic period. We describe the resulting variability as a stochastic perturbation to the periodicity,
\begin{equation}
    \sigma_{\ln P} \approx \frac{\delta P}{P} \sim \sqrt{\left(\frac{\delta R_{\mathrm{jet}}}{R_{\mathrm{jet}}}\right)^2 + \left(\frac{\delta v_A}{v_A}\right)^2}.
\label{eq:sigma_mhd}
\end{equation}
Magnetohydrodynamic simulations of kink modes \citep[e.g.,][]{Mizuno2012, Dong2020,Penil2025Kink} indicate that fluctuations in jet radius or magnetic field strength are typically at the 10--20\% level ($\delta R/R \sim 0.1$). As a result, even when the mean period appears stationary on average (i.e., small net drift $\varepsilon$), the cycle-to-cycle variance remains large, corresponding to $\sigma_{\ln P} \gtrsim 0.1$. Moreover, such features are often intermittent \citep{Bromberg2018}.

It is important to recognize that Eq.~(\ref{eq:sigma_mhd}) refers to the intrinsic period jitter of the emitting system. The measured volatility introduced in Sect.~\ref{subsec:parameter_extraction} recovers this variability with some attenuation, and the theoretical expectation $\sigma_{\ln P}\gtrsim0.1$ cannot be transferred directly to an observational threshold. We quantify how the measured values respond to injected jitter through injection--recovery tests (Sect.~\ref{subsec:calibration}), and interpret the volatility boundaries of Sect.~\ref{subsec:thresholds} in the light of that response.

\subsection{Observable properties of the physical scenarios}
\label{subsec:observables}

The three scenarios above are separated by four properties of the modulation. These are the quantities that the framework of Sect.~\ref{subsec:parameter_extraction} is built to measure:
\begin{itemize}
    \item Strength. Any mechanism must produce a modulation that stands above the red-noise background, which sets a minimum amplitude significance.
    \item Longevity. Gravitationally driven and large-scale accretion geometries persist far longer than the observing baseline. Plasma processes are limited by the damping time of the shock or the crossing time of the magnetic fluctuation, and are therefore expected to be intermittent \citep{Bromberg2018}. The fraction of the baseline over which the modulation survives is thus a discriminant in its own right, independent of its instantaneous strength.
    \item Secular evolution. The characteristic timescale is fixed for a rigid orbital clock (Scenario~A) but drifts for an evolving geometric configuration (Scenario~B).
    \item Cycle-to-cycle stability. The period is preserved from one cycle to the next for a gravitational driver but fluctuates for a turbulent one (Scenario~C).
\end{itemize}
No single number captures all four properties, and they are not redundant: a modulation can be strong but short-lived, or long-lived but drifting. Section~\ref{subsec:parameter_extraction} therefore defines four parameters that together quantify these properties: the cycle-weighted coherence $Q$, which combines strength and longevity, the persistence $\Pi$, which isolates longevity, the fractional drift rate $\varepsilon$ and the volatility $\sigma_{\ln P}$.

Section~\ref{subsec:robustness} then measures, by injection, how much of each property its measured parameter actually recovers on an $18$-yr baseline. This second step is part of the design. This validation is essential because a morphological parameter is useful only if it responds to the property it is intended to trace. The results are presented in Sect.~\ref{subsec:calibration}.


\section{Data and sample selection}
\label{sec:data}

\subsection{Blazar sample}
\label{subsec:blazar_sample}
Our analysis focuses on blazars detected by the \textit{Fermi}-LAT, a pair-conversion gamma-ray instrument operating in survey mode since 2008 and covering the energy range from $\sim$20~MeV to beyond 300~GeV \citep{Atwood2009}. The parent source list is drawn from the \textit{Fermi}-LAT Fourth Source Catalog, Data Release~2 (4FGL-DR2), which provides source classifications and spectral information based on long-term LAT observations \citep{Abdollahi2020, Ballet2020}.

From the 4FGL-DR2 catalog, we select sources classified as blazars (\texttt{CLASS1} of \texttt{bll}, \texttt{fsrq} or \texttt{bcu}, in either the identified or associated form). Of the 5788 catalog sources, 3436 satisfy this criterion. No additional cuts are applied at this stage. The selection is made on 4FGL-DR2, the release for which the Light Curve Repository products used here were generated. Source classes quoted in the tables and figures below follow the more recent 4LAC-DR3 \citep{Ajello2022}, with one exception discussed in Sect.~\ref{subsec:demographics}, and redshifts follow the compilation of Appendix~\ref{sec:appendix_z}. Redshifts are not taken from a catalog: as set out in Appendix~\ref{sec:appendix_z}, we assess each value against the primary spectroscopic literature, because the catalog entries for BL~Lacs frequently record intervening absorption systems or unconfirmed single-line detections instead of systemic redshifts.

The light curves used here are taken from the \textit{Fermi}-LAT Light Curve Repository (LCR; Sect.~\ref{subsec:lightcurves}), which provides products only for the 4FGL-DR2 sources flagged as variable (variability index $>21.67$), about a quarter of the catalog \citep{Abdollahi2023}. Of our 3436 blazars, 1415 meet this criterion, all of which have a light curve available, and these 1415 form the effective parent population. This is the appropriate starting point for a variability study. It also means that the fractions quoted in this paper, such as the proportion of sources reaching each rank and the subclass demographics of Sect.~\ref{subsec:demographics}, refer to that subset and should not be read as prevalences among blazars as a whole.

\subsection{Gamma-ray light curves}
\label{subsec:lightcurves}

The gamma-ray light curves used in this work are obtained from the \textit{Fermi}-LAT Light Curve Repository\footnote{\url{https://fermi.gsfc.nasa.gov/ssc/data/access/lat/LightCurveRepository/}} \citep{Abdollahi2023}, which provides uniformly generated time-domain data for sources in the 4FGL catalog.

The light curves span the full operational baseline of the \textit{Fermi} mission \citep{Atwood2009, Abdollahi2023} and provide photon flux measurements in the 0.1--100~GeV energy range. The total span is $T_{\mathrm{obs}} = 18.1$~yr, which we refer to as $\sim$18~yr throughout. Fluxes in each time bin are derived through a binned maximum-likelihood analysis using the standard \textit{Fermi}-LAT analysis framework \citep{Abdollahi2023}, accounting for the LAT instrument response function, diffuse background emission, and contributions from nearby cataloged sources within the region of interest. Source detections are quantified via the Test Statistic (TS; \citealt{Mattox1996}), and time bins that do not meet the detection criterion are reported as upper limits in the LCR products.

All light curves use a uniform 30-day binning, matched to our focus on year-scale QPOs. This resolves year-scale modulation while keeping adequate photon statistics per bin for most sources. Shorter bins would greatly increase the fraction of low-significance bins without adding information on multi-year morphology. Higher-cadence binning (e.g., 7 days) may be needed to resolve discrete intra-cycle bursts \citep[e.g.,][]{Maldonado2026}, whereas 30-day binning acts as a low-pass filter. It suppresses high-frequency stochastic flares and optimizes the sensitivity of the weighted wavelet Z-transform (WWZ) to the smooth, low-frequency modulations targeted by our time--frequency analysis. This choice sets the scope of the paper: the framework characterizes smooth, year-scale modulation, and burst-dominated or strongly non-sinusoidal QPOs are suppressed by construction, not absent from the data (Sect.~\ref{sec:summary}).

To ensure reliable flux measurements and sufficient temporal continuity, we apply bin-level quality criteria to every parent light curve and then select light curves on their sampling.

Individual time bins are treated as detections when ${\rm TS} \geq 4$, while lower-significance bins are excluded, since including upper limits could introduce artificial structure into the subsequent time--frequency analysis. This treatment preserves the long-term temporal coverage provided by the \textit{Fermi}-LAT monitoring while maintaining a uniform and well-defined criterion for variability assessment \citep[e.g.,][]{Rico2025}. We also apply the bin-level quality cuts of \citet{Maldonado2026} to remove unreliable likelihood-fit results. We discard bins for which the likelihood fit did not converge (\texttt{fit\_convergence} $\neq 0$), as well as detected bins with non-finite or non-positive flux uncertainties. Finally, we remove anomalous flux measurements satisfying
\begin{equation}
    \frac{F_i}{\sqrt{{\rm TS}_i}}
    >
    10\,{\rm median}
    \left(
    \frac{F}{\sqrt{\rm TS}}
    \right),
\end{equation}
where the median is evaluated independently for each source over bins with finite flux and ${\rm TS}>0$. Bins rejected by any of these criteria are treated as missing observations.

The sampling criteria are then evaluated on the remaining detections. We exclude light curves with an overall fraction of missing bins of 50\% or more, above which \citet{Penil2025Gaps} found a marked degradation in the reliability of the recovered period--significance relation for standard periodicity searches. We additionally require that no gap extends over more than three consecutive missing 30-day bins, counted over the stored bins of each light curve. These two criteria constrain complementary aspects of the sampling, namely the total amount of missing data and the maximum duration of an individual gap. Of the 1415 parent light curves, 430 pass the first and 120 pass both (Table~\ref{tab:sample_flow}); these 120 enter the analysis.


\section{Methodology}
\label{sec:methods}

\begin{figure}
\centering
\includegraphics[width=\columnwidth]{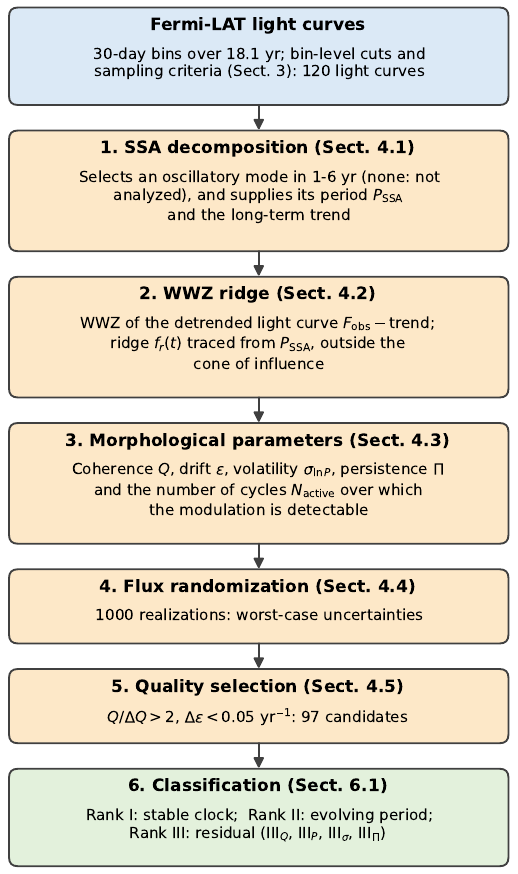}
\caption{Stages of the analysis pipeline. SSA selects the oscillatory mode and supplies its period and the long-term trend; every morphological parameter is measured on the WWZ ridge of the detrended light curve. The thresholds and red-noise rates of each stage are calibrated with red-noise controls and injected signals (Sect.~\ref{subsec:robustness}).}
\label{fig:pipeline}
\end{figure}

We employ a two-stage analysis pipeline to identify and characterize QPOs in the presence of stochastic red noise; Fig.~\ref{fig:pipeline} summarizes its stages. In the first stage, Singular Spectrum Analysis (SSA) decomposes each light curve (Fig.~\ref{fig:SSA}). It serves three purposes: it decides whether the light curve contains an acceptable oscillatory mode in the search band, it supplies the period of that mode, $P_{\mathrm{SSA}}$, and it supplies the long-term trend, which is subtracted from the light curve. In the second stage, the WWZ of the detrended light curve maps its power onto the time--period plane (Fig.~\ref{fig:WWZ}). The ridge is the continuous curve in this plane that follows the local maximum of the WWZ power of the dominant oscillatory feature, starting from $P_{\mathrm{SSA}}$. It therefore traces the characteristic period of the modulation as a function of time (red crosses in Fig.~\ref{fig:WWZ}). Every morphological parameter used for classification is measured on this ridge: SSA selects the modulation to be examined, and the detrended light curve itself provides the measurement.

The parameters then assign each candidate a rank, in order of decreasing compatibility with a stable clock. Rank~I contains coherent modulation that persists over the observable baseline with a stationary period. Rank~II contains equally coherent and persistent modulation whose period cannot be shown to be stationary, because it drifts or fluctuates. Rank~III contains all remaining candidates. The criteria are given in Sect.~\ref{subsec:hierarchy} and Table~\ref{tab:ranking_hierarchy}, and the physical scenarios each rank may correspond to are discussed in Sect.~\ref{sec:discussion}. Appendix~\ref{sec:appendix_tech} gives the technical details of the pipeline.

\subsection{Light curve decomposition via singular spectrum analysis}
\label{subsec:ssa}

To separate potential periodic drivers from the dominant stochastic variability of the blazar, we first use classical SSA as a data-adaptive filter. Unlike Fourier methods, which assume sinusoidal basis functions, SSA decomposes the time series into a sum of additive components through the eigen-decomposition of the light curve's trajectory matrix. Because classical SSA requires a continuous time series, observational gaps are first filled by linear interpolation. To prevent these artificial points from inflating the inferred coherence, the original gap mask is reapplied before any time-frequency analysis, so that only observed bins enter it.

The window length $L$ is the main free parameter of the decomposition. We adopt $L\approx0.4\,N$, where $N$ is the number of data points, following the first systematic SSA-based periodicity search in blazars \citep{Rico2025}. For an $18.1$-yr series this window, about 7~yr, spans several cycles of the longest period searched; Appendix~\ref{subsec:app_ssa} justifies the choice.

The key difficulty in SSA is to distinguish physically meaningful oscillatory components from stochastic fluctuations. Selecting components by signal-to-noise ratio (SNR) alone introduces a look-elsewhere bias, because red noise can produce transient high-power features in adaptive bases. We therefore apply a pairwise selection criterion. It is motivated by a known property of SSA: a stable, well-resolved oscillation is typically represented by a pair of reconstructed components (RCs), with similar eigenvalues and periods and a phase offset near $90^{\circ}$ \citep{Vautard1989, Ghil2002}. 

\begin{figure}[t!]
\centering
\includegraphics[width=\columnwidth]{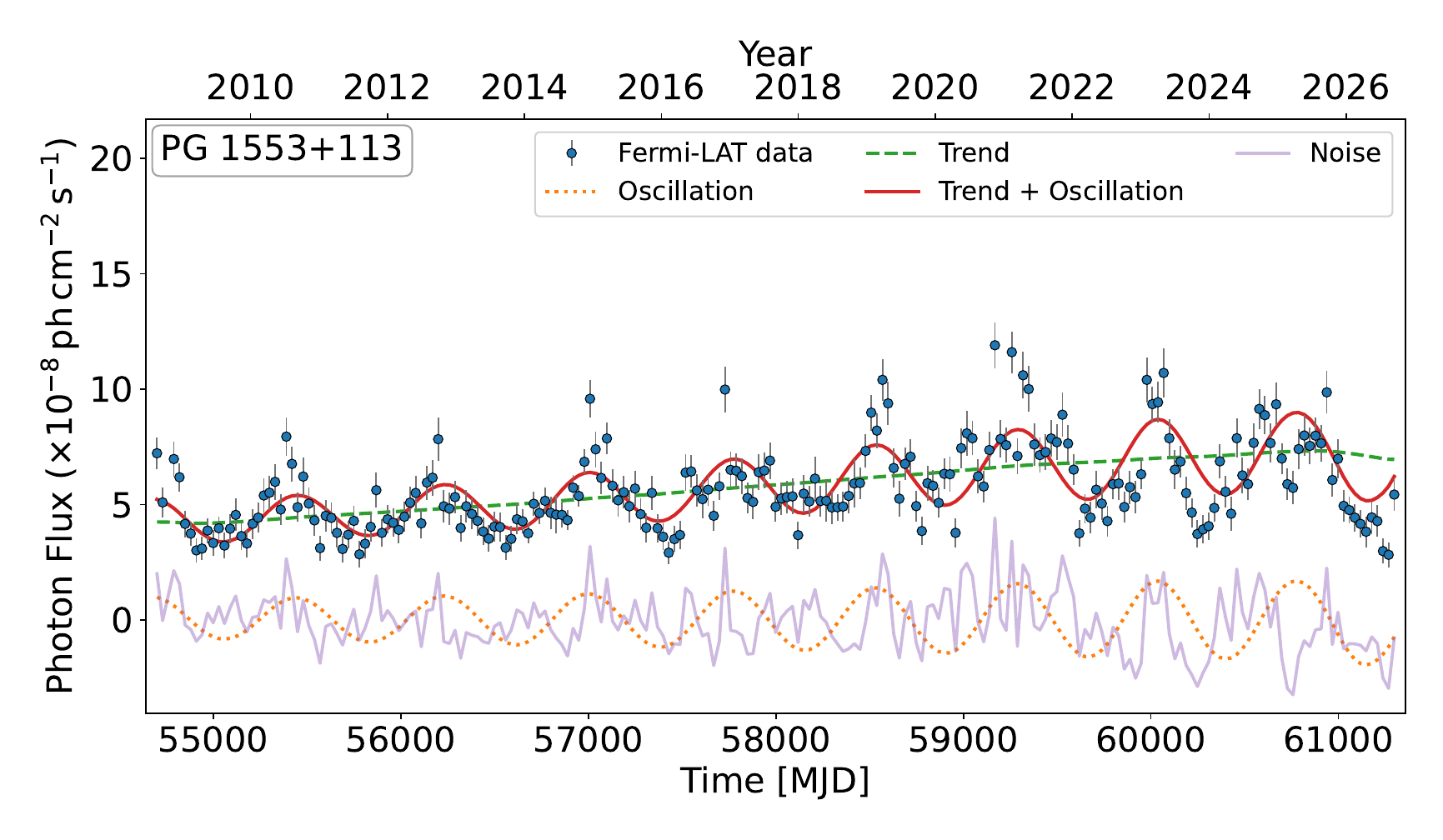}
\caption{SSA decomposition of the light curve of PG~1553+113. Blue circles are the observed 30-day binned \textit{Fermi}-LAT fluxes, with their uncertainties, before any filtering. The dashed green line is the SSA trend (RC~0), the dotted orange line is the oscillatory pair (RC~1 and RC~2) selected by the pairwise criterion, the red line is their sum, and the purple line is the sum of the remaining, noise-dominated components. Its period initializes the ridge traced on the detrended light curve in Fig.~\ref{fig:WWZ}.}
\label{fig:SSA}
\end{figure}

Our selection exploits this property in four steps, illustrated in Fig.~\ref{fig:ssa_flow} and detailed in Appendix~\ref{subsec:app_ssa}. First, components that vary too slowly to complete a cycle within the search band are classified as trend. Second, the remaining components whose periodogram has a significant peak inside the search band become candidates. Third, adjacent candidates are accepted as a pair, or as a block of pairs at a common period, only if they have consistent eigenvalues and periods and are strongly correlated in the weighted correlation (W-correlation) matrix of the reconstructed components. The correlation threshold is set by the noise-dominated components of each light curve. Fourth, the accepted mode with the largest mean eigenvalue per component is selected. The sum of the trend components is the trend subtracted from the light curve, and the period of the selected mode, $P_{\mathrm{SSA}}$, starts the WWZ ridge (Sect.~\ref{subsec:wwz}). Fig.~\ref{fig:SSA} shows the decomposition of the light curve of PG~1553+113, with the selected pair, RC~1 and RC~2, in orange.

Two consequences follow from this pre-filter.

First, the pairwise criterion is deliberately conservative. A genuine oscillation that does not split into a well-correlated pair, as happens when it is heavily noise-dominated, is rejected before the WWZ analysis (Sect.~\ref{subsec:false-positive}). We test whether this conservative pre-filter is responsible for excluding previously reported QPO candidates. Of the 52 previously reported periodicity candidates reanalyzed by \citet{Penil2026}, 20 appear among the 120 light curves that enter our analysis, and 19 of these pass the uncertainty-quality selection of Sect.~\ref{subsec:quality}. The 32 that do not enter are all removed by the sampling criteria of Sect.~\ref{subsec:lightcurves}, not by the pre-filter. Of these, 26 contain a gap longer than three consecutive 30-day bins, 5 miss half or more of their bins, and one, PMN~J0948+0022, is a narrow-line Seyfert~1 galaxy outside the blazar parent sample. Their median fraction of upper-limit bins is 13.6 percent, against 0.2 percent for those that enter, although the gap criterion also removes 12 candidates whose upper-limit fraction is below 10 percent. Every literature candidate that passes the sampling criteria yields an SSA mode and a trackable ridge, and the conservatism of the pre-filter is therefore not the reason previously reported candidates are absent from our sample.

Second, because we select the accepted mode with the largest mean eigenvalue per component, only the stronger of two comparable oscillatory modes in a source is characterized. The framework as implemented assigns one morphology per source and is not designed to detect multiple simultaneous drivers.

\subsection{Weighted wavelet Z-transform analysis and ridge extraction}
\label{subsec:wwz}

\begin{figure}[t!]
\centering
\includegraphics[width=\columnwidth]{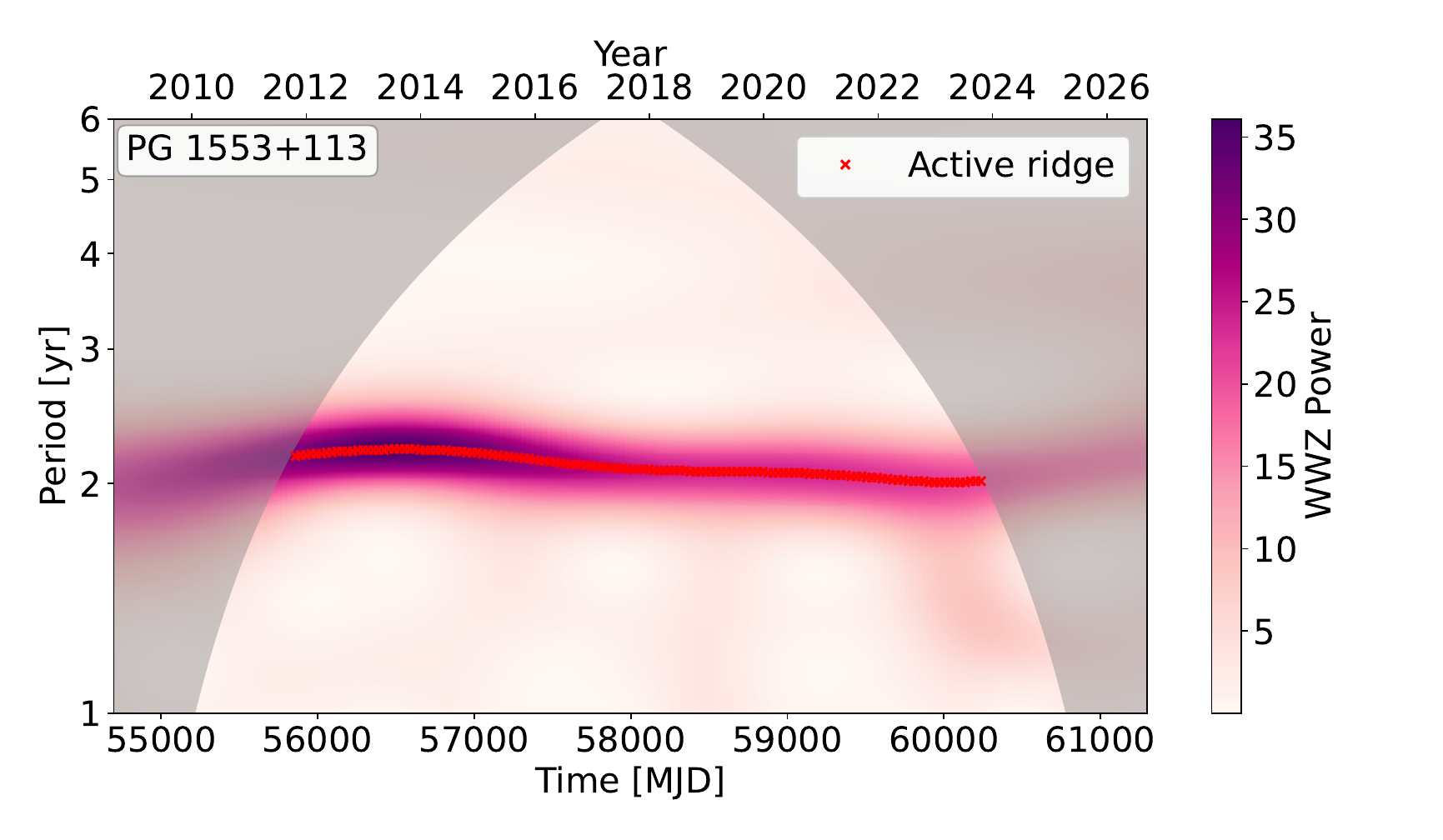}
\caption{WWZ ridge analysis of the detrended light curve of PG~1553+113 (Fig.~\ref{fig:SSA}). The color map is the WWZ power as a function of time and period. The gray shaded region is the cone of influence, where the finite length of the light curve distorts the transform; it is excluded from every ridge statistic. Red crosses mark the ridge points at which the modulation is detectable above the local noise floor (the active points of Sect.~\ref{subsec:parameter_extraction}); they trace the instantaneous period and are the points from which the morphological parameters are computed.}
\label{fig:WWZ}
\end{figure}

The WWZ is a wavelet transform specifically formulated for unevenly sampled astronomical data: unlike the continuous wavelet transform, it rescales the wavelet power to follow $\chi^2$ statistics. We compute the WWZ power map $S(\omega, t)$ on a grid of $500$ periods spaced logarithmically over $1$--$6$~yr and up to $200$ time steps across the light curve. We adopt the standard Morlet decay parameter $c = 0.0125$ \citep{Foster1996}, which sets the time--frequency localization of the transform and also fixes the extent of the cone of influence (Sect.~\ref{subsec:coi_ceiling}). The map is computed for the detrended light curve $F_{\mathrm{det}}$ defined below, not for the SSA reconstruction, because tracing the ridge on the reconstruction would smooth away exactly the period drift and jitter the parameters are meant to measure (Sect.~\ref{subsec:ablation}). The ridge introduced in Sect.~\ref{sec:methods} is shown in Fig.~\ref{fig:WWZ} for PG~1553+113. The WWZ is considerably more expensive than the frequency-domain methods used in large-scale periodicity searches, and has been set aside on those grounds when only a global period and significance are required \citep{Penil2025Gaps}. Here the time-resolved information it provides is precisely the quantity of interest, since the morphological parameters of Sect.~\ref{subsec:parameter_extraction} are defined on the time evolution of the ridge and not on a time-averaged spectrum, and we therefore accept the computational cost.

\subsubsection{Quasi-periodic oscillation frequency tracking}

Defining the instantaneous QPO frequency $f_r(t)$ simply as the global maximum of the WWZ power map at each time step is statistically unstable, as stochastic fluctuations can cause the peak location to jump discontinuously between widely separated frequencies. To extract a physically robust frequency trajectory, we employ a constrained ridge-tracking algorithm.

We restrict the search to characteristic periods $1~\mathrm{yr} \le P \le 6~\mathrm{yr}$. The lower bound has two purposes. It limits contamination from short-term red-noise flickering, which dominates sub-year timescales in monthly binned light curves. It also avoids spurious features near one year produced by the sampling pattern and the observational window function. \citet{Penil2025Gaps} show that the power spectral density (PSD) of gapped \textit{Fermi}-LAT light curves is dominated by a sharp $\sim$1~yr peak coinciding with strong structure in the spectral window. This artifact affects SSA-based analyses as well as the Lomb--Scargle periodogram and phase dispersion minimization.

The upper bound ensures that any detected modulation could in principle complete at least three cycles within the $\sim$18-yr baseline, the minimum needed to assess its stability and temporal evolution. However, $N_{\max}$ counts cycles over the full baseline, and fewer can actually be observed because the cone of influence blanks a period-dependent band at each end of the light curve. The search band is therefore deliberately wider than the range over which a three-cycle requirement can be met in practice. We derive the resulting ceiling in Sect.~\ref{subsec:coi_ceiling} and apply it explicitly to the classification in Sect.~\ref{subsec:hierarchy}.

The ridge, $f_r(t)$, is traced through the WWZ map starting from the SSA period. At each epoch it follows the local power maximum within a window of $\pm40\%$ around the starting period, and the raw track is smoothed with a short running median. The procedure, including how runs of active points are selected, is given in Appendix~\ref{subsec:app_ridge}.

\subsubsection{Analysis safeguards}
Crucially, we implement two safeguards designed to prevent methodological artifacts and to ensure that the inferred ridge properties reflect genuine source variability and not analysis-induced effects.

First, we explicitly exclude all regions lying within the Cone of Influence (COI) of the wavelet transform. Within the COI, the finite temporal extent of the light curve causes the wavelet kernel to extend beyond the observational window, leading to artificial suppression or distortion of power. Including these regions can bias both the inferred ridge trajectory and the associated morphological parameters, particularly near the beginning and end of the time series. By restricting all ridge identification and parameter estimation to times outside the COI, we ensure that the extracted periodic features are supported entirely by the data. As we show in Sect.~\ref{subsec:coi_ceiling}, this choice has a quantitative consequence for any persistence statistic that must be corrected for explicitly.

Second, the ridge and the background against which its strength is measured are taken from the same map. Both are computed on the detrended light curve,
\[
F_{\mathrm{det}} = F_{\mathrm{obs}} - F_{\mathrm{trend}},
\]
where $F_{\mathrm{trend}}$ is the sum of the SSA components classified as trend (Sect.~\ref{subsec:ssa}). The local noise floor entering the coherence of Sect.~\ref{subsec:parameter_extraction} therefore retains the full variance of the high-frequency stochastic variability, and $Q$ is anchored to the variability properties of the data, not to those of a smoothed reconstruction.

\subsection{Derivation of morphological parameters}
\label{subsec:parameter_extraction}

From the extracted ridge $f_r(t)$, corresponding to the time-dependent peak frequency of the WWZ power spectrum, we derive four parameters that quantify the four properties identified in Sect.~\ref{subsec:observables}: strength, longevity, secular evolution and cycle-to-cycle stability. The cycle-weighted coherence $Q$ combines strength and longevity, the persistence $\Pi$ isolates longevity over the observable baseline, the fractional drift rate $\varepsilon$ measures secular evolution, and the volatility $\sigma_{\ln P}$ measures cycle-to-cycle period stability. These metrics provide an operational link between the qualitative expectations outlined in Section~\ref{sec:background} and the diagnostic planes presented in the results. Each is defined below together with the property it is intended to measure; how faithfully each one recovers that property is a separate question, answered by the injection experiment of Sect.~\ref{subsec:robustness} and quantified in Sect.~\ref{subsec:calibration}.

Several of these parameters are computed only over the epochs in which the modulation is detectable. A ridge point is called active when the WWZ power of the detrended light curve $F_{\mathrm{det}}$ at the ridge period is at least the local noise floor $S_{\mathrm{noise}}$ of Eq.~(\ref{eq:snoise}), both evaluated from the same map. Because the ridge is traced on that map, this criterion asks directly whether the modulation is present in the data at each epoch. The longest contiguous run of active points outside the cone of influence is retained (Sect.~\ref{subsec:wwz}), and $\int_{\mathrm{active}}$ denotes integration over it.

\paragraph{a) Cycle-weighted coherence.} The coherence of the modulation, $Q$, measures the combined strength and longevity of the ridge and is defined as
\begin{equation}
Q = \int_{\mathrm{active}} \sqrt{\frac{S(t)}{S_{\mathrm{noise}}(P)}} \, \frac{\mathrm{d}t}{P(t)},
\label{eq:Q}
\end{equation}
where $S(t)$ denotes the WWZ spectral power evaluated along the ridge and $P(t)=1/f_r(t)$ is the instantaneous period. The integrand is an amplitude signal-to-noise ratio, $R(t) \equiv \sqrt{S(t)/S_{\mathrm{noise}}}$, so that $Q$ has the dimensions of a cycle count weighted by amplitude significance.

The background level $S_{\mathrm{noise}}$ is the local red-noise power at each epoch of the ridge, estimated empirically from the same WWZ map of the detrended light curve, outside the cone of influence. We adopt the largest of the four estimates, giving the most conservative local background (Appendix~\ref{subsec:app_noise}).

This local normalization is sufficient for the purposes of this work, as our goal is not to assign absolute detection significances but to quantify the relative strength and persistence of candidate quasi-periodic features in a consistent, data-driven manner. By estimating the background power at the same frequencies and timescales as the ridge, this approach naturally accounts for the frequency-dependent red-noise structure of the light curve, while avoiding assumptions about its global form.

Unlike a standard Fourier coherence measure (e.g.\ $f/\delta f$), $Q$ integrates over time and therefore penalizes short-lived or intermittent features. High values indicate modulation that is both strong and persistent, as expected for geometric drivers. For context, $Q$ grows roughly as the product of the average amplitude signal-to-noise ratio $R$ at the ridge and the number of cycles over which it is sustained. The gate adopted in Sect.~\ref{subsec:thresholds}, $Q-\Delta Q>3$, therefore corresponds to $R\simeq1$ sustained over three full cycles. On a smoothed reconstruction, whose local background is suppressed, the same rate of red-noise false positives requires $Q>10$, or $R\simeq3.3$ (a power ratio $S/S_{\mathrm{noise}}\simeq11$) over three cycles.

We stress that $Q$ is not a measure of phase coherence. $Q$ is exactly what Eq.~(\ref{eq:Q}) computes: a cycle-weighted integral of the ridge amplitude signal-to-noise, combining amplitude, local background and lifetime. It is not a measurement of phase coherence. A strong ridge that wanders in phase can still reach a high $Q$, particularly after the smoothing of Sect.~\ref{subsec:wwz}, and a large value does not by itself show that the modulation is phase-locked. Testing phase stability directly requires a per-cycle phase-residual ($O-C$) statistic on the unsmoothed ridge, a natural extension of this framework. We keep the name ``cycle-weighted coherence'' with this meaning throughout.

\paragraph{b) Fractional drift rate.} To characterize the long-term evolution of the characteristic timescale, we quantify the secular drift of the ridge period using a differential, power-weighted approach. We construct a weighted period track $P(t)$ from the local ridge points, using the WWZ ridge power as weights after all quality and COI cuts have been applied. A local slope $\mathrm{d}P/\mathrm{d}t$ is then computed via a numerical gradient of this weighted track. From this, we define the instantaneous fractional drift rate $\dot{P}(t)/P(t)$. The reported drift parameter $\varepsilon$ is obtained as the time-averaged fractional drift over the ridge lifetime, weighted by the local ridge power to emphasize high-significance intervals:
\begin{equation}
\varepsilon = \frac{\int_{\mathrm{active}} S(t) \, \frac{\dot{P}(t)}{P(t)} \, \mathrm{d}t}{\int_{\mathrm{active}} S(t) \, \mathrm{d}t},
\label{eq:eps}
\end{equation}
where $S(t)$ is the instantaneous WWZ power along the ridge and the integral runs over the ridge-active intervals. The drift $\varepsilon$ has units of $\mathrm{yr}^{-1}$ and measures the mean fractional change in period per unit time. Values consistent with zero indicate stationarity, as expected for rigid orbital motion, while $|\varepsilon| > 0.02\,\mathrm{yr}^{-1}$ indicates detectable secular evolution, as expected for viscously driven precession or evolving accretion geometry. The diagnostic planes show the sign of $\varepsilon$, which carries physical information (Sect.~\ref{subsec:mechanism_map}); the classification criteria use $|\varepsilon|$.

All periods and drift rates in this paper are observer-frame quantities. Since $P_{\mathrm{obs}}=(1+z)P_{\mathrm{rest}}$ and $t_{\mathrm{obs}}=(1+z)t_{\mathrm{rest}}$, the drift transforms as $\varepsilon_{\mathrm{obs}}=\varepsilon_{\mathrm{rest}}/(1+z)$. A common observer-frame threshold therefore corresponds to a rest-frame limit that is weaker by $(1+z)$ for distant sources. The stationarity and volatility criteria of Sect.~\ref{sec:discussion} are operational, observer-frame classifiers, not rest-frame physical limits, and any comparison with the model rates of Sect.~\ref{sec:background} must carry this factor. For PKS~1502+106 ($z=1.84$), a distant source with a resolved drift, the correction is substantial: its rest-frame drift is larger than the observer-frame value by a factor of $2.8$.

\paragraph{c) Volatility.} The intrinsic cycle-to-cycle variability around the secular trend is quantified by the volatility,
\begin{equation}
\sigma_{\ln P} = \left\{ \frac{1}{N_{\mathrm{active}}}\int_{\mathrm{active}} \left[\ln P(t) - \ln P_{\mathrm{trend}}(t)\right]^2 \frac{\mathrm{d}t}{P(t)} \right\}^{1/2},
\label{eq:vol}
\end{equation}
where $P_{\mathrm{trend}}(t)$ is the smooth secular component of the period evolution, obtained as the power-weighted log-linear trend $\ln P_{\mathrm{trend}} = a + \varepsilon\,(t - t_0)$ using the drift estimate of Eq.~(\ref{eq:eps}). The average is cycle-weighted, with the same $\mathrm{d}t/P$ measure used in Eqs.~(\ref{eq:Q}) and (\ref{eq:Nact}), so that $\sigma_{\ln P}$, $Q$ and $N_{\mathrm{active}}$ are all defined on a common footing. This parameter captures non-secular jitter in the characteristic timescale: low values correspond to a stiff, well-organized driver, while high values indicate stochastic or turbulent modulation, as expected for plasma-driven processes. The measured $\sigma_{\ln P}$ is about four times smaller than the jitter injected into simulated light curves (Sect.~\ref{subsec:calibration}), and it also includes the wander that red noise induces in the ridge. It is therefore an operational measure of period stability and not the intrinsic jitter.

\paragraph{d) Longevity.} We quantify the longevity of the modulation through the number of active cycles, $N_{\mathrm{active}}$, defined as the total number of oscillation cycles over which the ridge is active,
\begin{equation}
N_{\mathrm{active}} = \int_{\mathrm{active}} \frac{dt}{P(t)}.
\label{eq:Nact}
\end{equation}
By construction, this allows the coherence to be expressed in a form that highlights its dependence on lifetime,
\begin{equation}
Q = N_{\mathrm{active}} \, \left\langle \sqrt{\frac{S(t)}{S_{\mathrm{noise}}(t)}} \right\rangle,
\label{eq:QN}
\end{equation}
where the angle brackets denote the cycle-weighted average amplitude signal-to-noise ratio over the active intervals. Equation~(\ref{eq:QN}) is exact when $Q$ and $N_{\mathrm{active}}$ are evaluated over the same set of ridge points. In practice $N_{\mathrm{active}}$ is evaluated wherever the ridge is defined, while the integrand of Eq.~(\ref{eq:Q}) additionally requires a valid local noise estimate. Equation~(\ref{eq:QN}) therefore holds to the extent that the two masks coincide; the ratio $Q/N_{\mathrm{active}}$ is monitored for every source as a diagnostic of this agreement.

For reference, we define the maximum number of cycles that could be observed over the full baseline as
\begin{equation}
N_{\max} = \frac{T_{\mathrm{obs}}}{\langle P \rangle},
\label{eq:Nmax}
\end{equation}
where $\langle P \rangle$ is the mean ridge period evaluated over the active intervals, and the raw persistence as $p = N_{\mathrm{active}}/N_{\max}$.

\paragraph{e) Cone-of-influence correction to the persistence.}
\label{subsec:coi_ceiling}
Equation~(\ref{eq:Nmax}) counts cycles over the full baseline, whereas $N_{\mathrm{active}}$ can only accumulate outside the COI. The raw persistence $p$ therefore cannot reach unity even for a perfectly stationary, continuously detected signal, and the shortfall grows with period. For a Morlet kernel with decay parameter $c$, a period $P$ lies inside the COI whenever its distance to the nearest edge of the time series is smaller than $\lambda P$, with
\begin{equation}
\lambda = \frac{1}{2\pi\sqrt{c}} \simeq 1.42 \qquad (c = 0.0125).
\end{equation}
A band of width $\lambda P$ is thus blanked at each end of the light curve, and the maximum attainable persistence is
\begin{equation}
p_{\max}(P) = 1 - \frac{2\lambda P}{T_{\mathrm{obs}}} = 1 - \frac{2\lambda}{N_{\max}} \simeq 1 - \frac{2.85}{N_{\max}}.
\label{eq:pmax}
\end{equation}
For $T_{\mathrm{obs}} = 18.1$~yr this gives $p_{\max} = 0.84$ at $P = 1$~yr, $0.68$ at $P = 2$~yr, $0.53$ at $P = 3$~yr and only $0.21$ at $P = 5$~yr. A fixed cut on $p$ is therefore a period-dependent cut on the underlying physics. With a threshold of $p > 0.75$, no source with $P \gtrsim 1.5$~yr could satisfy it however rigid its clock, and Rank~I would be restricted to short periods by construction, not by measurement.

We therefore adopt the COI-corrected persistence
\begin{equation}
\Pi \equiv \frac{N_{\mathrm{active}}}{N_{\max} - 2\lambda} = \frac{p}{p_{\max}(P)},
\label{eq:Pi}
\end{equation}
which is the fraction of the observable baseline over which the modulation is present. Unlike $p$, its normalization does not depend on period, so that a given value carries the same meaning across the whole search band. Because $p_{\max}$ is evaluated at the mean ridge period while $N_{\mathrm{active}}$ integrates the instantaneous period, $\Pi$ can exceed unity marginally for a modulation whose period is not strictly constant; the largest value attained in the selected sample is $1.01$. All classification criteria in Sect.~\ref{sec:discussion} are expressed in terms of $\Pi$; both $p$ and $\Pi$ are reported in the source tables so that the two conventions can be compared directly. The practical impact is large. 
As a consistency check, PG~1553+113 has $P=2.09$~yr and $p=0.67$, for which $p_{\max}=0.67$ and hence $\Pi=1.00$: the modulation persists over the entire COI-free span, despite the much lower value of the raw persistence statistic. More generally, both Rank~I sources identified below have $p$ in the range $0.55$--$0.67$ and would have failed an uncorrected $p>0.75$ criterion, while both have $\Pi=1.00$--$1.01$. Both satisfy the persistence criterion only once the period-dependent loss of observable cycles within the cone of influence is accounted for.

\paragraph{f) The same ceiling constrains the cycle count.}
The argument just made for $p$ applies verbatim to any absolute requirement on $N_{\mathrm{active}}$, and this has a direct consequence for the ranking scheme of Sect.~\ref{sec:discussion}, which imposes a minimum of three observed cycles on Ranks~I and II. Since $N_{\mathrm{active}}$ can only accumulate outside the COI, its maximum attainable value is $N_{\max}-2\lambda$, and a gate $N_{\mathrm{active}}>3$ can be satisfied only when $T_{\mathrm{obs}}/P - 2\lambda > 3$, that is for
\begin{equation}
P < P_{\mathrm{ceil}} \equiv \frac{T_{\mathrm{obs}}}{3 + 2\lambda} = \frac{18.1~\mathrm{yr}}{5.85} \simeq 3.1~\mathrm{yr}.
\label{eq:pceil}
\end{equation}
There is an important asymmetry between this ceiling and the one affecting $p$. The persistence shortfall is an artifact of normalizing to an unobservable baseline and is removed by Eq.~(\ref{eq:Pi}); the cycle shortfall is not, because it reflects a genuine absence of information. A $5$-yr modulation simply does not complete three cycles within the COI-free span of an $18.1$-yr record, however rigid its clock. No rescaling can manufacture the missing cycles: a statistic of the form $N_{\mathrm{active}}/p_{\max}$ would credit the source with cycles that were never observed, which is precisely what a cycle count must not do.

This limitation cannot be corrected away and is therefore treated as an explicit period restriction. We treat Eq.~(\ref{eq:pceil}) as an explicit period restriction on Ranks~I and II, not as a by-product of the cycle gate. It is stated with the classification criteria (Sect.~\ref{subsec:hierarchy} and Table~\ref{tab:ranking_hierarchy}), and the candidates placed in Rank~III because of it are identified individually (Sect.~\ref{subsec:rank3}). Left undeclared, such a gate would turn a longevity requirement into a hidden short-period selection, the same failure mode found above for the raw persistence. The upper edge of the Rank~I period distribution would then look like a measurement when it is a boundary. The search band of Sect.~\ref{subsec:wwz} is deliberately kept at $1$--$6$~yr: longer-period ridges are still tracked, characterized and reported; they are simply not eligible for a rigidity classification on this baseline.

Together, these parameters provide a compact and physically motivated description of QPO morphology. The cycle-weighted coherence $Q$ quantifies the overall strength and persistence of the modulation, and the drift rate $\varepsilon$ its secular evolution. The volatility $\sigma_{\ln P}$ measures the cycle-to-cycle variability around the mean period, and the persistence $\Pi$ the fraction of the observable baseline over which the modulation remains present.

\subsection{Uncertainty estimation via flux randomization}
\label{subsec:uncertainties}

To quantify the uncertainties on the morphological parameters ($Q$, $\varepsilon$, $\sigma_{\ln P}$, $N_{\mathrm{active}}$ and hence $\Pi$), we employ a flux randomization procedure that propagates measurement errors through the full analysis pipeline while preserving the temporal correlations essential for SSA. We generate $10^3$ synthetic light curves per source by perturbing the observed flux values $F_{\mathrm{obs}}(t)$ with Gaussian noise drawn from the measurement errors $\sigma_{\mathrm{err}}(t)$. The number of realizations yielding a valid oscillatory mode is recorded for each source. Over the $120$ analyzed sources the median recovery is $978/1000$, with a minimum of $430/1000$ for the least stable candidate, and the reported uncertainties are computed over the successful realizations only.

Each realization is processed through the same SSA and WWZ pipeline, with one additional condition on the SSA selection. The selected mode must have a period within $20\%$ of the period measured on the unperturbed light curve, so that every realization measures the same modulation. Among the accepted candidates that satisfy this condition, the one with the largest mean eigenvalue per component is selected, and proximity to the unperturbed period is used only as a tie-breaker. If no candidate satisfies it, the realization counts as a failed recovery.

Appendix~\ref{subsec:app_fr} discusses how this period condition, and the restriction to successful realizations, affect the intervals.

To avoid the attenuation bias induced by flux randomization, we report the parameter values derived from the unperturbed time series as central estimates \citep{Peterson1998}. The uncertainties are the corresponding percentiles of the realization distribution, re-anchored to the unperturbed value: for a parameter $X$ with unperturbed value $X_0$ we quote
\begin{equation}
\Delta X_{\mathrm{lo}} = \max\left(X_0 - X_{16},\,0\right), \qquad
\Delta X_{\mathrm{hi}} = \max\left(X_{84} - X_0,\,0\right),
\end{equation}
where $X_{16}$ and $X_{84}$ are the 16th and 84th percentiles of the distribution. These bounds are asymmetric by construction and are equivalent to a $1\sigma$ interval only in the Gaussian limit. Two scalar summaries are derived from them and used consistently throughout. The mean half-width, $\Delta X = (\Delta X_{\mathrm{lo}} + \Delta X_{\mathrm{hi}})/2$, measures the overall width of the interval and is used in the precision cuts of Sect.~\ref{subsec:quality}. The maximum half-width, $\Delta X = \max(\Delta X_{\mathrm{lo}}, \Delta X_{\mathrm{hi}})$, is a one-sided worst case and is used in the classification bounds of Sect.~\ref{sec:discussion}. The two conventions differ, and the distinction matters for sources lying near a threshold.

\subsection{Common quality selection}
\label{subsec:quality}

Before assigning morphological ranks, we apply a common uncertainty-quality selection. We require finite central values of $Q$, $\varepsilon$ and $\sigma_{\ln P}$, together with finite lower and upper uncertainties derived from the flux-randomization realizations. We additionally require the coherence measurement to be adequately constrained,
\begin{equation}
\frac{Q}{\Delta Q} > 2,
\label{eq:q_precision_cut}
\end{equation}
which corresponds to a mean fractional uncertainty below $50\%$ in $Q$. For the drift rate, we impose the absolute precision criterion
\begin{equation}
\Delta\varepsilon < 0.05~\mathrm{yr}^{-1}.
\label{eq:epsilon_precision_cut}
\end{equation}
An absolute uncertainty is used for $\varepsilon$ because the relative uncertainty becomes ill-defined for stationary candidates with $\varepsilon\simeq0$: dividing by a value close to zero makes the fractional error diverge, which would preferentially reject the best-behaved clocks. In Eqs.~(\ref{eq:q_precision_cut}) and (\ref{eq:epsilon_precision_cut}) the mean half-width convention of Sect.~\ref{subsec:uncertainties} is used. No precision cut is imposed on the period itself. Measured on the detrended light curve, the period is less precise than on a smoothed reconstruction, but it remains well determined for every selected candidate, with $P/\Delta P > 3$ on the mean half-width convention. The reference curves of Eqs.~(\ref{eq:Nmax}) and (\ref{eq:pmax}) are therefore evaluated at reliable periods.

We also considered requiring nearly symmetric flux-randomization intervals, through the asymmetry statistic
\begin{equation}
A_X = \frac{\left|\Delta X_{\mathrm{hi}}-\Delta X_{\mathrm{lo}}\right|}{\Delta X_{\mathrm{hi}}+\Delta X_{\mathrm{lo}}},
\label{eq:uncertainty_asymmetry}
\end{equation}
with $X\in\{Q,\varepsilon,\sigma_{\ln P}\}$ and $A_X\le0.8$. Asymmetry is, however, an expected property of a well-measured parameter near a bound. A volatility close to its measurement floor, the smallest value the pipeline returns for a perfectly periodic signal (Sect.~\ref{subsec:volatility_regime}), can only scatter upward, $\Pi$ close to unity is capped, and the distribution of $Q$ is skewed. An asymmetry cut therefore preferentially rejects the best-behaved clocks, the same pathology that motivates the absolute precision criterion on $\varepsilon$. We settled the question on simulations alone, before ranking the observed sample. Three schemes were defined in advance: S0 applies all three asymmetry cuts, S1 omits the one on $\sigma_{\ln P}$, and S2 applies none. For each scheme we measured the Rank~I completeness for stationary injections below the period ceiling and the Rank~I rate of the red-noise controls (Sect.~\ref{subsec:robustness}). We adopted the scheme with the largest figure of merit for a search, the completeness divided by $1+\sqrt{B}$ \citep{Punzi2003}, where $B$ is the expected number of red-noise Rank~I sources among the 120 analyzed light curves.

Table~\ref{tab:schemes} gives the outcome. With the parameters measured on the detrended light curve, the asymmetry cuts strongly reduce completeness. They retain only $6\%$ of genuine stationary modulations, against $92\%$ without them, while lowering the fraction of red-noise controls that reach Rank~I only from $1.1\%$ to $0.8\%$. Scheme S2 has by far the largest figure of merit and is adopted throughout. The observed sample under the other schemes, computed after the choice, is a sensitivity check: S0 and S1 would leave Rank~I empty.

\subsection{From parent sample to QPO candidates}
\label{subsec:selection_flow}

The morphological parameters are defined only for sources in which an oscillatory mode can be isolated and a ridge tracked. Table~\ref{tab:sample_flow} records the number of sources surviving each stage: the 4FGL-DR2 parent sample, the continuity and detection cuts of Sect.~\ref{sec:data}, the SSA pairwise selection (Sect.~\ref{subsec:ssa}), the ridge-tracking requirement (Sect.~\ref{subsec:wwz}), and the uncertainty-quality selection (Sect.~\ref{subsec:quality}).

All 120 light curves that pass the data criteria yield an accepted SSA mode and a trackable ridge in the 1--6~yr range, and 97 pass the uncertainty-quality selection. These 97 form the QPO candidate sample, to which the population statistics of Sects.~\ref{sec:results} and \ref{sec:discussion} refer. The coherence-precision requirement $Q/\Delta Q>2$ is the most restrictive criterion and alone removes 22 of the 120 sources. The false-positive rates of Sect.~\ref{subsec:false-positive} are normalized to the 120 light curves that enter the analysis, the population the red-noise controls are built to reproduce.

\input{tables/table_sample}

\input{tables/table_schemes}

\subsection{Calibration of the morphological thresholds}
\label{subsec:thresholds}

The coherence and stationarity boundaries follow directly from the arguments above. The stationarity boundary, $|\varepsilon| < 0.02\,\mathrm{yr}^{-1}$, is the smallest drift distinguishable from zero over an 18-yr baseline. The coherence gate requires the integrated amplitude significance of the ridge to exceed the local red-noise background. Its numerical value depends on the map on which the ridge is measured: on the detrended light curve itself the background includes the full stochastic variance, and $Q$ is about an order of magnitude smaller than on a smoothed reconstruction. We set the gate to $Q-\Delta Q>3$, the value at which the red-noise controls reach Rank~I at about $1\%$ per light curve, the same rate as the reconstruction-based gate $Q-\Delta Q>10$ (Sect.~\ref{subsec:ablation}). The gate is fixed on the controls alone, before the data are ranked, and the drift and volatility boundaries are not tuned at all: they are the physical expectations of Sect.~\ref{sec:background}.

The volatility boundaries, $\sigma_{\ln P} = 0.06$ and $0.15$, are the expectations of Sect.~\ref{sec:background} for rigid and plasma-driven modulation. Because the ridge is measured on the detrended light curve, the measured volatility is only about four times smaller than the true jitter (Sect.~\ref{subsec:calibration}). Both boundaries therefore separate sources in practice: the rigid boundary decides the classification of several coherent candidates, and the plasma boundary is crossed by a substantial fraction of the sample (Sect.~\ref{subsec:volatility_regime}).

\subsection{Validation of the classification scheme}
\label{subsec:robustness}

Standard periodicity searches evaluate the false-alarm probability (FAP) of individual detections with source-specific noise simulations. Here we test the classification itself, by passing simulated light curves through the entire pipeline. Table~\ref{tab:simsets} lists the simulation sets.

\input{tables/table_simsets}

\paragraph{Red-noise controls.} We generate $10\,000$ red-noise light curves with the algorithm of \citet{Timmer1995}. Each control takes the time grid, detection mask and per-bin flux uncertainties of one of the 120 analyzed light curves, assigned in turn. Its intrinsic variability is lognormal red noise with the mean flux and the excess variance (the variance in excess of the measurement errors) of that light curve, and with a PSD slope drawn uniformly from $\beta\in[1.0,2.0]$. Gaussian measurement noise with the same per-bin errors is then added. The controls therefore reproduce the sampling, gaps, fractional variability and fractional flux errors of the sample, but contain no periodic signal. Every control that could reach Rank~I or II is flux-randomized and ranked exactly as the data. The Rank~I and Rank~II counts are therefore complete, and the Rank~I rate is determined to about $\pm0.2\%$ ($95\%$ confidence).

\paragraph{Injected signals.} Five families of synthetic modulation, matching the scenarios of Sect.~\ref{sec:background}, are embedded in the controls:
\begin{itemize}
    \item A: stationary sinusoids;
    \item B: sinusoids with a linear fractional period drift $\varepsilon_{\mathrm{inj}}=0.04\,\mathrm{yr}^{-1}$, twice the stationarity threshold;
    \item C: signals whose $\ln P$ follows a cycle-to-cycle random walk with steps $\sigma_{\ln P,\mathrm{inj}}=0.15$, at the plasma boundary;
    \item M: stationary sinusoids that vanish over one contiguous interval of the baseline, with cosine tapers of half a period, representing a rigid driver masked during part of the observation;
    \item N: pure red noise, as the null case.
\end{itemize}
The amplitude SNR is the amplitude of the injected modulation divided by the rms of the control light curve. A first set of injections is classified on nominal values and measures how each parameter responds. A second set goes through the flux randomization and the quality selection, and measures the completeness and purity of the classifier actually applied to the data. Injections at $P_{\mathrm{inj}}\geq3.0$~yr lie above the ceiling of Eq.~(\ref{eq:pceil}) and cannot reach Ranks~I and II; completeness is therefore quoted for $P_{\mathrm{inj}}<P_{\mathrm{ceil}}$.

\paragraph{Tests of the assumptions.} Three further sets test how completeness depends on period, how much flux-dependent censoring matters, and how the drift measurement responds to the injected drift. For censoring, matched injections are analyzed once with the fixed detection mask and once after removing every bin whose perturbed flux falls below twice its uncertainty, the approximate equivalent of TS~$<4$. Undetected bins cannot be restored, because the LAT products give no flux for them. This test therefore gives a pessimistic estimate of the effect. Section~\ref{subsec:ablation} also tests the choice of the map on which the morphology is measured. The results are reported in Sects.~\ref{subsec:false-positive}--\ref{subsec:ablation}.

\section{Results}
\label{sec:results}

We applied the pipeline of Sect.~\ref{sec:methods} to the 120 analyzed light curves and extracted the morphological parameters of each candidate; Figs.~\ref{fig:SSA} and \ref{fig:WWZ} show the two stages for PG~1553+113. The main results of this section are:
\begin{itemize}
    \item two sources, PG~1553+113 and B2~1215+30, show persistent, coherent and stationary modulation (Rank~I), and six show coherent modulation whose period cannot be shown to be stationary (Rank~II) (Sects.~\ref{subsec:mechanism_map}--\ref{subsec:volatility_regime});
    \item red noise processed identically reaches Rank~I at $1.1\%$ and Rank~II at $2.3\%$ per light curve; neither rank exceeds the red-noise expectation significantly, and only PG~1553+113 has a false-alarm probability of a few percent (Sect.~\ref{subsec:false-positive});
    \item stationary modulations below the period ceiling reach Rank~I with $63$--$97\%$ completeness at amplitude SNR of 1 and $100\%$ at SNR of 2, which limits stable 1--3~yr modulation to at most $7\%$ of the sample at SNR $\geq1$ and $5\%$ at SNR $\geq2$ (Sects.~\ref{subsec:calibration} and \ref{subsec:incidence});
    \item a drift injected into simulated light curves is measured about $1.2$ times smaller than it is, and an injected cycle-to-cycle jitter about 4 times smaller; both are therefore recovered, and drifting modulations are placed in Rank~II. Measured on the SSA reconstruction instead, they would come out $3.4$ and $44$ times smaller (Sects.~\ref{subsec:calibration} and \ref{subsec:ablation}).
\end{itemize}
We classify the candidates with two diagnostic planes: the Mechanism Map (Fig.~\ref{fig:mechanism_map}), which separates stationary clocks from drifting and turbulent variability, and the Persistence Plane (Fig.~\ref{fig:persistence_plane}), which distinguishes long-lived structures from transient behavior. A third construction, based on a rigidity parameter $k$, was designed to isolate binary candidates; it adds nothing to the drift and cycle count it combines, and we report it in Appendix~\ref{sec:appendix_sieve}.

\subsection{The mechanism map: coherence versus drift}
\label{subsec:mechanism_map}

\begin{figure}
\centering
\includegraphics[width=\columnwidth]{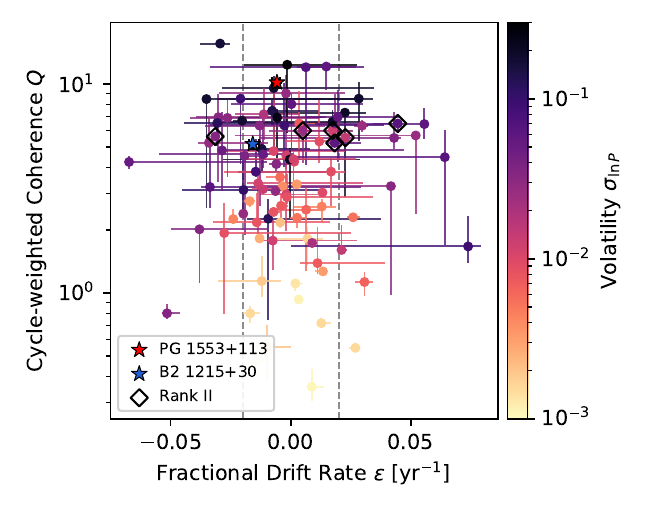}
\caption{Mechanism Map: cycle-weighted coherence $Q$ against the signed fractional drift rate $\varepsilon$, with points color-coded by volatility $\sigma_{\ln P}$. Both $Q$ and the color scale are logarithmic. The vertical dashed lines mark the stationarity threshold $|\varepsilon|=0.02\,\mathrm{yr}^{-1}$; the classification uses the worst-case bounds $|\varepsilon|+\Delta\varepsilon$ and $\sigma_{\ln P}+\Delta\sigma_{\ln P}$, so that the discriminating power is shared between the vertical axis (coherence), the horizontal axis (drift) and the color scale (volatility). Error bars show the flux-randomization uncertainties; for PG~1553+113 the horizontal ones are smaller than the symbol.}
\label{fig:mechanism_map}
\end{figure}

Fig.~\ref{fig:mechanism_map} presents the distribution of the 97 candidates in the plane of cycle-weighted coherence, $Q$, versus the signed fractional drift rate, $\varepsilon$, hereafter the Mechanism Map. The color scale encodes the volatility $\sigma_{\ln P}$.

With the ridge measured on the detrended light curve, the plane is populated in both directions. The coherence spans $Q=0.4$ to $15.6$, and the drift extends from $\varepsilon=-0.067$ to $+0.074\,\mathrm{yr}^{-1}$, well beyond the stationarity threshold. The typical drift uncertainty is $\Delta\varepsilon\simeq0.012\,\mathrm{yr}^{-1}$, and the median $|\varepsilon|/\Delta\varepsilon$ is $1.3$: for most candidates the drift is not resolved. Still, 27 candidates, including both Rank~I sources, have $|\varepsilon|/\Delta\varepsilon>3$. Rank~II is defined by worst-case drift bounds above the stationarity threshold, not by resolved drifts. The candidates separate into three groups:
\begin{enumerate}
\item A coherent, long-lived and stationary group, the two Rank~I sources.
\item A coherent, long-lived group that fails the rigidity test, either through a drift bound above the stationarity threshold or through intermediate volatility: the six members of Rank~II (Sect.~\ref{subsec:no_rank2}), which populate the ``Precession Wings'' at $|\varepsilon|\gtrsim0.02\,\mathrm{yr}^{-1}$.
\item A group failing the coherence gate (57 sources), the cycle-count gate (6 sources, all beyond the period ceiling), the plasma volatility boundary (15 sources), or the rigidity test without the persistence of Rank~II (11 sources).
\end{enumerate}

Because the drift is plotted with its sign retained, Fig.~\ref{fig:mechanism_map} also bears on the helical-knot prediction of Sect.~\ref{sec:background}, which requires $\varepsilon>0$. The distribution shows no preference for positive drift. Of the 97 candidates, 42 have $\varepsilon>0$ and 55 have $\varepsilon<0$, and the mean drift, $\langle\varepsilon\rangle=(-0.5\pm2.4)\times10^{-3}\,\mathrm{yr}^{-1}$, is consistent with zero ($t$-test $p=0.85$; Wilcoxon $p=0.45$). The skewness, $0.48\pm0.25$, hints at a tail toward positive drift at less than $2\sigma$. Among the Rank~II members, five drift toward longer periods and one toward shorter ones, a preference that a sample of six cannot establish. The expanding-helix scenario is therefore not favored as a population property, though it remains available for individual sources.

\subsection{The persistence plane: observation baseline versus longevity}
\label{subsec:persistence_plane}

\begin{figure}
\centering
\includegraphics[width=\columnwidth]{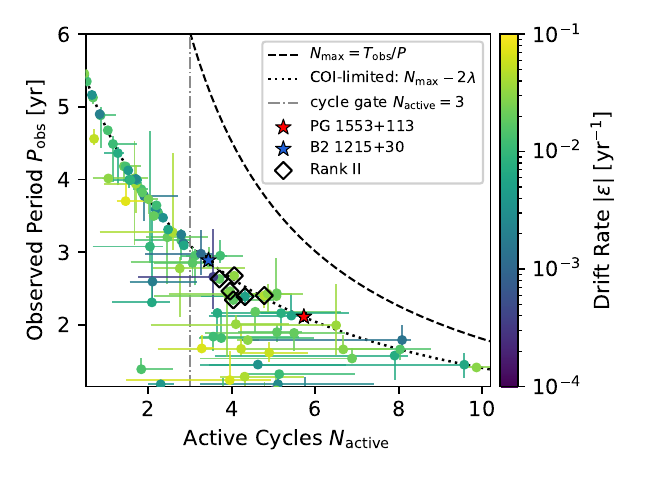}
\caption{Persistence Plane: number of active cycles, $N_{\mathrm{active}}$, against the observed period $P_{\mathrm{obs}}$. The dashed curve marks the maximum number of cycles in the $\sim$18-year \textit{Fermi}-LAT baseline, $N_{\max}=T_{\mathrm{obs}}/P_{\mathrm{obs}}$ ($T_{\mathrm{obs}}=18.1$~yr). The dotted curve marks the maximum observable once the cone of influence is excluded, $N_{\max}-2\lambda$ with $2\lambda=2.85$ (Eq.~\ref{eq:pmax}); this is the attainable boundary, and candidates on it have $\Pi=1$. The dash-dotted vertical line marks the cycle gate $N_{\mathrm{active}}=3$. Points are color-coded by the absolute drift rate $|\varepsilon|$. Sources well below the dotted curve are detectable in the light curve over only part of the observable baseline. Error bars show the flux-randomization uncertainties; for PG~1553+113 they are smaller than the symbol.}
\label{fig:persistence_plane}
\end{figure}

In Fig.~\ref{fig:persistence_plane} we examine the temporal endurance of the signals by plotting $N_{\mathrm{active}}$ against the observed period. This diagnostic plane, the Persistence Plane, separates structural drivers, expected to persist throughout the observational baseline, from intrinsically short-lived phenomena. No cut is applied on $N_{\mathrm{active}}$ for this diagnostic.

Two reference curves are drawn: the full-baseline maximum $N_{\max}=T_{\mathrm{obs}}/P$ (upper) and the COI-limited maximum $N_{\max}-2\lambda$ of Eq.~(\ref{eq:pmax}) (lower). The gap between them grows toward long periods and is the graphical expression of the ceiling of Sect.~\ref{subsec:coi_ceiling}. Candidates cannot lie above the lower curve, and a source on it is present over the entire observable span. Reading the distance to the upper curve as intermittency, as Eq.~(\ref{eq:Nmax}) alone would suggest, would systematically misclassify long-period clocks as transient. The diagram also shows the cycle-count ceiling: the line $N_{\mathrm{active}}=3$ meets the COI-limited curve at $P=P_{\mathrm{ceil}}\simeq3.1$~yr (Eq.~\ref{eq:pceil}), and at longer periods the whole attainable region lies to the left of the gate.

Both Rank~I sources lie on the COI-limited boundary ($\Pi=1.00$--$1.01$), and so do most Rank~II members ($\Pi=0.83$--$1.04$): their modulation is detectable over essentially the whole observable baseline. The sources well below the boundary are those whose modulation is detected only intermittently, the behavior expected from local jet instabilities but also from a persistent modulation whose amplitude falls below the noise floor or whose detection is censored in low states. The Persistence Plane therefore acts as a veto: candidates compatible with a stable binary interpretation must populate its upper boundary. Appendix~\ref{sec:appendix_cases} works through three sources at the boundaries of the classification.

\subsection{The observed volatility regime}
\label{subsec:volatility_regime}

In the sample, the volatility measured on the detrended light curve ranges from $\sigma_{\ln P}=0.001$ to $0.28$, with a median of $0.025$. Two calibration numbers are needed to interpret these values (Sect.~\ref{subsec:calibration}). For a perfectly stationary modulation the measurement returns about $0.005$, its floor, and 18 candidates lie below this value. Injected jitter is measured about four times smaller than it is. The MHD expectation $\sigma_{\ln P}\gtrsim0.1$ of Sect.~\ref{subsec:plasma} therefore corresponds to a measured value of about $0.024$, five times the floor. The volatility boundaries can therefore be applied to the measured values: the rigid boundary decides the classification of the Rank~II members with intermediate volatility, and 15 candidates exceed the plasma boundary ($\sigma_{\ln P}+\Delta\sigma_{\ln P}\geq0.15$).

These high-volatility sources are not evidence for plasma-driven modulation. Red-noise controls processed identically exceed the plasma boundary at an estimated $12\%$ per light curve, and the observed fraction is $13\%$ (15 of 120); the median volatility of the candidates, $0.025$, is likewise close to that of the controls, $0.024$. A ridge that follows red noise wanders in period, and on the detrended light curve this wander is measured as volatility. The volatility distribution of the sample is therefore consistent with red noise, and the test of the MHD jitter prediction is limited by the noise of the measurement, not by its attenuation.

\subsection{Validation against red-noise artifacts}
\label{subsec:false-positive}

Red noise can produce high-power fluctuations that look like periodicities. The question is how often it also reproduces the stability of a rigid clock, and the control population of Sect.~\ref{subsec:robustness} answers it directly. Of the $10\,000$ controls, $9996$ yield a trackable ridge, and Table~\ref{tab:fpr} summarizes the results.

\input{tables/table_fpr}

Red noise reaches every rank. Of the $10\,000$ controls, 110 reach Rank~I and 225 Rank~II, rates of $1.10\%$ and $2.25\%$ per light curve ($95\%$ confidence intervals $0.90$--$1.32\%$ and $1.97$--$2.56\%$). These rates are counted, not bounded, because every control that could reach these ranks was flux-randomized. Applied to the 120 analyzed light curves, they predict $1.3$ Rank~I and $2.7$ Rank~II sources from red noise alone, against 2 and 6 observed. Rank~I is therefore consistent with red noise ($p=0.38$), and Rank~II shows a marginal excess ($p=0.055$). Taken together, Ranks~I and II contain eight sources against $4.0$ expected ($p=0.049$), an excess carried by Rank~II. No rank shows a significant excess on its own, and a source-indexed null, which gives each light curve the false-positive rate of the controls built on it, leads to the same conclusion.

\begin{figure}
\centering
\includegraphics[width=\columnwidth]{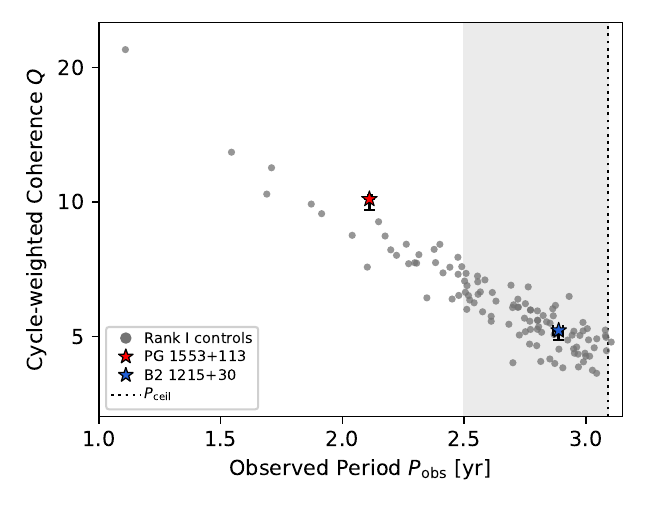}
\caption{Red-noise controls that reach Rank~I (gray; 110 of the $10\,000$), in the plane of observed period and cycle-weighted coherence, compared with the two Rank~I sources. The shaded band marks $2.5\,\mathrm{yr}\le P_{\mathrm{obs}}<P_{\mathrm{ceil}}$, which contains 81 of the 110 controls; the dotted line is the period ceiling of Eq.~(\ref{eq:pceil}). Only 4 controls reach the coherence of PG~1553+113.}
\label{fig:rank1_controls}
\end{figure}

Where the contamination occurs matters for individual sources (Fig.~\ref{fig:rank1_controls}). The Rank~I controls cluster just below the period ceiling: 81 of the 110 have $P=2.5$--$3.1$~yr. Near the ceiling, any ridge that passes the cycle gate spans essentially the whole COI-free baseline and therefore also passes the persistence gate, and long-period ridges are the ones red noise supplies most readily. B2~1215+30, at $2.89$~yr, lies in this band; PG~1553+113, at $2.11$~yr, lies below it, where the Rank~I rate is only $0.28\%$. The rate also grows with the variability of the light curve relative to its flux errors, from zero in the lowest quartile of that ratio to $2.9\%$ in the highest. It does not depend on the PSD slope ($1.16\%$ for $\beta\le1.5$ against $1.04\%$ for $\beta>1.5$).

The same controls give a false-alarm probability for each Rank~I source: the probability that red noise produces, among the 120 light curves searched, at least one Rank~I source with a coherence at least as high. It is $4.7\%$ for PG~1553+113 ($Q=10.2$; 4 of the $10\,000$ controls) and $60\%$ for B2~1215+30 ($Q=5.2$). Source-specific simulations, which use the sampling, errors, excess variance and fitted PSD slope of each source ($\beta=0.86$ for both), agree: none of $1000$ reaches Rank~I with $Q\geq Q_{\mathrm{obs}}$ for PG~1553+113, against 6 for B2~1215+30.

These numbers assume power-law red noise. A power law is the standard description of gamma-ray blazar variability on these timescales, and the source-specific simulations use each source's fitted slope. Some bright blazars, however, show breaks in their gamma-ray power spectra on timescales of about a year \citep{Ryan2019}. Red noise with a break inside the 1--6~yr search band would reach Ranks~I and II more often than our controls. The false-alarm probabilities of individual sources therefore hold only if no such break is present. Testing this requires fitting stochastic models to each light curve, which we leave to a dedicated study. The incidence limit of Sect.~\ref{subsec:incidence} does not share this dependence.

\subsection{Calibration: completeness, purity and measurement response}
\label{subsec:calibration}

The null test bounds contamination by noise. Whether the classifier recovers genuine modulation, and whether it separates the scenarios, is measured by the injections of Sect.~\ref{subsec:robustness}. Table~\ref{tab:confusion} gives the classification of injected signals by the production classifier, Table~\ref{tab:completeness} the Rank~I completeness as a function of period, and Table~\ref{tab:attenuation} the response of the drift and volatility measurements.

\input{tables/table_confusion}

\input{tables/table_completeness}

\paragraph{Completeness.} Stationary modulations reach Rank~I in $63$--$97\%$ of the injections at amplitude SNR of 1, depending on period, and in all of them at SNR of 2. Completeness is lowest at the shortest period tested, $1.2$~yr, not near the ceiling, where a stationary modulation most easily reaches full persistence. Above the ceiling, at $3.0$ and $4.0$~yr, stationary injections cannot reach Ranks~I and II, as Eq.~(\ref{eq:pceil}) requires. Pure red noise reaches Rank~I in about $1\%$ of the trials, consistent with the control rate.

\paragraph{Separation of scenarios.} Drifting injections reach Rank~II in $92\%$ of the trials and never Rank~I. Of the Rank~I injections drawn from the stationary, drifting and jittered families, $83\%$ are stationary; we call this fraction the injection purity of Rank~I. It applies to the families, amplitudes and periods tested, and it is not the probability that a Rank~I source is driven by a stationary process: phase diffusion, amplitude modulation or coherent non-sinusoidal signals could populate Rank~I differently. Jittered injections are the main contaminant: $18\%$ reach Rank~I, $41\%$ Rank~II and $39\%$ Rank~III. Masked clocks fall mostly in Rank~III ($63\%$) or Rank~II ($22\%$). A modulation that disappears for part of the baseline leaves a ridge that wanders in period while the signal is absent, and the volatility test catches it before the persistence test.

\paragraph{Censoring.} When the detection mask is recomputed after injection, the fraction of stationary injections in Rank~I drops from $92\%$ to $57\%$; the rest move to Rank~II ($18\%$) and Rank~III ($23\%$). Red noise is unaffected. Since this test can only remove bins (Sect.~\ref{subsec:robustness}), it is pessimistic, and the true completeness lies between the two values.

\paragraph{Drift and volatility.} We describe the response of each measurement by its attenuation factor, the median ratio of the injected value to the measured one over the injections (Table~\ref{tab:attenuation}). A factor of 1 means the quantity is recovered exactly, and a factor of 4 that it is measured four times smaller than it is. The measured drift follows the injected one almost linearly: an injected drift of $0.04\,\mathrm{yr}^{-1}$ is measured as $0.035\,\mathrm{yr}^{-1}$, an attenuation factor of $1.2$ that holds up to $0.06\,\mathrm{yr}^{-1}$ and rises to $1.4$ at $0.12\,\mathrm{yr}^{-1}$, with little dependence on period or amplitude. Half of the injections with $\varepsilon_{\mathrm{inj}}=0.029\,\mathrm{yr}^{-1}$, and $90\%$ of those with $0.038\,\mathrm{yr}^{-1}$, exceed the stationarity cut. Volatility is attenuated more: an injected jitter of $0.15$ is measured as $0.036$, a factor of 4, because the ridge smoothing removes part of the cycle-to-cycle fluctuation. Its floor, the value measured for stationary injections, is $\sigma_{\ln P}=0.005$. Both parameters therefore respond to the physical quantities at the amplitudes the scenarios of Sect.~\ref{sec:background} predict, which the classification needs in order to carry physical information.

\input{tables/table_attenuation}

\subsection{Rank~II: coherent, evolving modulation}
\label{subsec:no_rank2}

Six candidates satisfy the criterion for evolving modulation defined in Sect.~\ref{sec:discussion} and form Rank~II: coherent, long-lived modulation ($\Pi-\Delta\Pi>0.60$) that fails the rigidity test, either through a drift bound above the stationarity threshold ($|\varepsilon|+\Delta\varepsilon\geq0.02\,\mathrm{yr}^{-1}$) or through intermediate volatility ($0.06\leq\sigma_{\ln P}+\Delta\sigma_{\ln P}<0.15$). They are TXS~0141+268, 3C~66A, PKS~0537$-$441, Mkn~421, PKS~1502+106 and PKS~1509+022 (Table~\ref{tab:evolving}). All six fail the stationarity test: their worst-case drift bound, $|\varepsilon|+\Delta\varepsilon$, exceeds the threshold, from $0.024\,\mathrm{yr}^{-1}$ (PKS~1502+106) to $0.055\,\mathrm{yr}^{-1}$ (TXS~0141+268). The drift itself is resolved, with $|\varepsilon|>2\Delta\varepsilon$, in three of them: TXS~0141+268, PKS~0537$-$441 and PKS~1502+106. In 3C~66A, Mkn~421 and PKS~1509+022 it is not, and these three, together with TXS~0141+268, also have intermediate volatility: their periods cannot be shown to be stationary, without a secular drift being detected.

Rank~II exceeds the red-noise expectation only marginally (six against $2.7$, $p=0.055$), and its members should be read as candidates. A drift measured over a few cycles can also arise when a red-noise ridge shifts its dominant period. The red-noise rate of Rank~II, $2.3\%$ per light curve, includes exactly such cases, and about half of the six members are therefore expected to be interlopers. The excess is also sensitive to the threshold on volatility: with the last two monthly bins removed from every light curve, three further sources lie below the plasma boundary and join Rank~II (Sect.~\ref{subsec:degeneracies}). PKS~1502+106 is a borderline member, whose drift exceeds the stationarity threshold only marginally ($|\varepsilon|+\Delta\varepsilon=0.024\,\mathrm{yr}^{-1}$). Their drifts are not artifacts of the measurement: the drift injections show that the measurement is nearly unbiased ($\times1.2$), that stationary modulations almost never enter Rank~II ($3\%$), and that drifting ones almost always do. Read at face value, the three resolved drifts correspond to period-change timescales $|P/\dot{P}|$ of about 20--45~yr, comparable to the Lense--Thirring rates estimated in Sect.~\ref{sec:background} for viscous timescales of about a century; PKS~1509+022 is classified through its volatility.

\subsection{How common is stable year-scale modulation?}
\label{subsec:incidence}

The completeness and the red-noise rate together turn the Rank~I count into a limit on the incidence of stable modulation. With 2 sources observed against a red-noise expectation of $1.3$, the \citet{Feldman1998} $95\%$ upper limit on the number of genuine Rank~I sources is $5.4$, or $4.5\%$ of the 120 analyzed light curves. Dividing by the lowest completeness measured between $1.2$ and $2.95$~yr (Table~\ref{tab:completeness}), at most $7\%$ of the sample hosts a stable modulation with a period of 1--3~yr and an amplitude SNR $\geq1$, and at most $5\%$ one with SNR $\geq2$, at $95\%$ confidence.

This limit is robust in ways the individual false-alarm probabilities are not. A red-noise model that produces more Rank~I sources than our power-law controls, such as one with a spectral break inside the search band, raises the background and lowers the upper limit; the power-law limit is therefore the conservative one. The one assumption that loosens it is censoring: with the pessimistic completeness of the recomputed detection mask, the limit at SNR of 1--2 becomes $8\%$.

Four qualifications define the scope of the limit. It refers to modulation close to sinusoidal, because the completeness was measured with sinusoidal injections and the Morlet basis and 30-day binning suppress burst-dominated signals (Sect.~\ref{sec:summary}). Strongly non-sinusoidal modulation is not constrained, and even the clearest case, PG~1553+113, has cycles that are not strictly sinusoidal \citep{Penil2026b}. It refers to the analyzed population: variable, well-sampled gamma-ray blazars, which favors bright and persistently detected sources (Sect.~\ref{subsec:blazar_sample}). It applies below the period ceiling of Eq.~(\ref{eq:pceil}), and longer periods are not constrained. And stationarity is defined at the resolution of the measurement: a modulation drifting by less than about $0.03\,\mathrm{yr}^{-1}$ counts as stable. Within these limits, stable year-scale gamma-ray modulation is rare: at most a few percent of variable blazars host it.

\subsection{Why the morphology is measured on the detrended light curve}
\label{subsec:ablation}

We tested the choice of the map on which the ridge is measured by repeating the complete analysis, with its own controls and injections, in two alternative ways. In the first, the ridge is traced on the WWZ map of the SSA reconstruction; in the second, SSA is not used, and a cubic polynomial trend is removed instead. Both alternatives place the same two sources in Rank~I, recover stationary modulations with the same completeness, and give incidence limits of $7$--$8\%$ at SNR $\geq1$. With the ridge traced on the SSA reconstruction, however, the measured drift is $3.4$ times smaller than the injected one and the measured volatility $44$ times smaller, drifting injections are classified as stable clocks, and Rank~II is empty. The reconstruction is a sum of smooth components, and the period changes that drift and volatility measure are removed before they are measured. Without SSA, the results are close to those presented here. The morphology must therefore be measured on the detrended light curve; SSA is retained for its data-driven trend and its selection of the oscillatory mode.

\section{Physical interpretation and ranking of QPO candidates}
\label{sec:discussion}

The diagnostics of Sect.~\ref{sec:results} allow us to go beyond a two-way distinction between ``periodic'' and ``stochastic'' variability. We order the candidates by the morphology of the recovered ridge, applying in sequence the constraints of the Mechanism Map and the Persistence Plane, into three ranks of decreasing compatibility with a stable clock (Table~\ref{tab:ranking_hierarchy}).

The interpretation of these ranks is important: they describe the observed morphology, not the physical driver. Measured on the detrended light curve, the four parameters separate coherent, long-lived modulation from everything else, and within the coherent modulation they separate a stable period from an evolving one, with a calibrated completeness, purity and red-noise rate for each rank (Sect.~\ref{subsec:calibration}). They do not identify the physical driver: a stable period is expected from an orbital clock but also from slow precession, an evolving period from precession but also from red noise, and jittered modulation reaches Rank~I in a fraction of the injections. The rank names are therefore observational labels for the morphology of the ridge, and the scenarios of Sect.~\ref{sec:background} are the interpretations available for each morphology.

\subsection{Hierarchy of emission mechanisms}
\label{subsec:hierarchy}

Three provisions apply to all ranks.
\begin{enumerate}
    \item Cycle count. To guard against low-frequency red-noise flares that mimic short-lived periodicities, Ranks~I and II require $N_{\mathrm{active}} > 3$. A modulation that is detectable for fewer than three full cycles goes to Rank~III, whatever its amplitude or persistence.
    \item Period ceiling. As a direct corollary, Ranks~I and II are restricted to observed periods $P < P_{\mathrm{ceil}}\simeq3.1$~yr (Eq.~\ref{eq:pceil}). At longer periods the COI-free baseline does not contain three cycles, and no source can meet the cycle gate, however rigid its clock. We carry this restriction explicitly through the classification, because an undeclared version of it would reproduce the pathology found in Sect.~\ref{subsec:coi_ceiling} for uncorrected persistence statistics.
    \item Worst-case bounds. To avoid brittle boundary effects, assignments use the one-sided worst-case bounds from flux randomization (Sect.~\ref{subsec:uncertainties}), for example $Q-\Delta Q>3$. Candidates whose uncertainties allow them to cross into the stochastic regime are downgraded.
\end{enumerate}

\input{tables/table_class}

\subsubsection{Rank~I: stable clocks}

Sources in this rank show modulation whose observed morphology is compatible with a stable clock, with no resolved drift or volatility within the resolution of the measurements characterized in Sect.~\ref{subsec:calibration}. This morphology is expected from rigid geometric systems (Scenario~A) but is not specific to them (Sect.~\ref{subsec:degeneracies}). A Rank~I candidate must simultaneously satisfy:
\begin{enumerate}
    \item High coherence, $Q - \Delta Q > 3$, placing the modulation well above the local red-noise background;
    \item A cycle-count gate, $N_{\mathrm{active}} > 3$, which by Eq.~(\ref{eq:pceil}) confines this rank to $P<P_{\mathrm{ceil}}\simeq3.1$~yr;
    \item Dynamical rigidity, comprising stationary period evolution ($|\varepsilon| + \Delta\varepsilon < 0.02\,\mathrm{yr}^{-1}$) and low volatility ($\sigma_{\ln P} + \Delta\sigma_{\ln P} < 0.06$);
    \item High persistence, $\Pi - \Delta\Pi > 0.90$: the modulation is detectable in the light curve over essentially the whole observable baseline.
\end{enumerate}
The persistence threshold is expressed in the COI-corrected form of Eq.~(\ref{eq:Pi}). At $P\simeq2$~yr it is numerically equivalent to a raw criterion $p > 0.61$, and unlike the raw statistic it does not exclude longer-period clocks by construction (Sect.~\ref{subsec:coi_ceiling}).

\begin{figure*}
  \centering
  \includegraphics[width=0.49\textwidth]{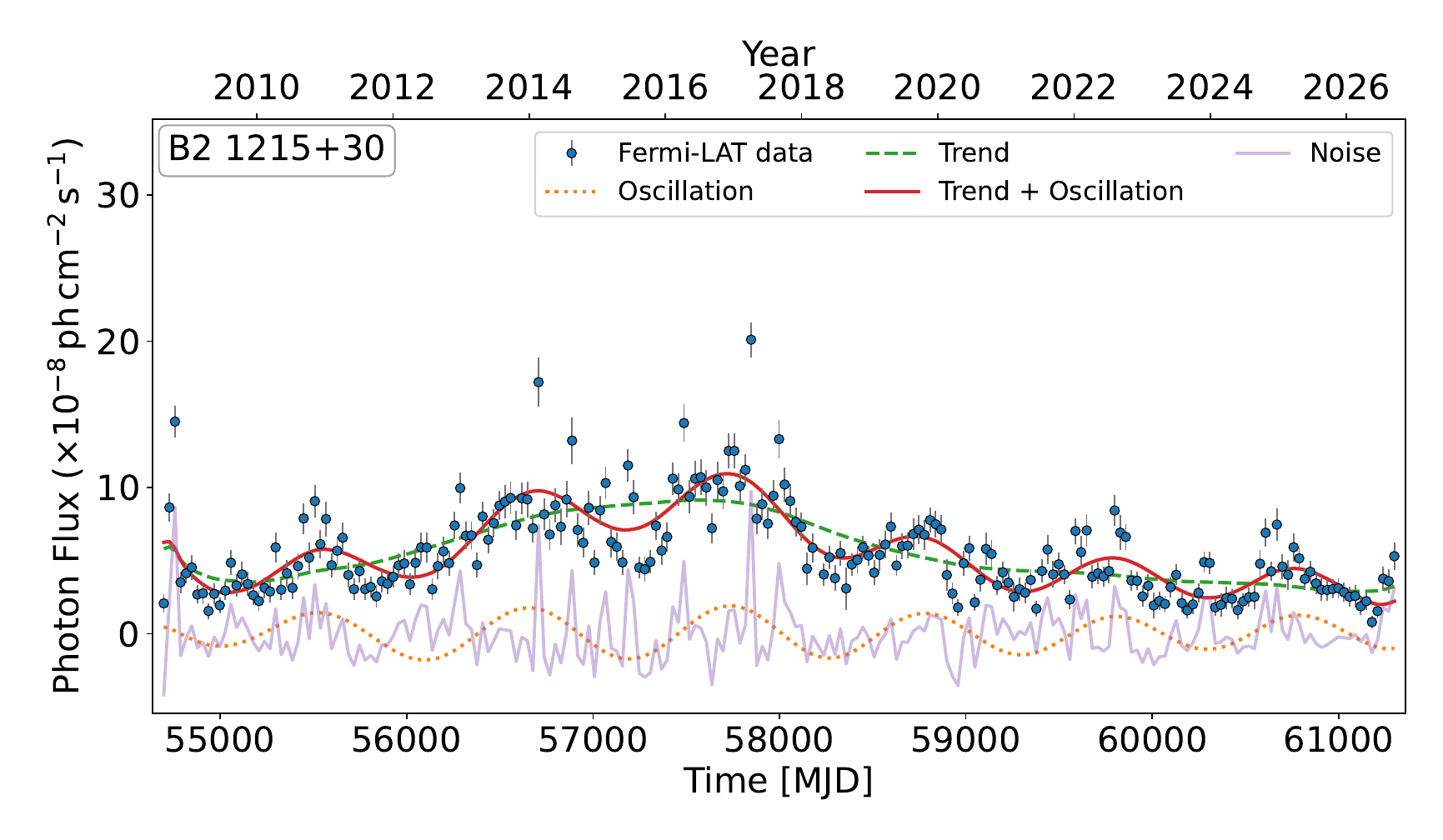}%
  \includegraphics[width=0.49\textwidth]{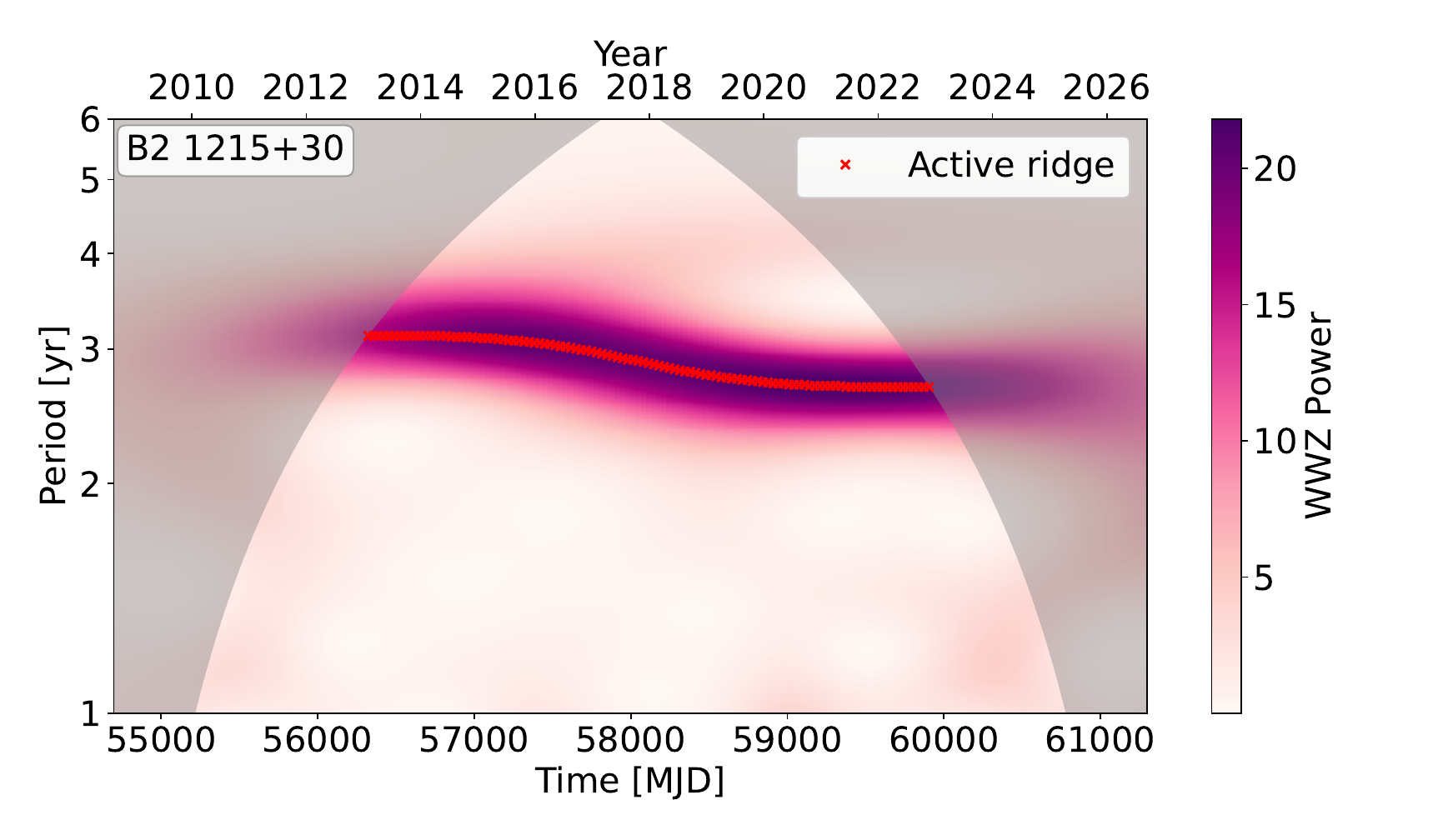}
  \caption{Rank~I source B2~1215+30 (4FGL~J1217.9+3007). Left: SSA decomposition of the 30-day binned \textit{Fermi}-LAT light curve, showing the data, the secular trend, the oscillatory component selected by SSA, their sum and the residual. Right: WWZ map of the detrended light curve, with the cone of influence in gray and the active ridge points as red crosses. The modulation, at $P_{\mathrm{obs}}=2.89$~yr and $Q=5.2$, is present over the whole observable baseline but lies just below the period ceiling ($3.09$~yr).}
  \label{fig:b2}
\end{figure*}

Two sources reach Rank~I (Table~\ref{tab:rank1_sources}). PG~1553+113, shown in Figs.~\ref{fig:SSA} and \ref{fig:WWZ}, has the highest coherence in the rank and a false-alarm probability of $4.7\%$ after trials (Sect.~\ref{subsec:false-positive}). B2~1215+30 (Fig.~\ref{fig:b2}) has a modulation present over the whole observable baseline, at a period of $2.89$~yr just below the ceiling, where the red-noise contamination of Rank~I concentrates; it is consistent with red noise ($60\%$). Both are stable under flux randomization, with the target mode recovered in all 1000 realizations, and both have resolved but sub-threshold drifts that bound their period stability (Table~\ref{tab:stability}). Neither would survive the asymmetry cuts of schemes S0 and S1 (Sect.~\ref{subsec:quality}).

\input{tables/table_ranki}

\subsubsection{Rank~II: coherent, evolving modulation}
\label{subsec:rank2}

Rank~II collects coherent, long-lived modulation that fails the rigidity test of Rank~I, either through a drift bound above the stationarity threshold ($|\varepsilon|+\Delta\varepsilon\ge0.02\,\mathrm{yr}^{-1}$) or through intermediate volatility ($0.06 \le \sigma_{\ln P}+\Delta\sigma_{\ln P} < 0.15$), with persistence $\Pi-\Delta\Pi>0.60$. This is the morphology predicted for Scenario~B, and the rank populates the ``Precession Wings'' of the Mechanism Map. The rank orders compatibility with a stable clock, not the quality of the measurement: a Rank~II modulation is as coherent and long-lived as a Rank~I one, but its period evolves.

Six sources form Rank~II (Table~\ref{tab:evolving}; Appendix~\ref{sec:appendix_atlas}): TXS~0141+268, 3C~66A, PKS~1509+022, PKS~0537$-$441, PKS~1502+106 and Mkn~421. Mkn~421 ($z=0.031$), the nearest bright gamma-ray blazar, is a member. With Mkn~501, it is among the sources most intensively monitored by imaging atmospheric Cherenkov telescopes. Its modulation at $P=2.63$~yr is present over $\Pi=0.92$ of the observable baseline, with a period that lengthens at $\varepsilon=+0.018\,\mathrm{yr}^{-1}$ and an intermediate volatility. All six fail the stationarity test, but the drift is resolved in only three, TXS~0141+268, PKS~0537$-$441 and PKS~1502+106; four members have intermediate volatility (Sect.~\ref{subsec:no_rank2}). The three resolved drifts correspond to period-change timescales of decades, too fast for gravitational-wave-driven orbital decay and of the order expected for Lense--Thirring precession with a viscous timescale of about a century, or for a helical trajectory. They are therefore the natural targets for tests of precession. Two cautions apply. About half of the six are expected to be red-noise interlopers, since a red-noise ridge that shifts its dominant period is measured as a drift. Two members also sit close to a rank boundary: PKS~1502+106, whose drift only marginally exceeds the stationarity threshold, and Mkn~421, whose volatility bound ($0.077$) lies between the rigid and plasma boundaries.

\subsubsection{Rank~III: residual candidates}
\label{subsec:rank3}

Rank~III contains all candidates that fail the criteria above. It is a residual, heterogeneous rank of 89 sources, and to keep its members apart in any population statement we divide it into four subclasses, according to the first test a source fails (Tables~\ref{tab:ranking_hierarchy} and \ref{tab:rank3_sources}):
\begin{itemize}
    \item III$_{Q}$, insufficient coherence (57 sources): genuine but weakly detected modulations, or quasi-periodic fluctuations of red noise. Mkn~501, the second nearest bright gamma-ray blazar, is a member (Appendix~\ref{sec:appendix_cases}).
    \item III$_{P}$, period beyond the ceiling (6 sources: ZS~0214+083, PKS~0426$-$380, PKS~0454$-$234, S5~1044+71, S4~1144+40 and TXS~1902+556). These are coherent, long-lived and in several cases stable modulations, demoted only because the COI-free baseline does not contain three cycles at their periods (Appendix~\ref{sec:appendix_cases}).
    \item III$_{\sigma}$, plasma-level volatility (15 sources). They are the morphology expected from turbulent drivers (Scenario~C), but red noise produces it at the same rate (Sect.~\ref{subsec:volatility_regime}), and the subclass is not evidence for plasma-driven modulation.
    \item III$_{\Pi}$, insufficient persistence (11 sources): coherent modulation that is detectable over too little of the baseline for Rank~I or II. A stable clock whose amplitude drops below the noise floor in low states would also belong here. In practice, a modulation that disappears from the light curve leaves a ridge that wanders in period while the signal is absent. It is therefore usually caught by an earlier test: masked injections fall mostly in Rank~III ($63\%$) or Rank~II ($22\%$), and only rarely remain clocks (Sect.~\ref{subsec:calibration}). The morphology alone therefore cannot single out a masked clock, and the masking interpretation of a given source has to be tested with spectral or multiwavelength diagnostics (Sect.~\ref{subsec:smbbh}).
\end{itemize}
III$_{P}$ sources would benefit directly from a longer baseline. At $T_{\mathrm{obs}}=25$~yr the ceiling moves to $P_{\mathrm{ceil}}\simeq4.3$~yr, and all six members would become eligible for a rigidity classification without any change to the pipeline. Mkn~501, also beyond the present ceiling, would in addition have to pass the coherence gate, which a persistent modulation does as its cycles accumulate.

\input{tables/table_stability}

\subsection{Relation to significance-based periodicity searches}
\label{subsec:significance}

The ranks defined above are morphological classes, not detection significances. The cycle-weighted coherence (Sect.~\ref{subsec:parameter_extraction}) is an amplitude signal-to-noise integrated along the ridge and normalized to a local, data-driven background. It measures how strong and long-lived a modulation is, given that a trackable ridge was found. It is not a false-alarm probability, carries no correction for the number of sources searched, and does not by itself establish that a source is periodic. At the population level, no rank contains significantly more sources than red noise would place there (Sect.~\ref{subsec:false-positive}). Membership of Rank~I is not a source-level detection either: about one of the two Rank~I sources is expected from red noise, and only PG~1553+113 has a false-alarm probability of a few percent once the 120 light curves searched are accounted for.

Analyses of a comparable \textit{Fermi}-LAT baseline reach conservative conclusions about individual candidates. \citet{Penil2026} reanalyze 52 previously reported gamma-ray periodicity candidates with 17.3~yr of data, using the generalized Lomb--Scargle periodogram, phase dispersion minimization and SSA, calibrated against power-law, bending power-law, autoregressive integrated moving-average (ARIMA) and fractionally integrated ARIMA (ARFIMA) models of the stochastic variability. Eight sources retain local test statistics $\ge2\sigma$ and two reach $\ge3\sigma$, but the global significance of even the strongest candidates is consistent with $\approx0\sigma$ once the trial factor of the parent survey is included. Nothing in the present work contradicts that assessment, and our framework orders candidates by the shape of their modulation; whether an individual source is periodic at a stated global confidence is a question for those searches.

The two views are mutually informative, and the source-level comparison is an external check on the ranking. Of the eight sources with test statistics $\ge2\sigma$ in \citet{Penil2026}, three (PKS~0215+015, S3~0458$-$02 and 87GB~164812.2+524023) are removed by the sampling criteria of Sect.~\ref{subsec:lightcurves}. The other five enter our analysis and all pass the quality selection:
\begin{itemize}
    \item PG~1553+113, their most significant source, is our highest-coherence Rank~I source, and our period of $2.11$~yr agrees with their $2.1\pm0.2$~yr;
    \item S5~1044+71, their second $\ge3\sigma$ source, and S4~1144+40 are in Rank~III, demoted by the period ceiling alone; a multiwavelength study confirms the $\sim$3-yr periodicity of S5~1044+71 and interprets it as a precessing jet \citep{Cerruti2027}, which is compatible with the stable modulation we measure (Appendix~\ref{sec:appendix_cases});
    \item TXS~0518+211 exceeds the plasma volatility boundary, and OJ~014 fails the coherence gate; OJ~014 is one of the three sources above $3\sigma$ in the source-specific template search of \citet{Maldonado2026}, before their correction for the number of light curves searched.
\end{itemize}

Our source-level result, that PG~1553+113 is the Rank~I source least compatible with red noise (Sect.~\ref{subsec:false-positive}), matches the state of the field. PG~1553+113 is the one gamma-ray blazar whose $\sim$2.2-yr modulation has been recovered consistently since its first report \citep{Ackermann2015}, and it has been confirmed with 15 years of \textit{Fermi}-LAT and multiwavelength data \citep{Abdollahi2024}. Systematic searches that account for red noise and the number of trials single it out as the strongest case while finding the other candidates consistent with stochastic variability. Its global significance is nonetheless modest: about $1.8\sigma$ \citep{Penil2025Extensive}, $2.2\sigma$ \citep{Rico2025}, and consistent with $0\sigma$ once the trials of the full literature sample are included \citep{Penil2026}. A morphology-aware template search of 7-day light curves reaches the same conclusion: PG~1553+113 has the highest significance after correcting for the 101 light curves searched, $2.75\sigma$, and no new periodicity is detected \citep{Maldonado2026}. Earlier analyses found that global significances are strongly reduced for all reported candidates \citep{AitBenkhali2020}, or that none survives at better than $95\%$ \citep{Covino2019}. A conservative multi-stage analysis of 41 TeV-selected blazars likewise finds no robust gamma-ray QPO, and under the least favorable noise slope the $p$-value of PG~1553+113 rises to $7.5\times10^{-3}$, or about $0.3$ after accounting for the full-sample search \citep{Yang2026}. Our false-alarm probability for PG~1553+113, $4.7\%$ after accounting for 120 light curves (about $1.7\sigma$, one-sided), is reached by an independent, morphology-based route and agrees with this picture. Like the significances of those searches, it is conditional on the stochastic model adopted for the noise. Its modulation is also not a strict sinusoid. A profile analysis of its individual cycles across wavebands finds a broad recurrent envelope whose substructure varies from cycle to cycle, including twin-peak features that keep a binary interpretation viable \citep{Penil2026b}. The ridge measured here follows the period of that envelope; the intra-cycle structure is what the 30-day binning and the Morlet basis suppress. PG~1553+113 stands out within the sample at modest significance, and no gamma-ray QPO candidate is yet individually established.

\subsection{Implications for SMBBH searches}
\label{subsec:smbbh}

This ranking addresses a long-standing issue of SMBBH searches, false positives \citep{Vaughan2003,Ackermann2015}. Previous studies have often mistaken evolving geometric modulation or red-noise fluctuations for binaries \citep[e.g.,][]{Sandrinelli2016,Bhatta2016}. Requiring candidates to satisfy coherence, longevity and rigidity simultaneously, under the worst-case bounds from flux randomization, reduces red-noise mimics to about $1\%$ per light curve, and the drift measurement separates evolving modulation from stable clocks. It does not remove all contamination: red noise is expected to supply about one of the two Rank~I sources, most likely near the period ceiling, and jittered modulation reaches Rank~I in a fraction of the injections (Sect.~\ref{subsec:calibration}).

Because the response of the drift measurement is known, the period stability of the Rank~I sources can be bounded (Table~\ref{tab:stability}). Correcting the worst-case measured drift for the measured attenuation, the intrinsic drift of PG~1553+113 is below $0.010\,\mathrm{yr}^{-1}$ and that of B2~1215+30 below $0.023\,\mathrm{yr}^{-1}$, corresponding to period-stability timescales $\tau_P=|P/\dot{P}|$ of at least 104 and 44~yr. The gravitational-wave decay timescale of a compact SMBBH at the fiducial values of Sect.~\ref{sec:background} is $\tau_{\mathrm{GW}}\simeq7\times10^{3}$~yr (Eq.~\ref{eq:tau_gw}), and our measurements are almost two orders of magnitude short of the precision needed to detect orbital decay. The Rank~I clocks are consistent with binary orbital motion but are not shown to be binaries: any mechanism whose period is stable over a century fits the data equally well, and discriminating among rigid drivers must rely on independent observables. Conversely, the drifts resolved in three Rank~II sources, with period-change timescales of decades, are too fast for gravitational-wave-driven decay and point to geometric evolution such as precession.

Interrupted detection can also be tested outside the morphology. For PKS~2155$-$304, the year-scale modulation fades during a secular decline of the baseline flux, after which the ridge wanders in period and the source exceeds the volatility boundary (Appendix~\ref{sec:appendix_cases}). This is the behavior expected from a stable clock whose amplitude drops below the noise floor in low states, through the rms--flux relation common in accretion flows \citep{Uttley2005}, but also from a transient modulation. PKS~2155$-$304 offers a spectral test. It is one of only two blazars with a significant gamma-ray spectral-hardening event in a systematic search of \textit{Fermi}-LAT flares \citep{Dinesh2025}, and that event is phase-locked to the trough of its $\sim$1.7-yr modulation. \citet{Dominguez2026} interpret it as geometric de-boosting of the soft envelope in a masked, two-component jet. Spectral changes locked to the modulation phase are expected for masking and not for a transient or drifting modulation, and they provide a diagnostic that is independent of the morphological parameters.

The ranking also has implications for multimessenger astrophysics. Candidates in the rigid regime are the most natural targets among our candidates for multiwavelength and pulsar timing array (PTA) follow-up, while remaining consistent with, and not diagnostic of, an SMBBH origin. PTA experiments have recently reported evidence for a nanohertz gravitational-wave background likely dominated by inspiraling SMBBHs \citep[e.g.,][]{Sesana2013,BurkeSpolaor2019,Agazie2024}. Our framework provides a physically motivated pre-selection of candidates for targeted follow-up, with a quantified probability, conditional on the noise model, that each is a red-noise interloper. For a candidate at $P\sim2$~yr, the corresponding gravitational-wave frequency is $f_{\mathrm{GW}}=2/P\sim3\times10^{-8}$~Hz, at the upper end of the PTA band, and the observed period is lengthened relative to the rest frame by $(1+z)$. The Rank~I sources are therefore useful to PTA searches primarily as sky-position and mass priors for continuous-wave searches, not as objects whose orbital periods can be matched directly to a resolved PTA signal.

This approach is also complementary to future space-based gravitational-wave observatories such as the Laser Interferometer Space Antenna (LISA), by identifying long-lived orbital modulators in blazars as potential pre-merger counterparts. The adopted sample selection favors bright and persistently detected blazars, which may overrepresent strongly Doppler-boosted or higher-mass systems and underrepresent fainter populations, motivating extensions of this framework to deeper datasets.

\subsection{Morphological dependence on the central engine and jet structure}
\label{subsec:demographics}

We next ask whether the morphology of the modulation is linked to the accretion mode of the central engine, by comparing the two main blazar subclasses. Flat spectrum radio quasars (FSRQs) have radiatively efficient accretion disks and dense external photon fields, particularly from a prominent broad line region. BL Lacertae objects (BL Lacs) have radiatively inefficient accretion flows and relatively clean circum-jet environments \citep{Urry1995,Ghisellini2005}. Observationally, FSRQs show broad emission lines with rest-frame equivalent widths above 5~\AA, while BL Lacs show weak or absent lines \citep{Urry1995}. The parent sample is defined on the CLASS1 column of 4FGL-DR2 (Sect.~\ref{subsec:blazar_sample}); the subclasses quoted here are those of 4FGL-DR3 \citep{Abdollahi2022}, which for these sources coincide with 4LAC-DR3 \citep{Ajello2022}. The one exception is PKS~0903$-$57, a blazar of uncertain type in both catalogs, which we classify as an FSRQ because its spectrum shows a broad H$\alpha$ line \citep{Goldoni2024}. The redshifts follow Appendix~\ref{sec:appendix_z}. Table~\ref{tab:demographics} summarizes the demographics.

\input{tables/table_demographics}

The two Rank~I sources are BL~Lacs, as are $72\%$ of the selected sample; two sources are too few to test any association. Rank~II contains four BL~Lacs and two FSRQs, PKS~1502+106 and PKS~1509+022, close to the sample proportions. With six members it cannot test whether evolving modulation prefers one accretion mode, a question that a longer baseline and a larger Rank~II population would address.

The subclasses of Rank~III are less evenly populated. Fourteen of the 15 members of III$_{\sigma}$ are BL~Lacs, whereas FSRQs make up a quarter of the rest of the sample; the excess is marginal (Fisher exact test $p=0.060$). Because the volatility of III$_{\sigma}$ is consistent with red noise (Sect.~\ref{subsec:volatility_regime}), this points to a difference in the stochastic variability of the two subclasses, not to plasma-driven modulation preferring BL~Lacs. A ridge that follows red noise wanders more in some light curves than in others, and the controls inherit the noise properties of their templates. Identifying which property is responsible requires comparing the noise spectra of the two subclasses, which we leave to future work. The III$_{P}$ subclass shows the opposite tendency: three of its six members are FSRQs, and five lie at $z\simeq0.7$--$1.2$, as expected if cosmological time dilation pushes distant modulations beyond the period ceiling (below), although six sources do not make the excess significant ($p=0.34$).

The redshift assessment of Appendix~\ref{sec:appendix_z} reveals a large asymmetry between the two subclasses. All FSRQs have redshifts from emission lines, one of them uncertain (PKS~0907$-$023), whereas only 33 of the 70 BL~Lacs have systemic redshifts. Of the other BL~Lacs, 9 are constrained only from below by intervening absorption systems, 12 only from above by the attenuation of their \textit{Fermi}-LAT spectra by the extragalactic background light (EBL; \citealt{Dominguez2024}), and 16 are unconstrained. Two catalog values are superseded outright by later spectroscopy. The $z=0.444$ long quoted for 3C~66A rests on a single tentative Mg\,\textsc{ii} detection that deep Gran Telescopio Canarias spectroscopy excludes; the source is instead placed at $z\simeq0.340$ by its host cluster \citep{Paiano2017,TorresZafra2018}. The $z=0.695$ quoted for PKS~0903$-$57 belongs to a neighboring object, and five narrow lines place the blazar at $z=0.2621$ \citep{Goldoni2024}. Among our Rank~I sources, B2~1215+30 has no confirmed redshift. Any statement about the redshift distribution therefore rests on a subset of the sample, and we quote medians over secure values only: $z=1.00$ for the FSRQs and $z=0.27$ for the BL~Lacs.

\begin{figure}
\centering
\includegraphics[width=\columnwidth]{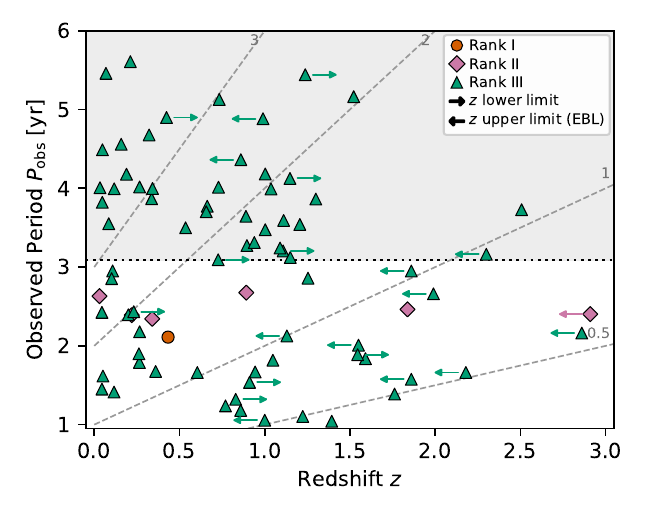}
\caption{Observed period against redshift for the selected candidates: Rank~I (circles), Rank~II (diamonds) and Rank~III (triangles). Right-pointing arrows mark lower limits on $z$ set by intervening absorption systems, and left-pointing arrows upper limits at $84\%$ confidence from extragalactic background light attenuation \citep{Dominguez2024}, plotted at the limit plus its upper uncertainty. Sources with featureless continua and no redshift constraint are omitted. The dotted line and shaded band mark the period ceiling of Eq.~(\ref{eq:pceil}), above which three cone-of-influence-free cycles cannot accumulate on an 18.1-yr baseline; the dashed diagonals are lines of constant rest-frame period, labeled in years. Cosmological time dilation moves a source of fixed rest-frame period up and to the right along these diagonals, so that a modulation classifiable locally crosses the ceiling at sufficient redshift.}
\label{fig:period_redshift}
\end{figure}

A selection effect complicates any subclass comparison, and the cone-of-influence analysis of Sect.~\ref{subsec:coi_ceiling} makes it sharper. BL Lacs dominate the local Universe ($z<0.5$), whereas FSRQs are typically more distant ($z>1.0$; \citealt{Urry1995,Abdollahi2020}). Cosmological time dilation stretches intrinsic year-scale variability and reduces the number of observable cycles. Fig.~\ref{fig:period_redshift} shows this for the candidates with a redshift or a limit, with the ceiling of Eq.~(\ref{eq:pceil}) and lines of constant rest-frame period overlaid. An intrinsic 2-yr modulation crosses the ceiling already at $z\simeq0.55$. By $z=1.5$ it appears at $P_{\mathrm{obs}}=5$~yr, where the attainable persistence is only $p_{\max}=0.21$, and it cannot enter Ranks~I and II whatever its intrinsic behavior. The ceiling therefore affects a substantial fraction of the FSRQs while leaving the nearby BL Lacs largely untouched. PKS~1502+106, the most distant Rank~II source ($z=1.838$), illustrates the point: its observed period of $2.46$~yr corresponds to a rest-frame period of $0.87$~yr.

EBL attenuation adds a second bias \citep[e.g.,][]{SaldanaLopez2021,Finke2022,Banerjee2026,Baxter2026}. The EBL acts as a static screen and does not alter the temporal dynamics, but the loss of high-energy photons lowers the amplitude signal-to-noise ratio and makes the coherence requirement harder to meet for distant sources. High-redshift FSRQs are thus suppressed in Ranks~I and II both by the geometry of the measurement and by EBL attenuation, and any intrinsic subclass trend is convolved with these biases.

Structured-jet models in which a geometrically amplified soft envelope alternately masks and reveals a compact high-energy core \citep{Madero2026,Dominguez2026,Dominguez2026RNAAS} provide a natural physical interpretation of interrupted modulation. With the ridge measured on the detrended light curve, such interruptions appear as volatility, and the morphology alone does not distinguish masking from a transient modulation (Sect.~\ref{subsec:rank3}); spectral diagnostics such as the phase-locked hardening of PKS~2155$-$304 are required.

\subsection{Remaining degeneracies and limitations}
\label{subsec:degeneracies}

The degeneracies discussed here are not resolved by the framework, and they bound the strength of the conclusions drawn above.

Rigid drivers are not uniquely binaries. Steady jet-nozzle or disk precession about a single black hole with a long viscous timescale produces $\varepsilon$ below our stationarity threshold and is morphologically indistinguishable from an SMBBH orbit over 18 years (Sect.~\ref{sec:background}). Rank~I therefore identifies persistent modulations with apparently stable periods, consistent with geometric scenarios such as compact binaries or long-lived precession, but not uniquely identifying them. Breaking this degeneracy requires either a longer baseline, since $\varepsilon_{\mathrm{GW}}\simeq1.4\times10^{-4}\,\mathrm{yr}^{-1}$ at the fiducial values is far below any plausible precessional drift, or independent evidence such as the phase-resolved spectral signatures predicted by the masking scenario.

Interrupted detection has more than one cause. A modulation can disappear from the light curve through geometric masking, through the rms--flux relation during low states, through censoring by the detection threshold, or because it is genuinely transient. With the ridge measured on the detrended light curve all of these appear as volatility or reduced persistence, and the censoring test of Sect.~\ref{subsec:calibration} shows that the detection threshold alone moves a third of stationary modulations out of Rank~I. The masking interpretation of any individual source should be read as a hypothesis consistent with its morphology.

Rank~I is not free of red noise. The power-law controls reach Rank~I at $1.1\%$ per light curve, implying that about one of the two Rank~I sources is expected from red noise, and the contaminating ridges cluster just below the period ceiling, where B2~1215+30 lies. At the source level, PG~1553+113 has a false-alarm probability of $4.7\%$ after trials and B2~1215+30 is consistent with red noise. Membership of the ranks is also sensitive to small changes in the data near the thresholds. Removing the last two monthly bins from every light curve leaves the rank of 106 of the 120 sources unchanged, and both Rank~I sources and all six Rank~II members keep their ranks. Every change occurs at a threshold or in a source whose SSA decomposition has two comparable modes: eleven sources cross the coherence-precision cut between Rank~III and the unselected sample, and three cross the plasma volatility boundary between Rank~II and Rank~III. One further source, 4C~+01.02, is analyzed only with the shorter light curves, where it is in Rank~I: its last five monthly bins are undetected, and with them its terminal gap exceeds the continuity criterion of Sect.~\ref{subsec:lightcurves}. The Rank~II excess is the population statement most affected, nine members against $2.5$ expected ($p<0.001$) with the shorter light curves and six against $2.7$ ($p=0.055$) with the full ones. The classification is equally robust to the detection threshold. Lowering it from TS~$\geq4$ to TS~$\geq2$ admits 181 light curves instead of 120, because fewer bins count as missing. PG~1553+113 and B2~1215+30 remain in Rank~I, joined by 4C~+01.02, whose recent bins are then detected, and all six Rank~II sources remain in Rank~II. The incidence limit tightens to $5.5\%$ at SNR $\geq1$ and $3.4\%$ at SNR $\geq2$. The larger sample contains 12 Rank~II sources against $4.1$ expected from red noise ($p=0.001$), a stronger excess carried by the same core of evolving modulations. In future work, a statistic based on the width of the ridge, defined before looking at the data, could help separate genuine clocks from red-noise ridges.

Volatility is limited by noise. The volatility measurement responds to jitter at the amplitudes the MHD models predict, but red noise produces comparable volatility (Sect.~\ref{subsec:volatility_regime}). Separating plasma-driven from stochastic variability therefore requires either a higher signal-to-noise ratio, as for the brightest flaring sources, or a statistic that distinguishes a jittering oscillation from a wandering noise ridge.

Ranks~I and II carry a hard period ceiling. The requirement of three observed cycles restricts Ranks~I and II to $P<P_{\mathrm{ceil}}\simeq3.1$~yr (Eq.~\ref{eq:pceil}). This is a property of the baseline and of the wavelet kernel, not of the sources: any clock with a longer observed period, including an intrinsically shorter one redshifted into that range, is unclassifiable here. Unlike the persistence ceiling, this one cannot be corrected for; it can only be declared.

The uncertainties are conditioned on the measured period. The flux-randomization intervals of Sect.~\ref{subsec:uncertainties} are computed over realizations that recover a period within $20\%$ of the unperturbed value, and over successful realizations only. The injections processed through the same flux randomization calibrate the classification that results, but the coverage of the individual intervals has not been established.

The red-noise controls assume a single noise family. The controls use lognormal power-law noise with slopes drawn uniformly in $[1,2]$, matched to the mean and excess variance of each template, and the source-specific simulations use each source's fitted slope. A spectral break inside the search band, as found in some bright blazars \citep{Ryan2019}, would raise the false-positive rates and the false-alarm probabilities of individual sources, while tightening the incidence limit. Fitting stochastic models such as continuous-time autoregressive moving-average processes to each light curve, and drawing the controls from the fitted models, would remove this dependence.

One morphology per source. The SSA selection retains only the accepted mode with the largest mean eigenvalue per component. Systems in which a geometric driver coexists with a comparable plasma-driven modulation are characterized by whichever dominates the variance.

\section{Broader applicability of the framework}
\label{sec:applications}

Although this work concerns year-scale modulation in \textit{Fermi}-LAT light curves, the morphological description is largely scale-invariant. It rests on dimensionless diagnostics of signal quality ($Q$ and $\sigma_{\ln P}$) and on fractional measures of temporal evolution ($\varepsilon$ and $\Pi$). Changing the absolute timescale only rescales the time axes: $\varepsilon$ takes whatever units suit the phenomenon, and $Q$, $\sigma_{\ln P}$ and $\Pi$ are unchanged.

One element does not rescale trivially. The cone-of-influence correction of Eq.~(\ref{eq:pmax}) depends on the number of cycles in the record: $\Pi$ must be recomputed with the appropriate $N_{\max}$, and the cycle ceiling of Eq.~(\ref{eq:pceil}) must be re-derived for each application. For records containing few cycles, such as an individual flare, the ceiling is more severe and the corrected statistic becomes essential.

We briefly consider two possible extensions of the framework.

The first is a comparison across wavelength bands. For sources monitored in the optical, X-ray and gamma-ray bands, the morphological parameters can be computed independently in each band. A modulation that is coherent and long-lived in every band would favor a driver common to all emitting regions, as expected for a large-scale geometry. One that is stable in one band and transient in another would instead point to energy-dependent stratification within the jet, for example a stable outer sheath coupled to a turbulent inner spine \citep{Ghisellini2005,Tavecchio2008}.

PG 1553+113 illustrates why such a comparison could be useful. Its gamma-ray modulation is the most coherent in our sample, whereas targeted \textit{Swift}-XRT monitoring finds a much less clear-cut X-ray periodicity \citep{Penil2026Xray}. A difference between wavelength bands would itself be informative, because it could indicate that the modulation is produced differently across the emitting regions. Such a comparison requires light curves of comparable cadence and duration in each band, which at present exist for very few objects.

A second application is to other accreting systems. Because the morphological parameters we discuss here are not specific to blazars, the same analysis could be applied to other sources with quasi-periodic variability, such as X-ray binaries and microquasars \citep{Remillard2006,Belloni2011}. Mapping such populations onto common morphological planes may reveal scaling relations governing the transition from stable, geometry-dominated modulation to turbulent, plasma-driven variability across a wide range of mass and accretion regime \citep[see e.g.,][]{McHardy2006}.

We do not extend the framework here to intra-flare quasi-periodic pulsations, which are reported on timescales of minutes to days \citep[e.g.,][]{Bhatta2016,Ackermann2016,Shukla2020}. Such an application is not simply a change of units: the 30-day binning adopted in Sect.~\ref{subsec:lightcurves} removes that variability by construction, and the attenuation measured in Sect.~\ref{subsec:calibration} was calibrated for year-scale ridges, not for the few-cycle records a single flare provides. Applying these diagnostics at that scale would require its own binning choice, its own injection calibration and its own cone-of-influence ceiling, and we regard it as a separate study and not a corollary of this one.

\section{Summary and conclusions}
\label{sec:summary}

We have developed and calibrated a morphological framework for year-scale QPOs in gamma-ray blazars and applied it to 120 \textit{Fermi}-LAT light curves spanning $\sim$18~yr. SSA selects the oscillatory mode, its period and the long-term trend. WWZ ridge tracking on the detrended light curve then measures four parameters: the coherence $Q$, the drift $\varepsilon$, the volatility $\sigma_{\ln P}$ and the persistence $\Pi$. The classification is calibrated with $10\,000$ red-noise controls and about $20\,000$ injected signals, and the quality selection was fixed on simulations before the data were ranked. Our main conclusions are:

\begin{enumerate}
\item Of the 97 sources that pass the quality selection, two show persistent, coherent and stationary modulation (Rank~I: PG~1553+113 and B2~1215+30). Six show coherent modulation whose period cannot be shown to be stationary (Rank~II: TXS~0141+268, 3C~66A, PKS~0537$-$441, Mkn~421, PKS~1502+106 and PKS~1509+022). The remaining 89 form Rank~III.

\item Red-noise simulations show that false candidates are expected at a low but non-negligible rate. The controls reach Rank~I in 1.1\% and Rank~II in 2.3\% of light curves, corresponding to 1.3 and 2.7 expected false positives among the 120 analyzed sources. We observe 2 Rank~I and 6 Rank~II sources; neither rank individually shows a significant excess over the red-noise expectation ($p=0.38$ and $p=0.055$, respectively). Thus, roughly one of the two observed Rank~I sources is expected from red noise.

\item Injection--recovery tests quantify how efficiently genuine modulations are recovered and whether the classification responds to their temporal evolution. For stationary modulations with periods of 1--3~yr, Rank~I completeness ranges from 63--97\% at amplitude SNR $=1$, depending on period, and reaches 100\% at SNR $=2$. Drifting modulations are classified as Rank~II in 92\% of the trials. Combining the Rank~I recovery efficiency with the observed Rank~I count and the expected red-noise background limits stable, approximately sinusoidal 1--3~yr modulation to at most 7\% of the analyzed sample at SNR $\geq 1$ and 5\% at SNR $\geq 2$ (95\% confidence).

\item The injection tests also quantify how the measured parameters respond
to the underlying variability. On the detrended light curve, an injected
drift is recovered at about 85\% of its true value, while injected
cycle-to-cycle jitter is recovered at about one quarter of its true value.
The calibrated parameters distinguish stable from evolving periods, but
they do not uniquely identify the physical driver; the observed volatility
distribution is also consistent with red noise.

\end{enumerate}

The analysis also identifies four methodological considerations. First, the cone of influence limits both persistence statistics and cycle counts. Persistence can be corrected for this loss, but a minimum-cycle requirement cannot: it imposes a hard maximum period. For our 18.1-yr baseline, requiring three observable cycles gives $P_{\mathrm{ceil}}\simeq3.1$~yr (Eq.~\ref{eq:pceil}). Second, morphology should be measured from the WWZ ridge of the detrended light curve rather than from a smoothed SSA reconstruction. On the reconstruction, the recovered drift and volatility are suppressed by factors of $3.4$ and $44$, even though the same candidate signals are detected. Injection tests are therefore essential for revealing this smoothing bias. Third, each morphological parameter must be calibrated with injections before it is given a physical interpretation. This includes the response to observational selection: in our tests, flux-dependent censoring alone moves about one third of stationary signals out of Rank~I. Fourth, the most robust population result is the upper limit on the incidence of stable modulation. Individual false-alarm probabilities and claims of a population excess depend on the assumed red-noise model. If a more complex noise model produces more Rank I false positives, fewer of the observed candidates can be attributed to genuine stable modulation, and the incidence upper limit becomes tighter. The ranks therefore describe morphology, not statistical significance.

The pipeline favors smooth, sinusoidal modulation, because of the Morlet basis and the 30-day binning (Sect.~\ref{subsec:lightcurves}). Burst-dominated modulation is better searched with templates on finer binning, as in \citet{Maldonado2026}, who show that the recovered significance can depend strongly on the assumed intra-cycle shape. The two approaches are complementary.

PG~1553+113 is the strongest candidate for multiwavelength follow-up and a useful prior for continuous gravitational-wave searches. B2~1215+30, consistent with red noise and close to the period ceiling, is a weaker candidate, and the Rank~II sources are the natural targets for tests of precession. A longer \textit{Fermi}-LAT baseline will raise the period ceiling, make the six ceiling-limited Rank~III sources eligible for a rigidity classification, and tighten the drift limits toward the regime where precession and orbital decay become distinguishable.

Because red noise can mimic periodicity, detecting a candidate modulation is not sufficient to establish a stable QPO. Its strength, persistence, and period stability must also be tested against red-noise expectations. By these criteria, stable year-scale modulation is rare among gamma-ray blazars. Whether any blazar hosts a clock that keeps time over decades remains an open question.

\section*{Data availability}

Data are available in the \textit{Fermi}-LAT Light Curve Repository. The morphological parameters and rank assignments for all 97 selected candidates are provided in Tables~\ref{tab:rank1_sources}, \ref{tab:evolving} and \ref{tab:rank3_sources}.

\section*{Declaration of competing interest}
The authors declare that they have no known competing financial interests or personal relationships that could have appeared to influence the work reported in this paper.


\section*{Acknowledgements}
A. Dinesh acknowledges support from grant PID2022-138132NB-C42 (MCIN/AEI/10.13039/501100011033/FEDER, EU), a UCM-Santander fellowship and UCM FEI-24-0. A. Dom\'inguez acknowledges support from MCIN/AEI under grant PID2025-171816NB-I00. This work has made use of public Fermi data obtained from the High Energy Astrophysics Science Archive Research Center (HEASARC), provided by NASA Goddard Space Flight Center. We specifically acknowledge the Fermi Science Support Center (FSSC) for the provision of the Fermi-LAT Light Curve Repository. This work was also supported by the European Research Council, ERC Starting grant MessMapp, S.B. Principal Investigator, under contract no. 949555, and by the German Science Foundation DFG, research grant “Relativistic Jets in Active Galaxies” (FOR 5195, grant No. 443220636).

\appendix

\section{Technical details of the pipeline}
\label{sec:appendix_tech}

This appendix collects the technical details of the pipeline summarized in Sect.~\ref{sec:methods}: the selection of the oscillatory SSA mode, the tracking of the WWZ ridge, the estimate of the local noise floor, and the conditioning of the flux-randomization intervals.

\subsection{Selection of the oscillatory SSA mode}
\label{subsec:app_ssa}

\begin{figure}
\centering
\includegraphics[width=\columnwidth]{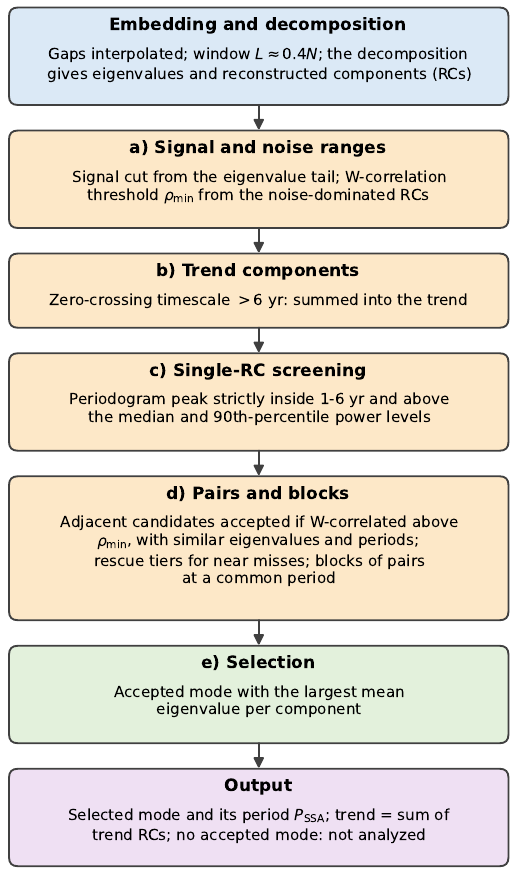}
\caption{Selection of the oscillatory SSA mode (Sect.~\ref{subsec:ssa}). The steps are described in the text of this appendix; a light curve with no accepted mode is not analyzed.}
\label{fig:ssa_flow}
\end{figure}

The window length must be long enough to resolve the structures of interest, short enough for the grouping of components to remain interpretable without overfitting noise, and no longer than half the series \citep{Golyandina2023}. The value $L\approx0.4\,N$ adopted in Sect.~\ref{subsec:ssa} is supported by the controlled tests of \citet{Penil2025Gaps}. Comparing windows of $20$, $30$ and $40$ percent of the series length, they found that longer windows capture larger-scale oscillations better and give higher detection significance for genuine signals when the fraction of missing data is moderate. Shorter windows are preferable only in the heavily gapped regime ($\gtrsim70$ percent), where the lagged embedding vectors contain too few valid entries and the trajectory matrix becomes ill-conditioned. The continuity requirement of Sect.~\ref{subsec:selection_flow} keeps our sample far from that regime. We do not test the phase offset of the conjugate pair (Sect.~\ref{subsec:ssa}) directly. Instead, candidate pairs must be strongly W-correlated, have comparable eigenvalues, and agree in period both individually and in their summed reconstruction. The W-correlation criterion uses $|w_{ij}|$ and therefore accepts both positively and negatively correlated pairs. These conditions are consistent with an oscillatory SSA pair but do not by themselves establish phase quadrature.
\begin{enumerate}
    \item 
We first estimate the number of signal-dominated reconstructed
components from the SSA eigenvalue spectrum. The final 20\% of the eigenvalue sequence, with a minimum of five components, is used as a proxy for the noise-dominated tail. We define the corresponding noise threshold as
\begin{equation}
\lambda_{\rm noise}=\mu_{\rm tail}+3\sigma_{\rm tail},
\end{equation}
where $\mu_{\rm tail}$ and $\sigma_{\rm tail}$ are the mean and standard deviation of the tail eigenvalues. The preliminary signal range extends
to the last RC whose eigenvalue exceeds this threshold, excluding RC0 from determining the cutoff. To prevent pathological under- or over-selection in particularly noisy decompositions, the final cutoff is constrained to lie between RC8 and RC20, subject to the number of available components. Components beyond this cutoff are treated as noise dominated and are used to calibrate the correlation threshold.

\item An adaptive W-correlation threshold is then derived from the noise-dominated RCs. We compute the distribution of the absolute off-diagonal W-correlations, $|W_{ij}|$, among these components and set
\begin{equation}
\rho_{\min}={\rm clip}
\left[P_{95}\left(|W_{ij}|\right),\,0.5,\,0.8
\right],
\end{equation}
where $P_{95}$ denotes the 95th percentile. RC0 is excluded from this noise estimate, and at most the 30 highest-index noise components are used. If the threshold cannot be estimated because too few noise-dominated components are available, we adopt the conservative fallback value $\rho_{\min}=0.8$.

\item Before searching for oscillatory modes, we identify slowly varying baseline components using their zero-crossing timescales. For each of the first eleven reconstructed components, RC0--RC10, we estimate
\begin{equation}
T_{{\rm RC},k} = \frac{2\,(t_{{\rm last},k}-t_{{\rm first},k})}{N_{{\rm cross},k}-1},
\end{equation}
where $N_{{\rm cross},k}$ is the number of sign changes of the median-centered reconstruction and $t_{{\rm first},k}$ and $t_{{\rm last},k}$ are the interpolated times of the first and last of them, that is, twice the mean interval between crossings. Unlike the simpler estimate $2T_{\rm obs}/N_{{\rm cross},k}$, this timescale varies continuously with the data, and a small change of the baseline cannot by itself move a component across the trend threshold. Components satisfying
$T_{{\rm RC},k}>P_{\max}$ are treated as baseline-trend candidates, where $P_{\max}=6$ yr is the upper boundary of the QPO search. This test is applied independently to each component, and the resulting trend set may be empty or non-contiguous and RC0 is not automatically classified as the trend. 
Although RC0 is not forced to be a trend component, it is excluded from the normal oscillatory scan because its usually dominant eigenvalue could otherwise control the variance-based selection. A dedicated rescue test, described below, allows RC0 to enter only when it forms a valid in-band pair with RC1 or RC2.

\item The remaining signal-range components are subjected to an initial single-RC Lomb--Scargle screening. For each RC we calculate a periodogram over $1\leq P\leq6$ yr. Its dominant peak is the highest local maximum strictly inside the period grid; a maximum on the edge of the grid only marks power rising toward the boundary and is not counted as a peak. The RC is retained as a provisional candidate only when this peak satisfies all of the following:
\begin{equation}
1.05~{\rm yr} < P_{\rm peak}<5.95~{\rm yr},
\end{equation}
\begin{equation}
\frac{S_{\rm peak}}{{\rm median}(S_{\rm LS})}>3,
\qquad
\frac{S_{\rm peak}}{P_{90}(S_{\rm LS})}>1.5.
\end{equation}
Here $S_{\rm LS}$ is the Lomb--Scargle power distribution of the component and $P_{90}$ is its 90th percentile. These ratios are heuristic screening quantities, not false-alarm probabilities.

\item The provisional single-RC candidates are sorted by RC index, and adjacent candidates in this ordered list are tested as possible
oscillatory pairs. Restricting the normal pair search to neighboring candidates reflects the tendency of SSA conjugate components to occupy
nearby eigenvalue ranks and reduces the number of pair trials. Additional nearby combinations can enter through the rescue procedures described below.

\item A normal pair $(i,j)$ is accepted when it satisfies the following conditions:
\begin{enumerate}
\item its absolute W-correlation exceeds the adaptive threshold,
      $|W_{ij}|\geq\rho_{\min}$;
\item its eigenvalues are sufficiently similar,
\begin{equation}
\frac{\min(\lambda_i,\lambda_j)}
     {\max(\lambda_i,\lambda_j)}
\geq0.2;
\end{equation}
\item the individual Lomb--Scargle peak periods agree within
\begin{equation}
\frac{|P_i-P_j|}
     {(P_i+P_j)/2}
\leq0.5;
\end{equation}
\item the Lomb--Scargle period $P_{ij}$ of the summed pair
reconstruction is consistent with the mean component period,
\begin{equation}
\frac{|P_{ij}-(P_i+P_j)/2|}
     {(P_i+P_j)/2}
\leq0.5;
\end{equation}
\item the summed reconstruction has a peak-to-median periodogram ratio greater than 2.
\end{enumerate}
The reconstructed pair peak is also required to remain at least 0.05 yr from either boundary of the period-search interval. Accepted normal pairs are retained as oscillatory candidates. The implementation also contains a conservative test for multi-component blocks, requiring every pair of components within the proposed block to
satisfy the pair-consistency conditions.

\item To reduce incompleteness when one member of an otherwise plausible SSA pair fails the initial single-component screening, we apply a restricted pair-rescue procedure to unpaired candidate RCs. For each unpaired component, neighboring RCs within two component indices are examined.

The first rescue tier retains the W-correlation requirement but relaxes
the eigenvalue-similarity threshold to
\begin{equation}
\frac{\min(\lambda_i,\lambda_j)}
     {\max(\lambda_i,\lambda_j)}
\geq0.10.
\end{equation}
The member-period and reconstructed-pair period-consistency conditions are retained, while no minimum pair peak-to-background ratio is required at this stage.

If this rescue fails only because the W-correlation lies below $\rho_{\min}$, a second, harmonic-companion rescue is attempted. In this tier the W-correlation gate is omitted, but the normal eigenvalue and
period-consistency thresholds are restored and the joint pair peak-to-background ratio must exceed 2. This route is intended to
recover quasi-conjugate pairs whose temporal correlation has been reduced by noise, interpolation, or period evolution. Because this tier does not require $|W_{ij}|\geq\rho_{\min}$, the final pair sample is not restricted exclusively to highly W-correlated components. A separate rescue is performed for oscillations divided between RC0 and either RC1 or RC2. RC0 is admitted only when it is not a baseline-trend candidate, its individual periodogram has a peak inside the search interval, and the resulting pair passes either the standard or harmonic-companion rescue criteria. The tier without the W-correlation requirement is not applied to RC0. This protects against the automatic selection of a slowly varying RC0 while allowing
the recovery of cases in which the dominant oscillatory variance is carried partly by the first component.
\item All accepted normal and rescued pairs are combined into a common candidate pool, from which we select the non-single candidate with the largest mean SSA eigenvalue per component,
\begin{equation}
\Lambda_{\rm group} = \frac{1}{n_{\rm group}}\sum_{k\in{\rm group}}\lambda_k,
\end{equation}
where $n_{\rm group}$ is the number of components in the candidate. Normalizing by the number of components prevents a block of several weaker components, such as a modulation spread over two adjacent pairs, from being preferred to a single, more concentrated pair merely because it contains more components.
The period of the corresponding summed RC reconstruction initializes the WWZ ridge analysis, and the trend components define the baseline subtracted from the light curve before it. The selector returns at most one oscillatory mode per source.

\end{enumerate}

\begin{figure}[t!]
   \centering
   \includegraphics[width=\columnwidth]{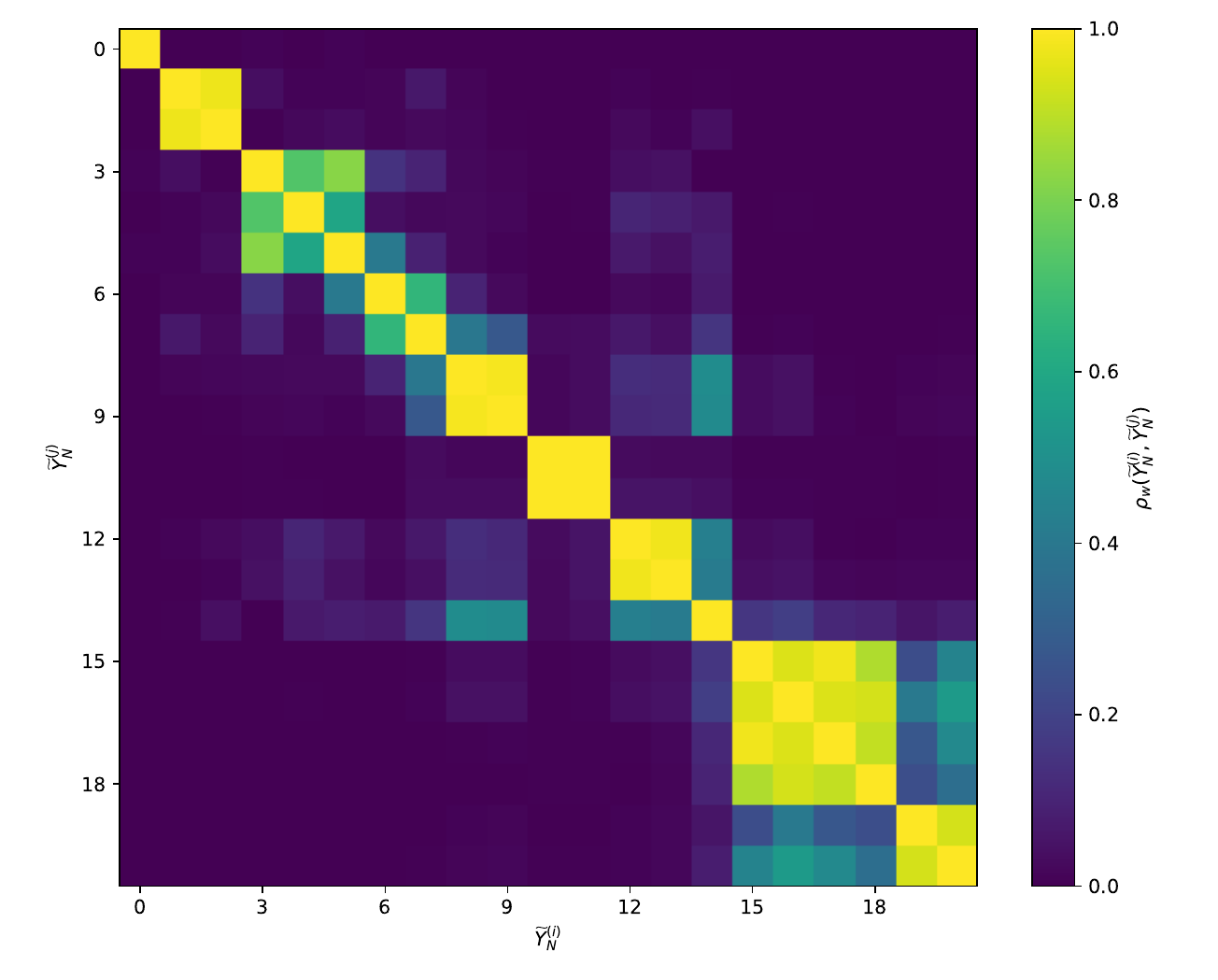}
   \caption{W-correlation matrix of the SSA decomposition of PG 1553+113 (4FGL J1555.7+1111), the worked example used throughout this paper. Each cell gives the weighted correlation $w_{ij}$ between reconstructed components $i$ and $j$; the diagonal is unity by construction. A well-resolved oscillation appears as a $2\times2$ block of strongly correlated off-diagonal cells, isolated from the trend and from the noise floor. Here RC~0 carries the secular trend, RC~1 and RC~2 form the pair selected as the oscillatory mode, and the components above them mix with one another and constitute the noise floor. This is the diagnostic on which the pairwise selection described in this section operates; it is shown once, for this source, and is not repeated for the remaining candidates.}
   \label{fig:wcorr}
\end{figure}

Fig.~\ref{fig:wcorr} illustrates the structure that this procedure is designed to find, using PG 1553+113, and Fig.~\ref{fig:SSA} shows the resulting decomposition of its light curve. The block-diagonal appearance of the matrix is the visual counterpart of the criteria above. Components of a common oscillatory mode are mutually correlated at $w_{ij}\ge\rho_{\min}$ and decorrelated from the trend and from the higher-index components. The pair can therefore be extracted without dragging in either the baseline or the stochastic flickering. Where no such block is present, no pair is accepted and the source does not reach the ridge stage. We show this diagnostic for a single source only; for the remaining candidates the outcome of the decomposition is summarized by the reconstructed light curves and ridge maps of Appendices~\ref{sec:appendix_cases} and \ref{sec:appendix_atlas}.

\subsection{Ridge tracking}
\label{subsec:app_ridge}

The ridge-tracking procedure proceeds as follows:
\begin{enumerate}
\item We start the ridge at the period of the selected SSA mode, $P_{\mathrm{global}}=P_{\mathrm{SSA}}$, within the period range defined above.
\item At each epoch, the raw ridge is the maximum of the WWZ power within a fixed window of $\pm40\%$ around $P_{\mathrm{global}}$. The window is the only constraint on the track. It prevents the ridge from locking onto noise features at distant periods, but within it the ridge follows whichever feature is strongest at each epoch, and it can move between features from one epoch to the next.
\item The raw track is smoothed with a running median of width five time steps, which suppresses pixel-scale quantization noise and is more robust against outliers than a Gaussian kernel. The smoothed ridge is the one shown in the diagnostic plots.
\end{enumerate}

Ridge points are then classified as active or not (Sect.~\ref{subsec:parameter_extraction}). When the active points form more than one contiguous run in time, the longest run is retained; if two runs are of equal length, the one carrying the larger integrated WWZ power is selected. Only the retained run enters the morphological parameters. Within it, the period can change abruptly where the ridge moves between features, and such changes are measured as volatility. PKS~2155$-$304 (Appendix~\ref{sec:appendix_cases}) is an example: its retained run spans 2011--2024, and the move of the ridge from about $1.7$~yr to features near $1.0$ and $1.4$~yr after 2018 produces $\sigma_{\ln P}=0.17$.

The window and the median filter stabilize the drift estimate, but they also suppress genuine high-frequency period fluctuations. The volatility parameter of Sect.~\ref{subsec:parameter_extraction} must therefore be read as a measure of the residual jitter that survives this smoothing, whose attenuation is quantified in Sect.~\ref{subsec:calibration}.

\subsection{Local noise floor}
\label{subsec:app_noise}

To avoid model-dependent fitting of a global red-noise continuum or the use of Monte Carlo simulations, the background power level $S_{\mathrm{noise}}$ is estimated empirically from the data, using the detrended light curve $F_{\mathrm{det}}$ defined in Sect.~\ref{subsec:wwz}. At each time step where the ridge is identified we form up to four independent estimates of the local background from the corresponding column of the WWZ map, always excluding the COI:
\begin{itemize}
    \item $\hat{S}_{\mathrm{side}}$, the median power in frequency sidebands adjacent to the ridge, in the ranges $0.8f$--$0.9f$ and $1.1f$--$1.2f$, offsets of about $\pm10\%$ from the instantaneous ridge frequency $f$;
    \item $\hat{S}_{\mathrm{broad}}$, the median power over the broad continuum, all periods with $|\ln(P/P_{\mathrm{ridge}})| > 0.15$;
    \item that broad-continuum median raised by twice its median absolute deviation (MAD);
    \item $\hat{S}_{25}$, the 25th percentile of the broad-continuum power.
\end{itemize} We adopt
\begin{equation}
S_{\mathrm{noise}} = \max\left\{ \hat{S}_{\mathrm{side}},\; \hat{S}_{\mathrm{broad}},\; \hat{S}_{\mathrm{broad}} + 2\,\mathrm{MAD},\; \hat{S}_{25} \right\},
\label{eq:snoise}
\end{equation}
that is, the most pessimistic of the four noise estimates. Taking the maximum instead of an average makes the reported coherence conservative (Sect.~\ref{subsec:parameter_extraction}), in an operational sense: this prescription systematically returns a smaller $Q$ than the alternative background prescriptions would, but it is not a statistical confidence bound. The broad-continuum estimates also ensure that $S_{\mathrm{noise}}$ remains defined at long periods, where the narrow sidebands can fall entirely inside the COI and estimate (i) becomes unavailable.

\subsection{Conditioning of the flux-randomization intervals}
\label{subsec:app_fr}

Two properties of this design matter, because the resulting uncertainties feed directly into the quality selection and the classification bounds. First, requiring the recovered period to lie within $20\%$ of the unperturbed value conditions the distribution on the original solution and suppresses realizations in which the tracker locks onto a different mode. Second, the intervals are computed over successful realizations only, and the recovery fraction depends on morphology, from $978/1000$ in the median to $430/1000$ for the least stable candidate.

Both effects can narrow the reported intervals relative to an unconditioned resampling, and most for the least stable candidates. The intervals should therefore be read as the stability of a given solution under measurement noise, not as marginal posteriors on the morphological parameters. A coverage test against injections, repeating the selection without the period condition, would show how much the conditioning matters. We have not performed it here and list it in Sect.~\ref{subsec:degeneracies} as a needed development.

\section{Case studies at the classification boundary}
\label{sec:appendix_cases}

The Rank~I sources are presented in Figs.~\ref{fig:SSA}, \ref{fig:WWZ} and \ref{fig:b2}, and the Rank~II sources in Appendix~\ref{sec:appendix_atlas}. Here we examine four sources of Rank~III, one for each subclass, chosen because each isolates one way in which a modulation fails to reach the coherent ranks. Each figure shows the SSA decomposition of the light curve (left) and the WWZ map of the detrended light curve (right), with the cone of influence in gray and the active ridge points as red crosses.

\newcommand{\srcpanel}[4]{%
  \begin{figure*}
    \centering
    \includegraphics[width=0.49\textwidth]{Figures/#1_comps_#3_ssa_lc.pdf}%
    \includegraphics[width=0.49\textwidth]{Figures/#1_comps_#3_wwz_ridge_fit.pdf}
    \caption{#2. Left: SSA decomposition of the 30-day binned \textit{Fermi}-LAT light curve, showing the data, the secular trend, the oscillatory component selected by SSA, their sum and the residual. Right: WWZ map of the detrended light curve, with the cone of influence in gray and the active ridge points, from which the morphological parameters are computed, as red crosses. #4}
    \label{fig:panel_#1}
  \end{figure*}}

\subsection{A strong modulation that fades: PKS 2155$-$304 (III$_{\sigma}$)}
\label{subsec:app_volatility}

PKS~2155$-$304 (4FGL~J2158.8$-$3013, BL~Lac, $z=0.116$) carries the most coherent ridge in the sample ($Q=15.6$). SSA selects a pair (RC~1 and RC~2) at $1.63$~yr, and a modulation at about $1.7$~yr dominates the light curve from 2010 to 2018 (Fig.~\ref{fig:panel_4FGL+J2158.8-3013}). It then fades, together with a secular decline of the baseline flux, and in the second half of the record the ridge moves between weaker features near $1.0$ and $1.4$~yr. On the light curve this is measured as volatility, $\sigma_{\ln P}=0.17$, above the plasma boundary. The morphology alone cannot say whether a stable clock has dropped below the noise floor in a low state or a transient modulation has ended; the phase-locked spectral hardening discussed in Sect.~\ref{subsec:smbbh} is the kind of evidence that can.

\srcpanel{4FGL+J2158.8-3013}{PKS~2155$-$304 (4FGL~J2158.8$-$3013), Rank~III$_{\sigma}$}{1_2}{The $\sim$1.7-yr modulation dominates until 2018; afterwards the ridge follows weaker features at shorter periods.}

\subsection{A coherent but intermittent modulation: 4C +47.08 (III$_{\Pi}$)}
\label{subsec:app_persistence}

4C~+47.08 (4FGL~J0303.6+4716, BL~Lac, $z<0.63$ from EBL attenuation) shows a coherent modulation at $P=1.06$~yr ($Q=9.0$, $N_{\mathrm{active}}=5.8$; Fig.~\ref{fig:panel_4FGL+J0303.6+4716}), one of the more coherent ridges in Rank~III. The modulation is present in the light curve over only $\Pi=0.40$ of the observable baseline, and its coherence is not robust under flux randomization ($\Delta Q=5.1$). It is the morphology expected from a transient quasi-periodic episode, or from a clock detected only in high states; without independent information the two cannot be distinguished, and the source falls in III$_{\Pi}$.

\srcpanel{4FGL+J0303.6+4716}{4C~+47.08 (4FGL~J0303.6+4716), Rank~III$_{\Pi}$}{3_4}{The $\sim$1-yr modulation is detectable over only part of the baseline.}

\subsection{A stable modulation beyond the ceiling: S5 1044+71 (III$_{P}$)}
\label{subsec:app_ceiling}

S5~1044+71 (4FGL~J1048.4+7143, FSRQ, $z=1.15$) is one of the two sources above $3\sigma$ in the reanalysis of \citet{Penil2026}. Its modulation, at $P=3.12$~yr, is as stable as any in the sample: the drift is $\varepsilon=0.0015\pm0.0010\,\mathrm{yr}^{-1}$, the volatility $\sigma_{\ln P}=0.010$, and the modulation is present over $\Pi=0.97$ of the observable baseline (Fig.~\ref{fig:panel_4FGL+J1048.4+7143}). Its period lies $0.03$~yr above the ceiling, and the COI-free span contains only $N_{\mathrm{active}}=2.8$ cycles: the source satisfies the coherence, rigidity, and persistence criteria, but the COI-free baseline contains fewer than three cycles at its period. It is the clearest example of a candidate that a longer baseline would admit to the rigidity classification. A multiwavelength study confirms its $\sim$3-yr gamma-ray periodicity after a further cycle and finds all the available data consistent with a precessing jet \citep{Cerruti2027}. Our measurements are compatible with that picture: a jet precessing steadily, with a long viscous timescale, produces exactly a stable, persistent modulation, although over 18~yr such a modulation cannot be distinguished from an orbital one (Sect.~\ref{subsec:degeneracies}).

\srcpanel{4FGL+J1048.4+7143}{S5~1044+71 (4FGL~J1048.4+7143), Rank~III$_{P}$}{1_2}{A stable $3.1$-yr modulation present over the whole observable baseline, just beyond the period ceiling.}

\subsection{A long, weak modulation: Mkn 501 (III$_{Q}$)}
\label{subsec:app_coherence}

Mkn~501 (4FGL~J1653.8+3945, BL~Lac, $z=0.0336$) is, with Mkn~421, one of the two nearest bright gamma-ray blazars and among the sources most intensively monitored by imaging atmospheric Cherenkov telescopes, which makes the behavior of its year-scale variability of particular interest. Its light curve carries a modulation at $P=4.01$~yr that is present over the entire observable baseline ($\Pi=1.02$) with a stable period ($|\varepsilon|=0.003\,\mathrm{yr}^{-1}$, $\sigma_{\ln P}=0.005$; Fig.~\ref{fig:panel_4FGL+J1653.8+3945}). It fails the coherence gate ($Q=2.3$), but it could not reach the coherent ranks in any case: at this period the COI-free span contains fewer than two cycles ($N_{\mathrm{active}}=1.7$, the maximum the span allows). A 25-yr baseline, with $P_{\mathrm{ceil}}\simeq4.3$~yr, would bring its period below the ceiling; it would also have to pass the coherence gate, as a persistent modulation does as its cycles accumulate.

\srcpanel{4FGL+J1653.8+3945}{Mkn~501 (4FGL~J1653.8+3945), Rank~III$_{Q}$}{3_4}{The modulation is present over the entire cone-of-influence-free span, which at $P=4.01$~yr contains fewer than two cycles.}

\section{The rigidity parameter $k$}
\label{sec:appendix_sieve}

A third construction combines a rigidity parameter with the volatility. The rigidity parameter is defined as
\begin{equation}
k = \frac{N_{\mathrm{active}}}{\max\left(|\varepsilon|,\,\varepsilon_{\mathrm{floor}}\right)},
\label{eq:k}
\end{equation}
with a fixed floor $\varepsilon_{\mathrm{floor}}=10^{-3}\,\mathrm{yr}^{-1}$ imposed because the ratio diverges as $\varepsilon\to0$. The intent was that $k$ should measure the number of cycles over which stability has been demonstrated, with rigid, long-lived clocks separating toward large $k$ and low $\sigma_{\ln P}$.

Applied to the present sample, this diagnostic adds nothing to the parameters it combines. The floor is rarely active: $|\varepsilon|>\varepsilon_{\mathrm{floor}}$ for 95 of 97 sources. The value of $k$ is set mainly by the drift: $\log k$ correlates with $\log|\varepsilon|$ at Spearman $\rho=-0.75$, more strongly than with the quantity it was meant to track ($\rho=+0.52$ with $\log N_{\mathrm{active}}$). Because the drift is measured on the detrended light curve, $k$ is essentially an inverse drift scaled by the cycle count, and it carries no information that the ranks do not already use. Rank~II, defined by its drift bound, has a median $k$ of $188$, close to that of Rank~III ($206$), and the two Rank~I sources ($610$ in the median) are too few for any comparison to have statistical power. Replacing the fixed floor by the per-source uncertainty, $k=N_{\mathrm{active}}/\max(|\varepsilon|,\Delta\varepsilon)$ or $k=N_{\mathrm{active}}/(|\varepsilon|+\Delta\varepsilon)$, does not change this. The physical content of $k$ is better expressed directly as a limit on drift, equivalently on the period-stability timescale (Table~\ref{tab:stability}), which the measured response of the drift measurement makes possible. We retain the definition of Eq.~(\ref{eq:k}) for reference, but do not use it to rank or compare candidates.

\section{Atlas of Rank II}
\label{sec:appendix_atlas}

\begin{figure*}
  \centering
  \includegraphics[width=0.49\textwidth]{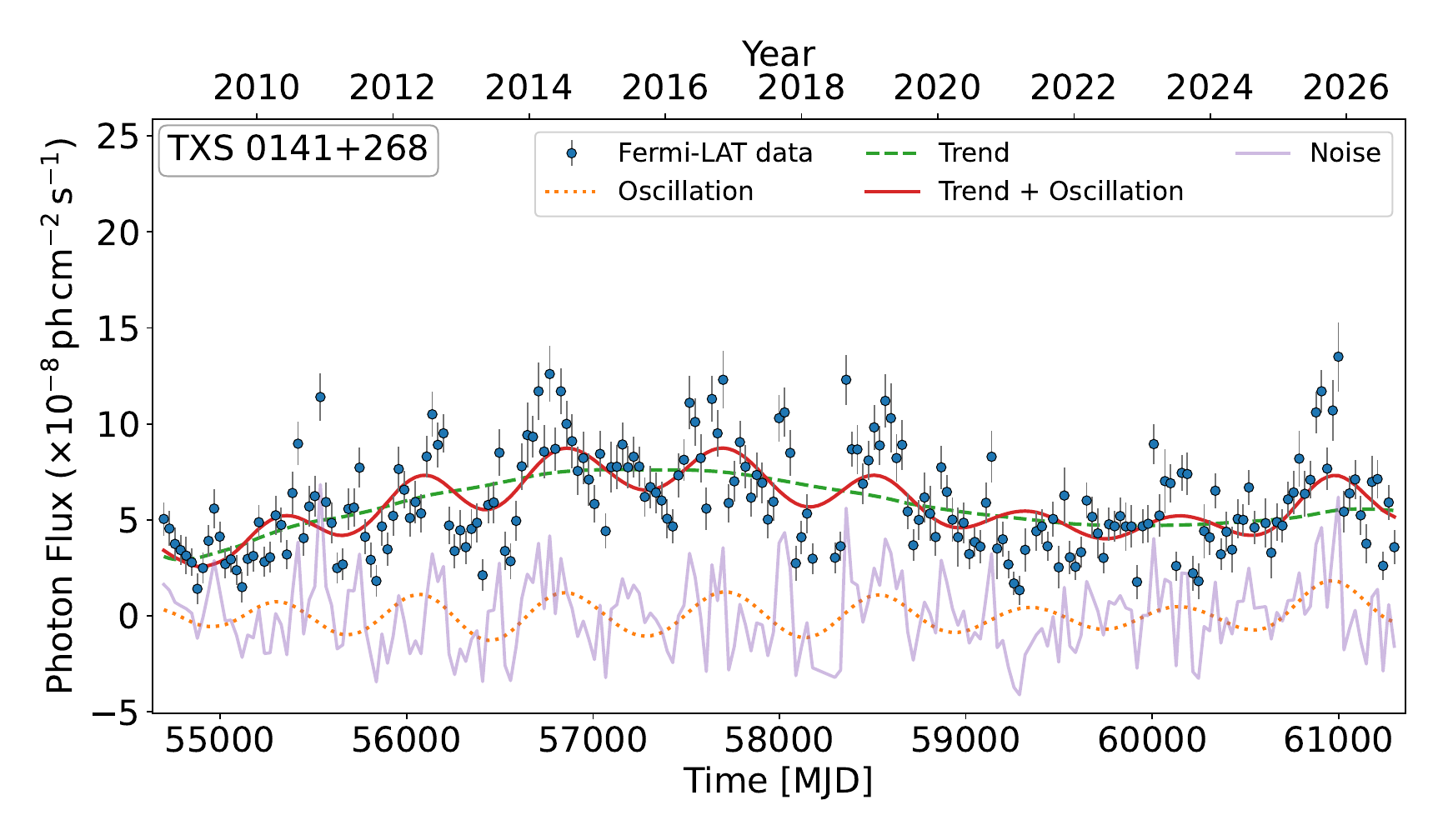}%
  \includegraphics[width=0.49\textwidth]{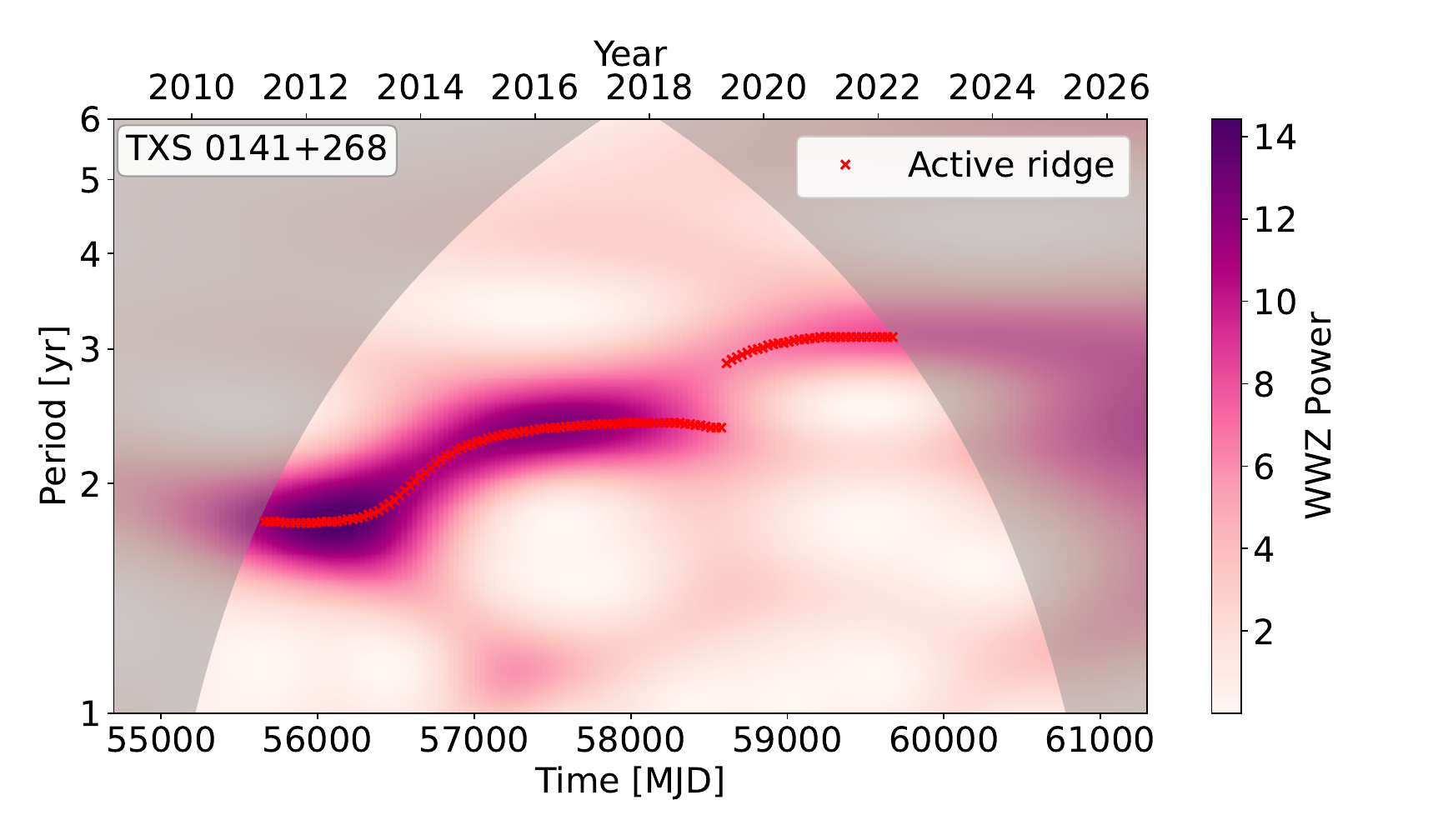}\\[2pt]
  \includegraphics[width=0.49\textwidth]{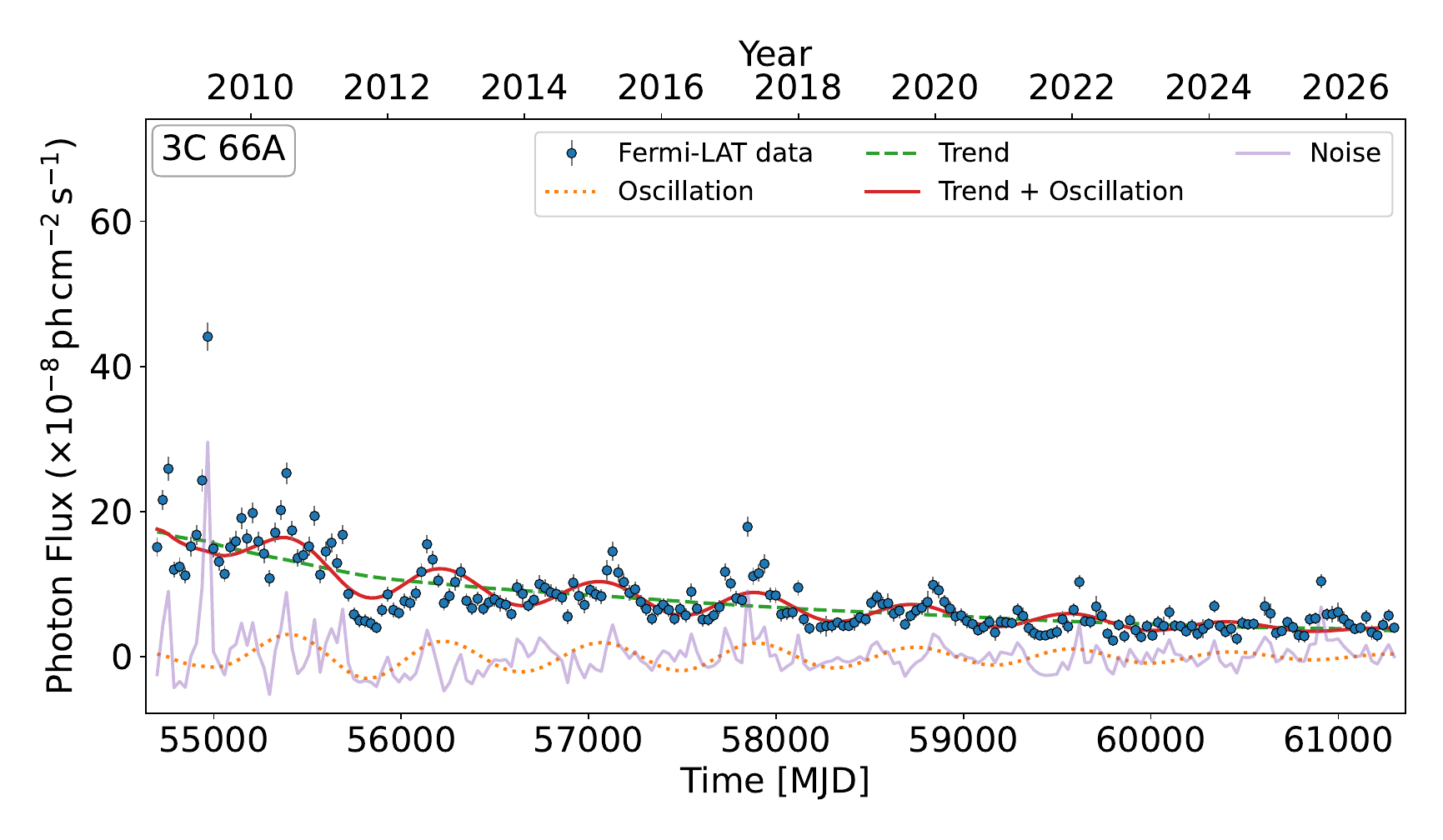}%
  \includegraphics[width=0.49\textwidth]{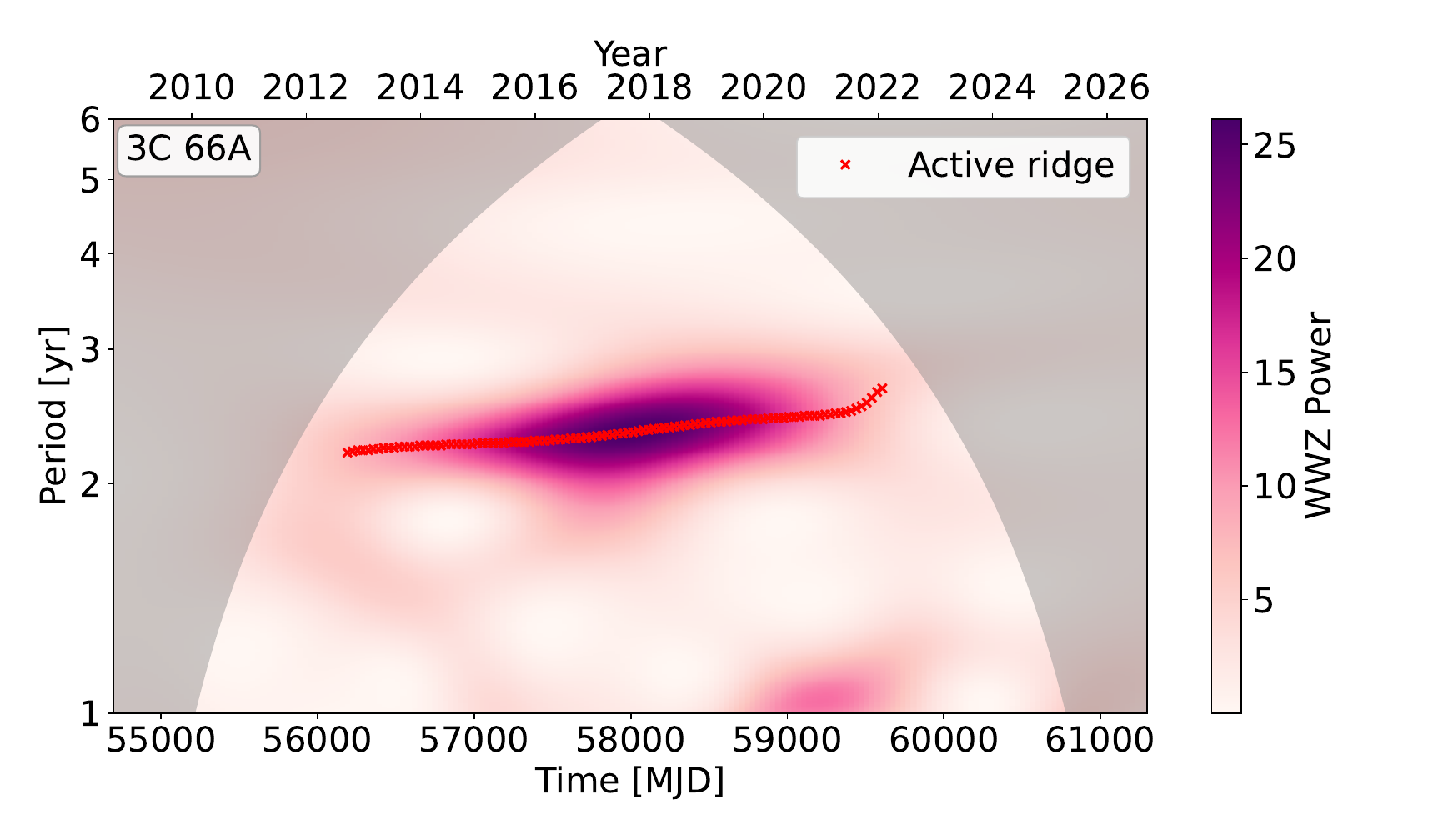}\\[2pt]
  \includegraphics[width=0.49\textwidth]{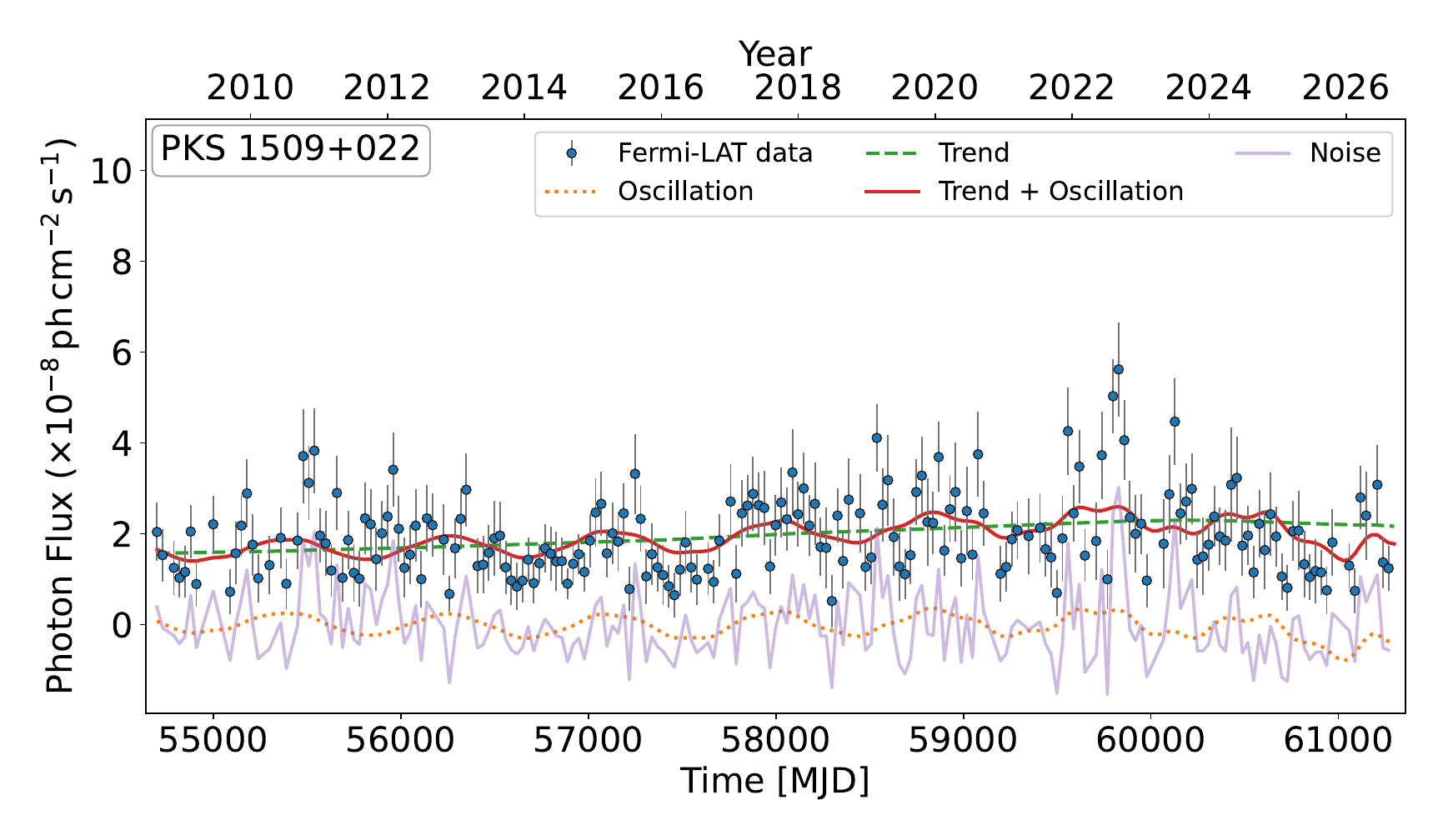}%
  \includegraphics[width=0.49\textwidth]{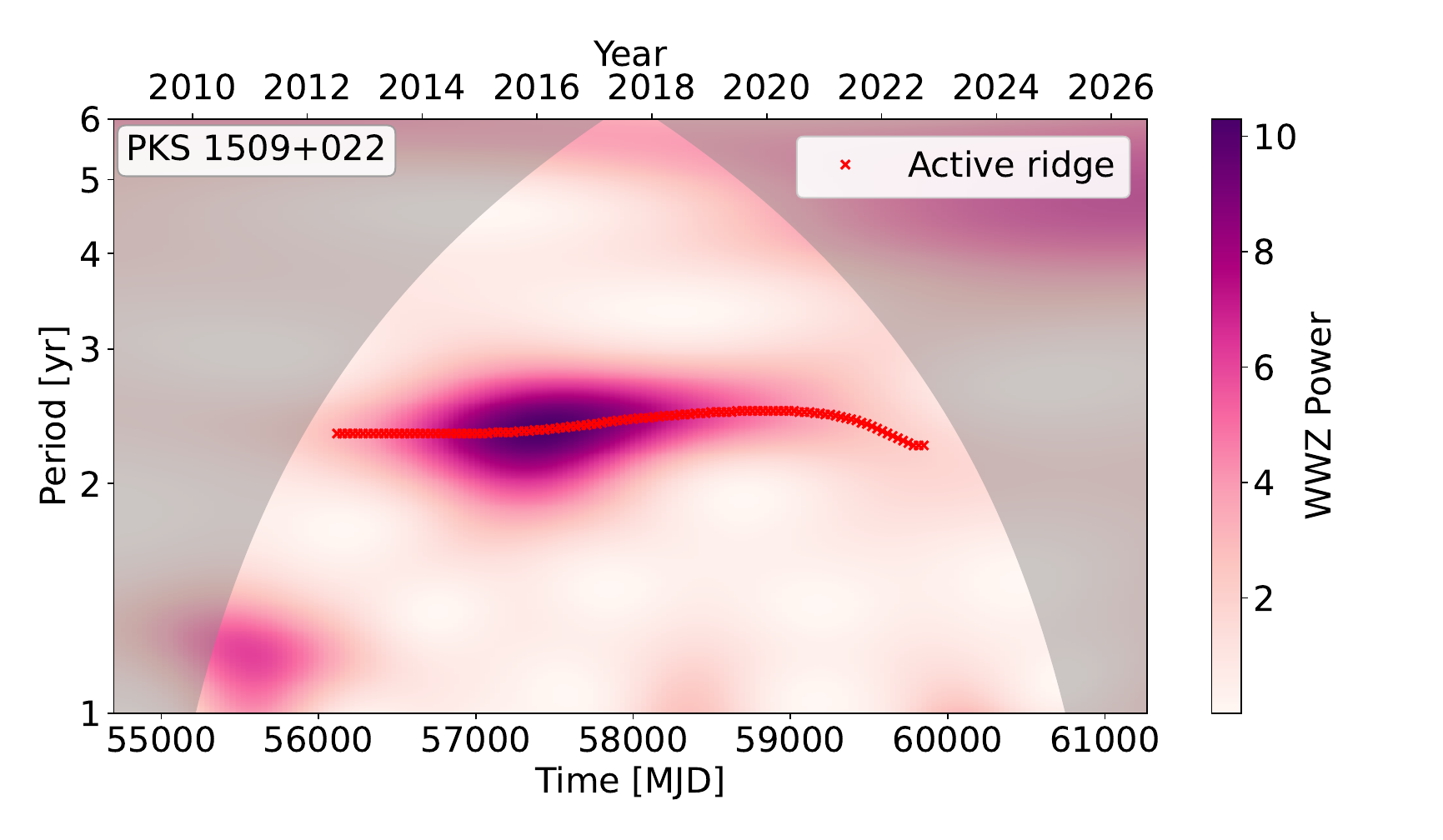}
  \caption{Rank~II sources, from top to bottom TXS~0141+268 (4FGL~J0144.6+2705; $P_{\mathrm{obs}}=2.40$~yr, $\varepsilon=+0.045\,\mathrm{yr}^{-1}$), 3C~66A (4FGL~J0222.6+4302; $2.34$~yr, $+0.017\,\mathrm{yr}^{-1}$) and PKS~1509+022 (4FGL~J1512.2+0202; $2.39$~yr, classified through its volatility). Left: SSA decomposition of the 30-day binned \textit{Fermi}-LAT light curve. Right: WWZ map of the detrended light curve, with the cone of influence in gray and the active ridge points as red crosses.}
  \label{fig:atlas}
\end{figure*}

\begin{figure*}
  \centering
  \includegraphics[width=0.49\textwidth]{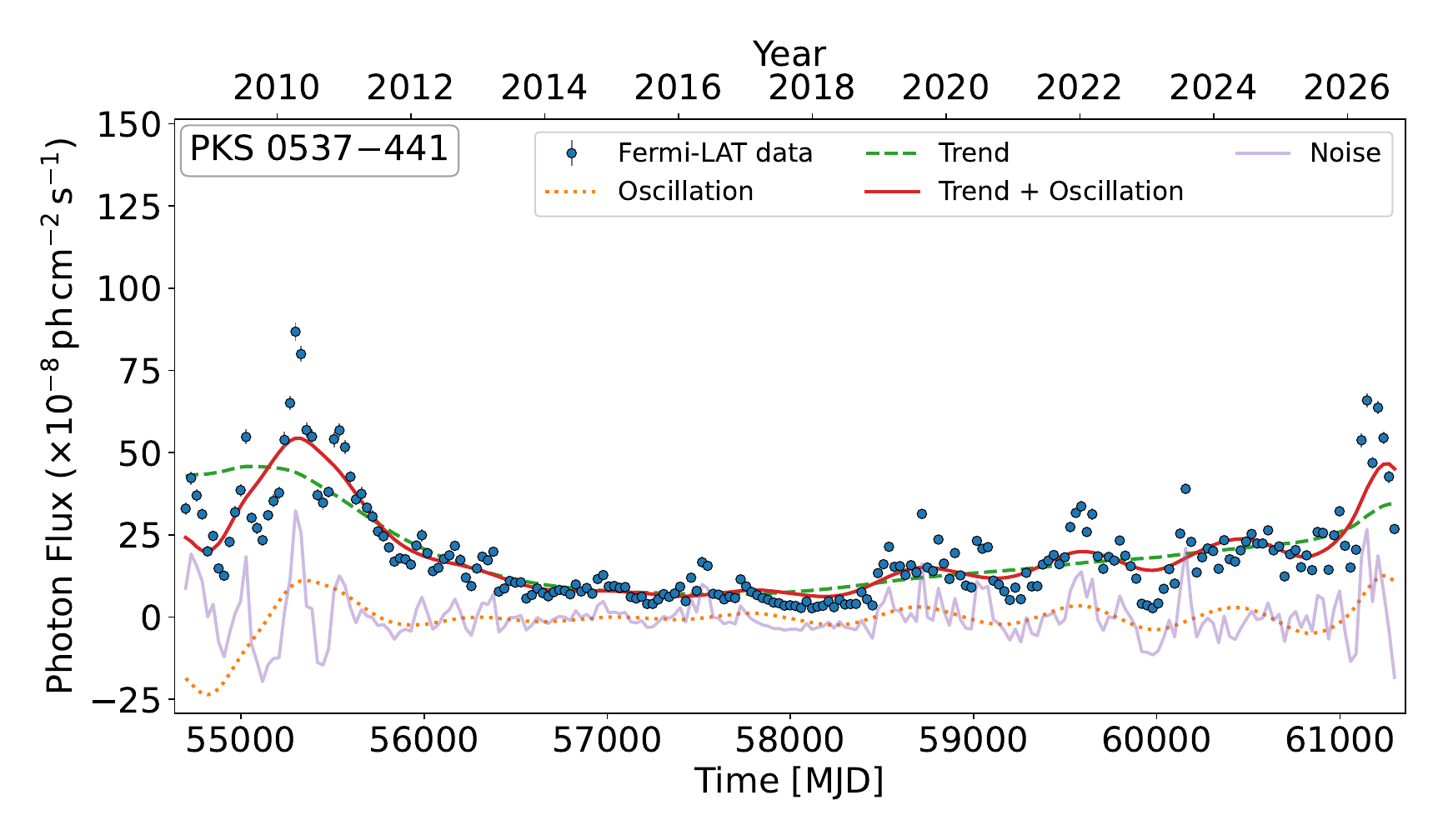}%
  \includegraphics[width=0.49\textwidth]{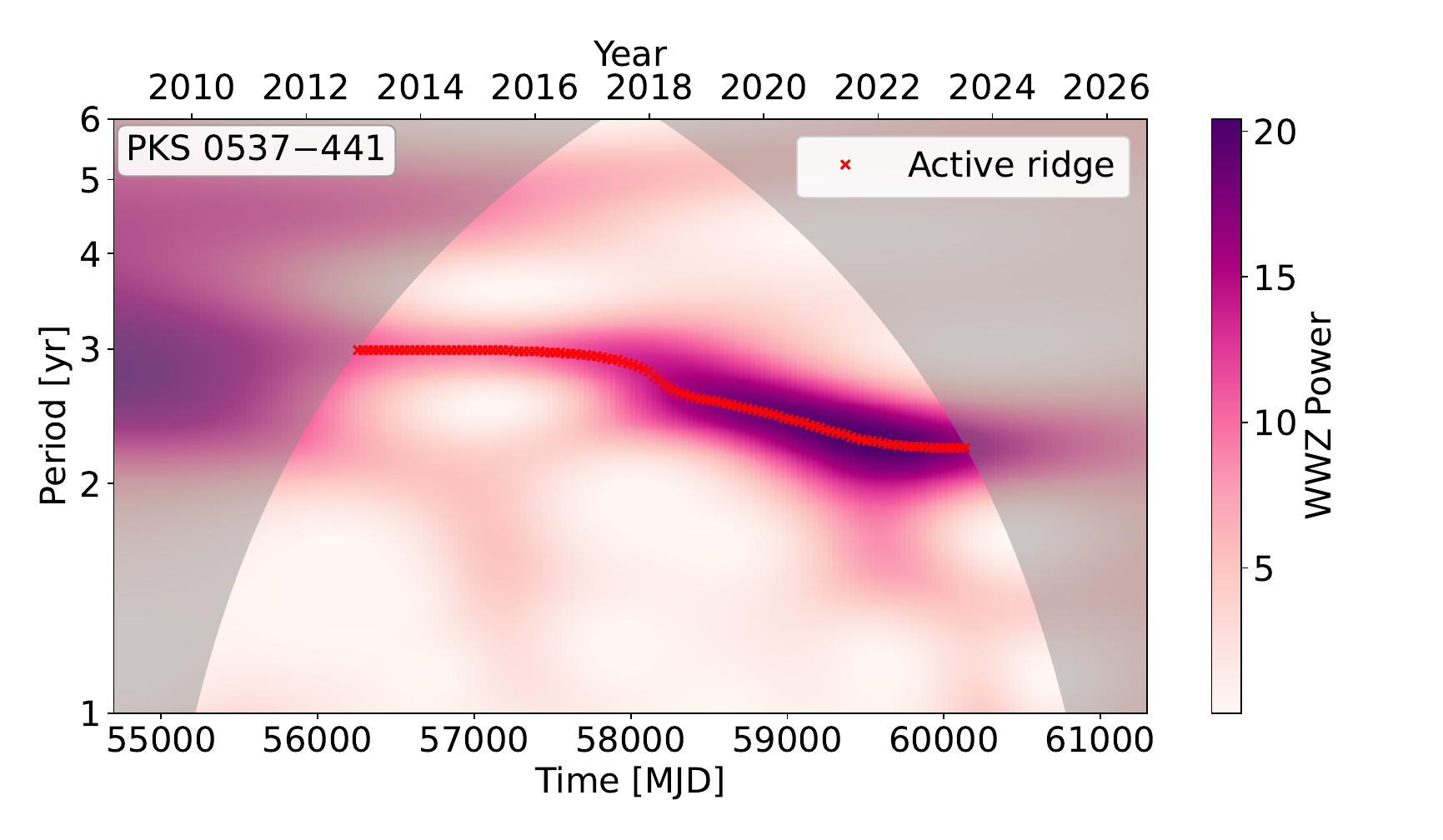}\\[2pt]
  \includegraphics[width=0.49\textwidth]{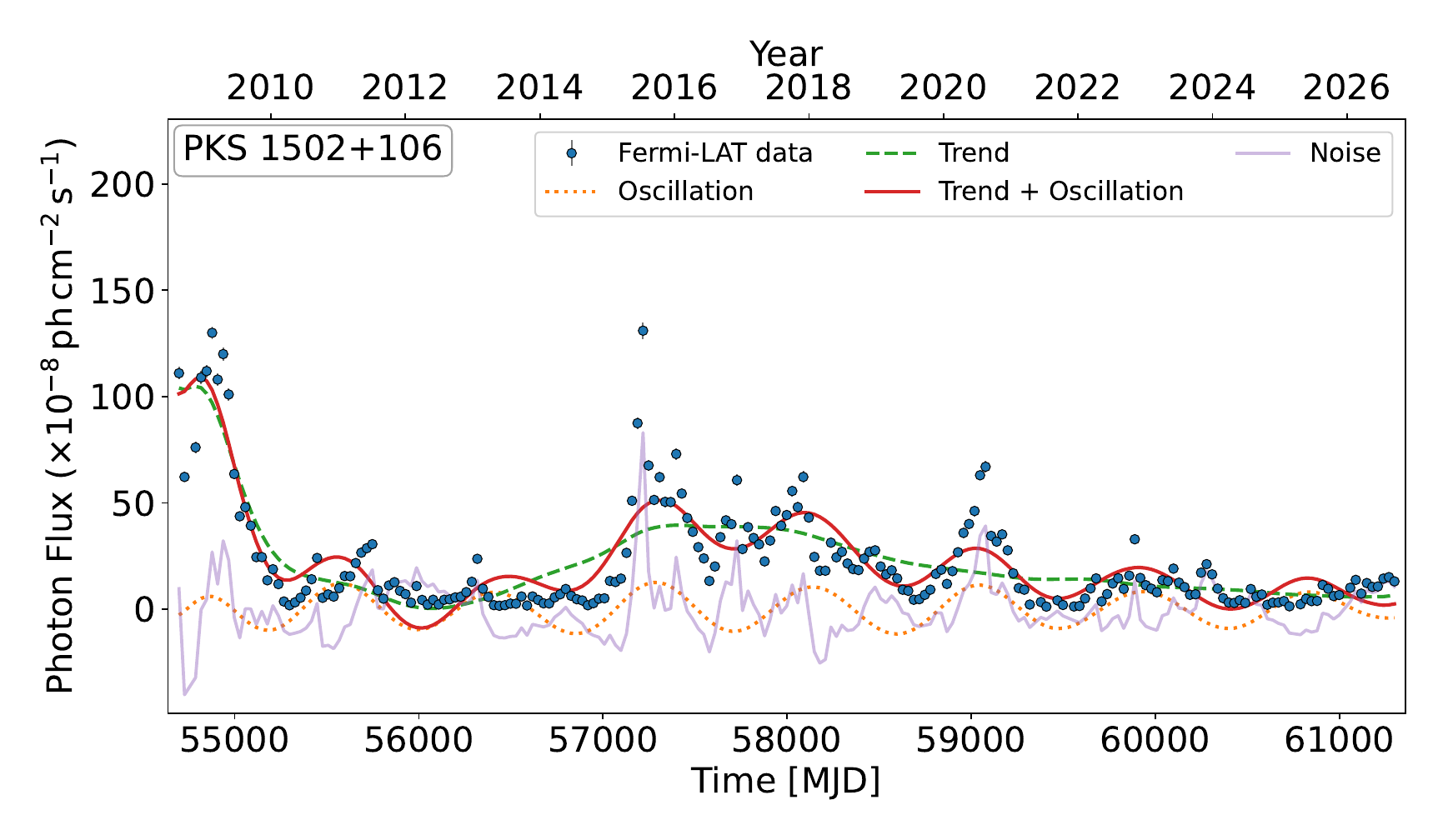}%
  \includegraphics[width=0.49\textwidth]{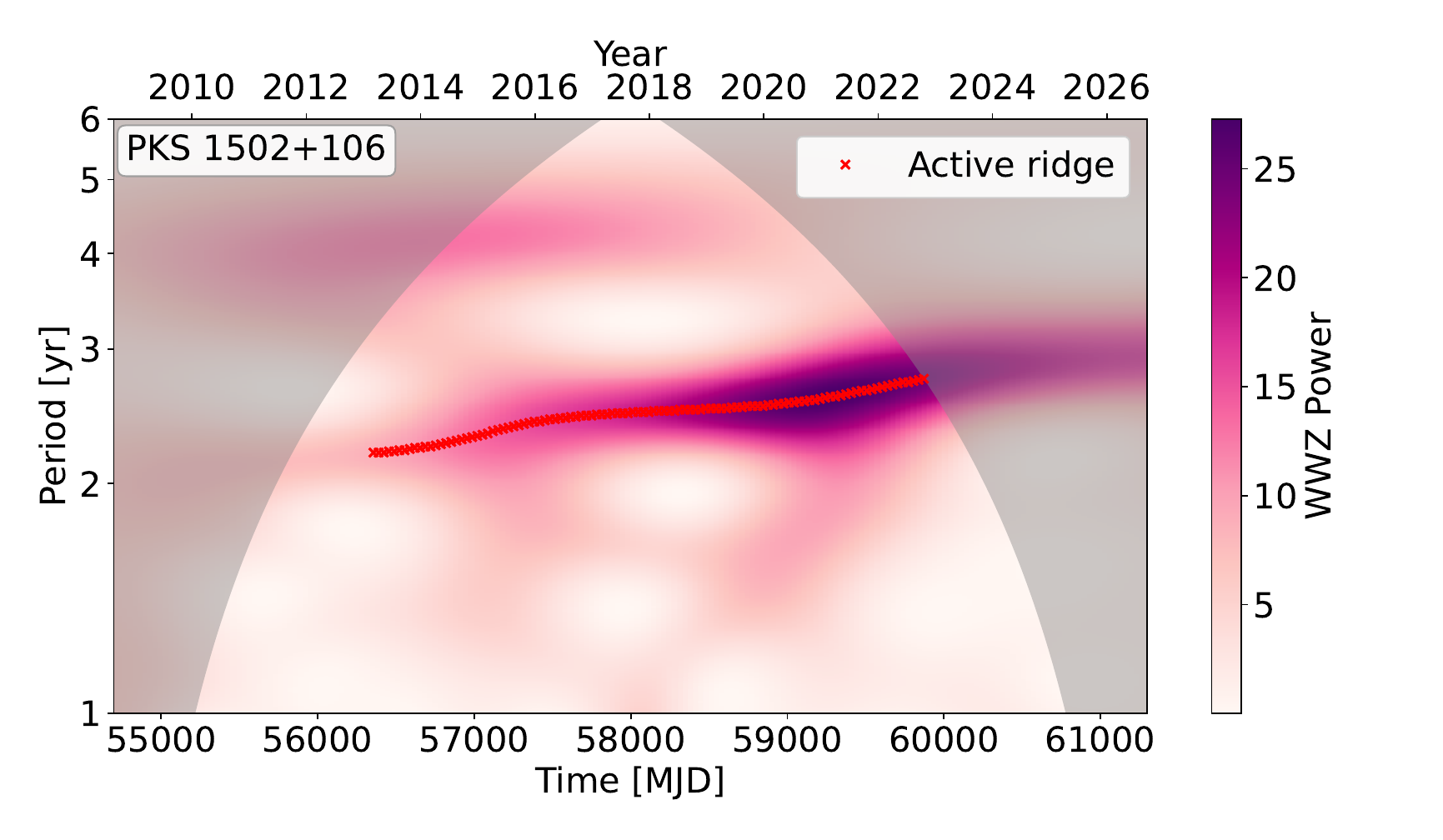}\\[2pt]
  \includegraphics[width=0.49\textwidth]{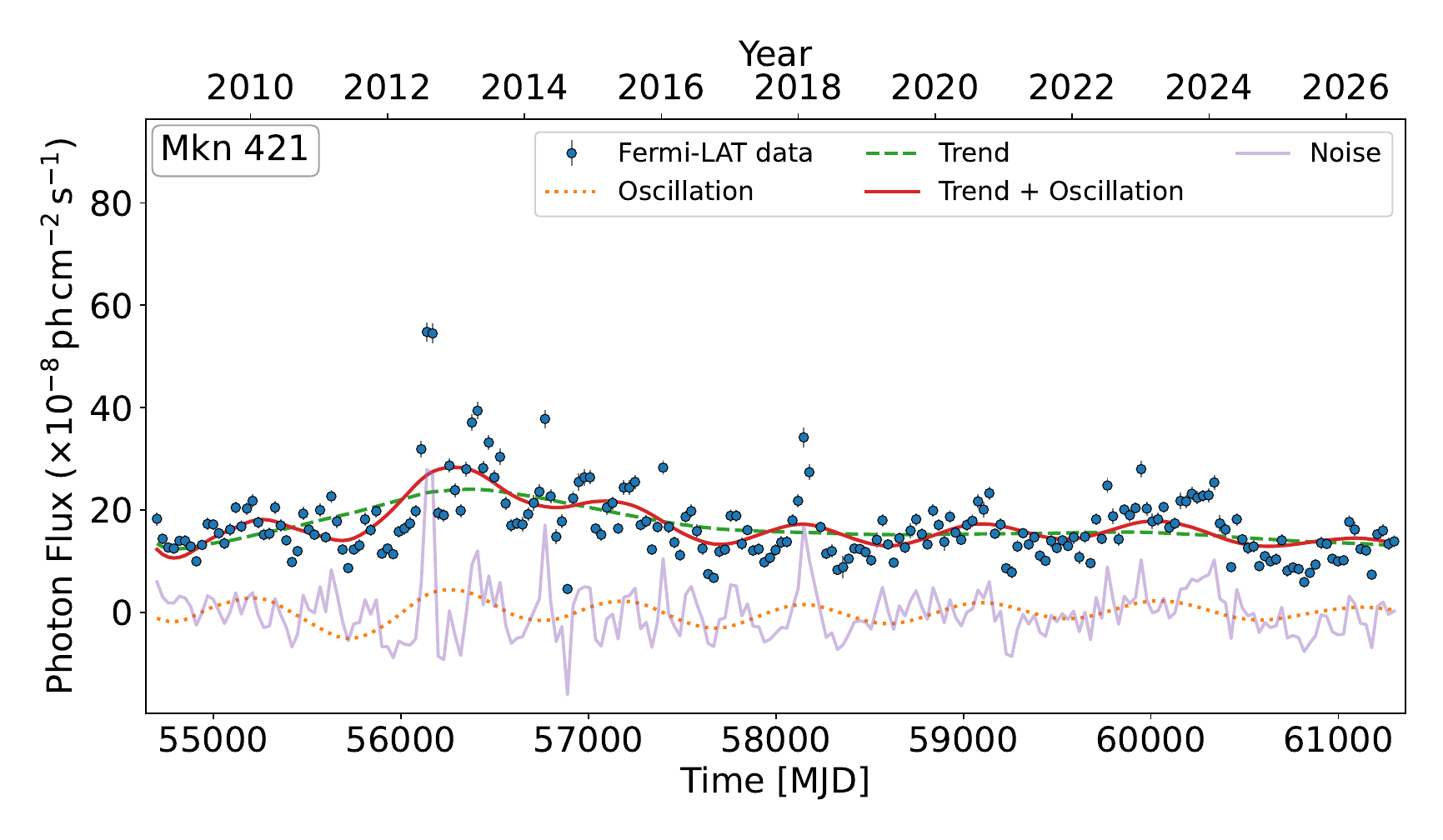}%
  \includegraphics[width=0.49\textwidth]{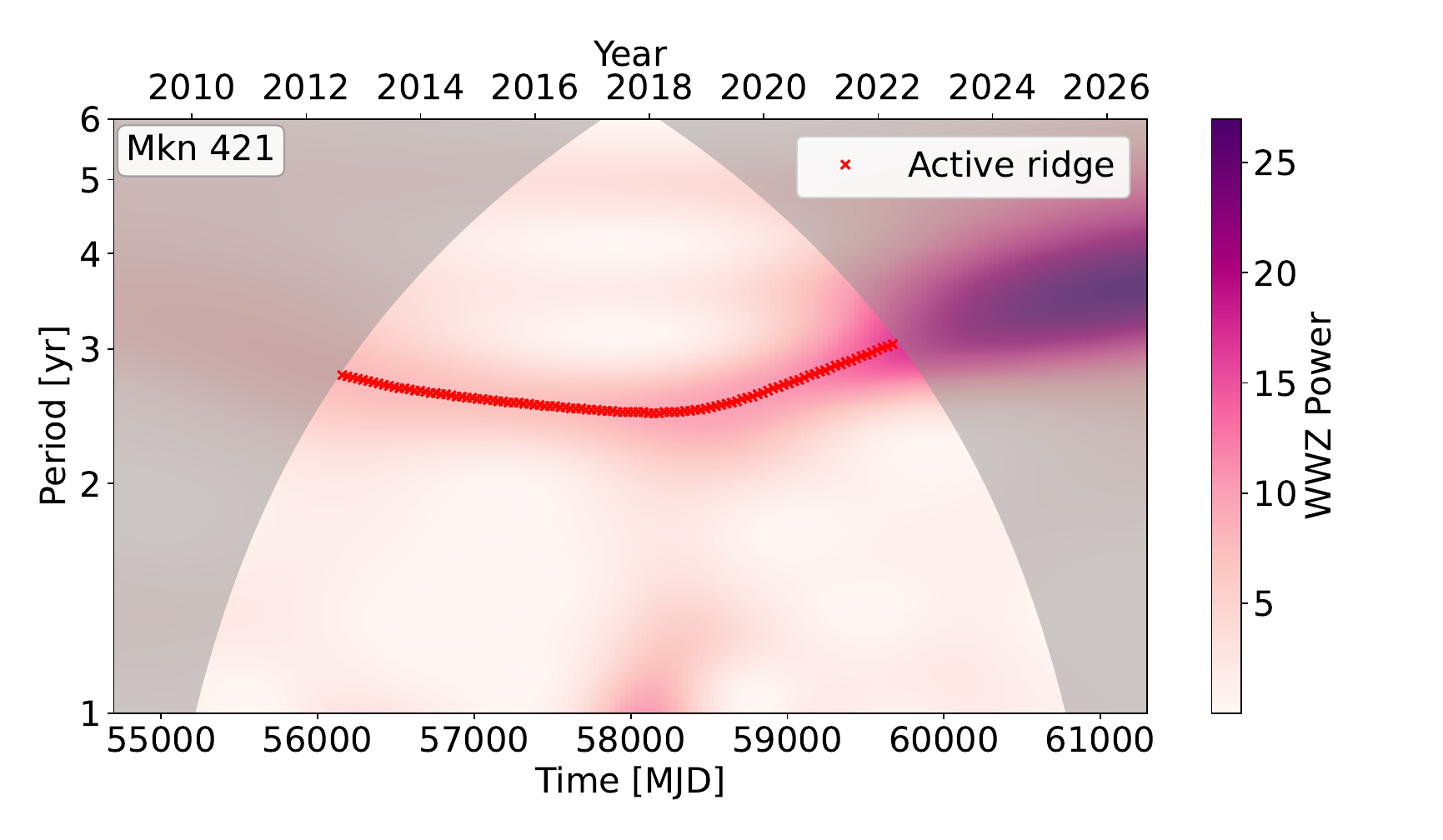}
  \caption{Rank~II sources, continued: PKS~0537$-$441 (4FGL~J0538.8$-$4405; $P_{\mathrm{obs}}=2.68$~yr, $\varepsilon=-0.031\,\mathrm{yr}^{-1}$), PKS~1502+106 (4FGL~J1504.4+1029; $2.46$~yr, $+0.023\,\mathrm{yr}^{-1}$, a borderline member) and Mkn~421 (4FGL~J1104.4+3812; $2.63$~yr, $+0.018\,\mathrm{yr}^{-1}$ with intermediate volatility). Panels as in Fig.~\ref{fig:atlas}. Measured on the SSA reconstruction instead of the light curve, the modulation of PKS~1502+106 appears stationary ($\varepsilon=0.006\,\mathrm{yr}^{-1}$), the clearest illustration of the smoothing discussed in Sect.~\ref{subsec:ablation}.}
  \label{fig:atlas_b}
\end{figure*}

Figs.~\ref{fig:atlas} and \ref{fig:atlas_b} present the SSA decomposition and the WWZ ridge analysis of the six Rank~II sources, ordered by decreasing cycle-weighted coherence. Their parameters are given in Table~\ref{tab:evolving}.

\section{Spectroscopic redshift compilation}
\label{sec:appendix_z}

The redshifts used in this paper are not taken from a catalog. Catalog compilations for BL~Lacertae objects frequently record values that are not systemic: the non-thermal continuum is strongly Doppler-boosted and swamps the host-galaxy stellar absorption, and the features that do appear are often imprinted by intervening gas. We have therefore assessed each of the 97 selected candidates against the primary optical, ultraviolet and near-infrared spectroscopic literature, and Table~\ref{tab:zcomp} records the outcome together with the method and reference for each value.

We treat as systemic only those redshifts established by host-galaxy stellar absorption, by broad or narrow emission lines, or by a spectroscopically confirmed group or cluster association. Values derived from intervening Mg\,\textsc{ii} or Fe\,\textsc{ii} doublets constrain the foreground absorber and are quoted as lower limits on the source. Where a high signal-to-noise spectrum is featureless to stringent equivalent-width limits, no systemic redshift is established and none is assumed; where available we quote the upper limit set by extragalactic background light attenuation of the \textit{Fermi}-LAT spectrum at $84\%$ confidence \citep{Dominguez2024}. Redshifts estimated from the host-galaxy luminosity in imaging, or assigned by automated spectroscopic pipelines without confirmed features, are not adopted.

Of the 97 candidates, 60 have systemic redshifts, one of them uncertain. Nine are bounded from below by intervening absorption and 12 from above by extragalactic background light attenuation. The remaining 16 are unconstrained, either because their spectra are featureless or because no redshift has been published. Two widely used catalog values do not survive. The $z=0.444$ adopted for 3C~66A rests on a single tentative detection excluded by later deep spectroscopy; its host cluster places the source at $z\simeq0.340$ instead. The $z=0.695$ adopted for PKS~0903$-$57 was measured on a neighboring object, and the blazar lies at $z=0.2621$ \citep{Goldoni2024}. The redshift of PKS~1424+240, long uncertain, is set by weak [O\,\textsc{ii}] and [O\,\textsc{iii}] emission at $z=0.6047$ \citep{Paiano2017b}, confirmed at higher signal-to-noise ratio in a low state \citep{DAmmando2024}. Two sources carry both kinds of limit: B3~0609+413 and TXS~1902+556 have intervening Mg\,\textsc{ii} absorbers at $z=1.108$ and $0.727$ and EBL upper limits of $z<1.51$ and $z<0.83$, which together bracket their redshifts; Table~\ref{tab:zcomp} lists the spectroscopic lower limits. A further case matters directly for Rank~I: the values of $0.130$ and $0.237$ reported for B2~1215+30 are both unconfirmed, and high-sensitivity spectroscopy finds a featureless continuum.

None of the classification results of this paper depends on a redshift. The morphological parameters are measured in the observer frame, and the ranking uses no distance-dependent quantity. The compilation matters for interpretation, in two places: the comparison of observer-frame with rest-frame drift rates in Sect.~\ref{subsec:parameter_extraction}, and the redshift dependence of the period ceiling discussed in Sect.~\ref{subsec:demographics} and shown in Fig.~\ref{fig:period_redshift}.

\input{tables/table_zcomp}

\section{Extended catalog of QPO candidates}
\label{sec:appendix_tables}

Tables~\ref{tab:evolving} and \ref{tab:rank3_sources} provide the morphological parameters of the Rank~II and Rank~III candidates that pass the quality selection of Sect.~\ref{subsec:quality}. The subclass demographics are given in Table~\ref{tab:demographics}.

\input{tables/table_evolving}
\input{tables/table_rankiii}

\end{document}

%% file: tables/table_sample.tex
\begin{table}
\caption{Sample selection. The first block gives cumulative counts down to the light curves that enter the analysis. The middle block gives the number of those 120 passing each uncertainty-quality criterion individually. The last block gives their intersection and its morphological ranks.}
\label{tab:sample_flow}
\centering
\small
\setlength{\tabcolsep}{6pt}
\begin{tabular}{lc}
\hline\hline
\noalign{\smallskip}
Selection stage & $N_{\mathrm{sources}}$ \\
\noalign{\smallskip}
\hline
\noalign{\smallskip}
4FGL-DR2 catalog & 5788 \\
Blazar classes (\texttt{bll}, \texttt{fsrq}, \texttt{bcu}) & 3436 \\
Variability index $>21.67$ (LCR light curve available) & 1415 \\
Missing-bin fraction $<50\%$ & 430 \\
At most three consecutive missing bins & 120 \\
SSA pair and WWZ ridge, $1\le P\le6$~yr & 120 \\
\noalign{\smallskip}
\hline
\noalign{\smallskip}
\multicolumn{2}{l}{Uncertainty-quality selection (Sect.~\ref{subsec:quality}), applied individually:} \\
\quad Finite flux-randomization intervals & 120 \\
\quad $Q/\Delta Q > 2$ & 98 \\
\quad $\Delta\varepsilon < 0.05$~yr$^{-1}$ & 116 \\
\noalign{\smallskip}
\hline
\noalign{\smallskip}
Final selected sample & 97 \\
\quad Rank I & 2 \\
\quad Rank II & 6 \\
\quad Rank III & 89$^{a}$ \\
\noalign{\smallskip}
\hline
\end{tabular}
\tablefoot{Missing-bin statistics are computed over the stored 30-day bins. Counts in the middle block are the number passing each individual criterion out of 120; the final sample of 97 is the intersection of all of them. The coherence-precision requirement $Q/\Delta Q>2$ dominates (22 failures). The quality scheme is the one selected on simulations alone (Table~\ref{tab:schemes}). The false-positive expectations of Sect.~\ref{subsec:false-positive} are normalized to the 120 light curves that enter the analysis. $^{a}$Of the 89 Rank~III sources, 57 fail the coherence gate, 6 the cycle gate (6 of them period-ceiling limited, $P\geq P_{\mathrm{ceil}}$, Eq.~\ref{eq:pceil}), 15 the plasma volatility boundary, and 11 the rigidity test without meeting the persistence requirement of Rank~II (Sect.~\ref{subsec:rank3}).}
\end{table}

%% file: tables/table_schemes.tex
\begin{table*}
\caption{Choice of the uncertainty-quality scheme, made on simulations alone before the observed sample was ranked (Sect.~\ref{subsec:quality}).}
\label{tab:schemes}
\centering
\small
\setlength{\tabcolsep}{5pt}
\begin{tabular}{lccccccc}
\hline\hline
\noalign{\smallskip}
Scheme & Asymmetry cuts & Completeness & Red-noise rate & Expected & Rank~I & FOM & Observed \\
 & & (stationary) & (Rank~I) & red-noise Rank~I & inj.\ purity & & selected / Rank~I \\
\noalign{\smallskip}
\hline
\noalign{\smallskip}
S0 & $A_Q$, $A_\varepsilon$, $A_{\sigma_{\ln P}}$ & $6\%$ & $0.76\%$ & $0.91$ & $71\%$ & $0.03$ & 46 / 0 \\
S1 & $A_Q$, $A_\varepsilon$ & $6\%$ & $0.81\%$ & $0.97$ & $74\%$ & $0.03$ & 65 / 0 \\
S2 & none & $92\%$ & $1.10\%$ & $1.32$ & $83\%$ & $0.43$ & 97 / 2 \\
\noalign{\smallskip}
\hline
\end{tabular}
\tablefoot{All schemes require finite flux-randomization intervals, $Q/\Delta Q>2$ and $\Delta\varepsilon<0.05$~yr$^{-1}$, and all rank with the worst-case bounds of Table~\ref{tab:ranking_hierarchy}. Asymmetry cuts require $A_X\le0.8$ (Eq.~\ref{eq:uncertainty_asymmetry}). Completeness is the fraction of stationary injections (scenario~A, $P_{\mathrm{inj}}=1.5$ and $2.1$~yr, amplitude SNR $1$ and $2$, 360 trials) assigned to Rank~I by the production classifier; the red-noise rate is that of the $10\,000$ controls; injection purity is the fraction of Rank~I injections of scenarios A, B and C that are stationary. FOM is the figure of merit of \citet{Punzi2003}, completeness divided by $1+\sqrt{B}$, with $B$ the expected number of red-noise Rank~I sources among 120 light curves. The last column, computed after the choice, is a sensitivity check.}
\end{table*}

%% file: tables/table_simsets.tex
\begin{table*}
\caption{Simulation sets used to calibrate and test the classification (Sect.~\ref{subsec:robustness}).}
\label{tab:simsets}
\centering
\small
\setlength{\tabcolsep}{5pt}
\begin{tabular}{llccccl}
\hline\hline
\noalign{\smallskip}
Set & Families & $P_{\mathrm{inj}}$ [yr] & SNR & Per cell & Total & Classification \\
\noalign{\smallskip}
\hline
\noalign{\smallskip}
Red-noise controls & \dots & \dots & \dots & \dots & $10\,000$ & as the data$^{a}$ \\
Parameter response & A, B, C, N & 1.5, 2.1, 3.0, 4.0 & 0.5, 1, 2, 4 & 50 & $9600$ & nominal values \\
 & M$^{b}$ & 1.5, 2.1 & 1, 2, 4 & 50 & $1800$ & nominal values \\
Completeness and purity & A, B, C, M$^{b}$, N & 1.5, 2.1 & 1, 2 & 30 & $2160$ & as the data$^{c}$ \\
Completeness vs. period & A & 1.2--2.95$^{d}$ & 1, 2 & 30 & $1440$ & as the data$^{c}$ \\
Censoring & A, M$_{25\%}$, N & 1.5, 2.1 & 1, 2 & 30 & $2160^{e}$ & as the data$^{c}$ \\
Drift response & B$^{f}$ & 1.5, 2.1, 2.5 & 1, 2, 4 & 20 & $3240$ & nominal values \\
\noalign{\smallskip}
\hline
\end{tabular}
\tablefoot{All injection sets use PSD slopes $\beta\in\{1.0,1.5,2.0\}$, and the controls slopes drawn uniformly from $[1.0,2.0]$. A: stationary sinusoids; B: linear period drift; C: cycle-to-cycle jitter; M: masked stationary sinusoids; N: pure red noise. $^{a}$The 1215 controls whose central values could reach Rank~I or II, and a random $10\%$ of the others (853), are flux-randomized with 200 realizations. $^{b}$Masked over $25\%$ or $50\%$ of the baseline. $^{c}$Flux randomization with 100 realizations and the quality selection. $^{d}$Eight periods: $1.2$, $1.5$, $1.8$, $2.1$, $2.4$, $2.6$, $2.8$ and $2.95$~yr. $^{e}$Matched pairs analyzed twice, with the fixed and with the recomputed detection mask. $^{f}$Drift rates $\varepsilon_{\mathrm{inj}}=0.01$, $0.02$, $0.04$, $0.06$, $0.08$ and $0.12$~yr$^{-1}$.}
\end{table*}

%% file: tables/table_fpr.tex
\begin{table*}
\caption{Red-noise false-positive rates from $10\,000$ control light curves processed through the identical pipeline, including the flux randomization and the uncertainty-quality selection, compared with the observed sample.}
\label{tab:fpr}
\centering
\small
\setlength{\tabcolsep}{5pt}
\begin{tabular}{lccccc}
\hline\hline
\noalign{\smallskip}
Stage / rank & $N/N_{\rm trial}$ & Rate per input & 95\% interval & Expected among 120 & Observed \\
\noalign{\smallskip}
\hline
\noalign{\smallskip}
No trackable ridge & $4/10\,000$ & $0.04\%$ & \dots & \dots & 0 \\
Not assigned to a rank$^{a}$ & $8281/10\,000$ & $82.8\%$ & \dots & \dots & 23 \\
\noalign{\smallskip}
\hline
\noalign{\smallskip}
Rank I & $110/10\,000$ & $1.10\%$ & $0.90$--$1.32\%$ & $1.32$ & 2 \\
Rank II & $225/10\,000$ & $2.25\%$ & $1.97$--$2.56\%$ & $2.70$ & 6 \\
Rank III$^{b}$ & $1380/10\,000$ & $\ge13.80\%$ & \dots & \dots & 89 \\
\noalign{\smallskip}
\hline
\end{tabular}
\tablefoot{Control light curves inherit the time grid, detection mask and per-bin uncertainties of the 120 light curves that enter the analysis, with PSD slopes drawn uniformly from $\beta\in[1.0,2.0]$ (Sect.~\ref{subsec:robustness}). Intervals are Clopper--Pearson at $95\%$ confidence. The expected number is the per-input rate multiplied by 120. The binomial probability of observing at least the number in the last column under the red-noise null is $0.381$ for Rank~I, $0.055$ for Rank~II and $0.049$ for Ranks~I and II combined. $^{a}$Removed by the uncertainty-quality selection or, for controls whose central values cannot reach Rank~I or II, not passed to the flux randomization. $^{b}$Lower limit: Rank~III assignments are counted only among the 2068 controls passed to the flux randomization. The Rank~I and Rank~II counts are complete.}
\end{table*}

%% file: tables/table_confusion.tex
\begin{table*}
\caption{Classification of injected modulations by the production classifier (flux randomization, uncertainty-quality selection and worst-case bounds). Entries give the percentage of injections assigned to each class, for periods below the ceiling ($P_{\mathrm{inj}}=1.5$ and $2.1$~yr).}
\label{tab:confusion}
\centering
\small
\setlength{\tabcolsep}{4pt}
\begin{tabular}{lccccc}
\hline\hline
\noalign{\smallskip}
Injected scenario & Rank I & Rank II & Rank III & Not selected & $N$ \\
\noalign{\smallskip}
\hline
\noalign{\smallskip}
A: stationary, SNR $=1$ & 83.9 & 5.6 & 9.4 & 1.1 & 180 \\
A: stationary, SNR $=2$ & 100.0 & 0.0 & 0.0 & 0.0 & 180 \\
B: drifting ($\varepsilon_{\mathrm{inj}}=0.04$~yr$^{-1}$) & 0.0 & 91.9 & 7.8 & 0.3 & 360 \\
C: jittered ($\sigma_{\ln P,\mathrm{inj}}=0.15$) & 18.3 & 40.6 & 39.4 & 1.7 & 360 \\
M: masked, $25\%$ of the baseline & 12.5 & 33.3 & 51.9 & 2.2 & 360 \\
M: masked, $50\%$ of the baseline & 0.6 & 11.7 & 74.2 & 13.6 & 360 \\
N: pure red noise & 0.6 & 5.6 & 75.8 & 18.1 & 359 \\
\noalign{\smallskip}
\hline
\end{tabular}
\tablefoot{Each row pools amplitude SNR $1$ and $2$ and PSD slopes $\beta\in\{1.0,1.5,2.0\}$ unless stated, with 30 injections per cell; 2159 injections in total. The injection purity of Rank~I, the fraction of Rank~I injections of scenarios A, B and C that are stationary, is $83\%$; it is conditional on the families, amplitudes and periods injected. On the full injection grid with the nominal classifier ($9600$ trials, $P_{\mathrm{inj}}=1.5$--$4.0$~yr, SNR $0.5$--$4$), stationary modulations at $3.0$ and $4.0$~yr almost never reach Ranks~I and II, as required by the period ceiling (Eq.~\ref{eq:pceil}).}
\end{table*}

%% file: tables/table_completeness.tex
\begin{table}
\caption{Rank~I completeness for stationary modulations as a function of period, production classifier.}
\label{tab:completeness}
\centering
\small
\setlength{\tabcolsep}{5pt}
\begin{tabular}{lcc}
\hline\hline
\noalign{\smallskip}
$P_{\mathrm{inj}}$ [yr] & SNR $=1$ & SNR $=2$ \\
\noalign{\smallskip}
\hline
\noalign{\smallskip}
$1.20$ & $63\%$ & $100\%$ \\
$1.50$ & $71\%$ & $100\%$ \\
$1.80$ & $81\%$ & $100\%$ \\
$2.10$ & $87\%$ & $100\%$ \\
$2.40$ & $94\%$ & $100\%$ \\
$2.60$ & $90\%$ & $100\%$ \\
$2.80$ & $97\%$ & $100\%$ \\
$2.95$ & $97\%$ & $100\%$ \\
\noalign{\smallskip}
\hline
\end{tabular}
\tablefoot{Fraction of stationary injections assigned to Rank~I, with 90 injections per cell spanning PSD slopes $\beta\in\{1.0,1.5,2.0\}$, and the detection mask held fixed. The minimum over periods, $63\%$ at SNR $=1$ and $100\%$ at SNR $=2$, sets the conservative incidence limit of Sect.~\ref{subsec:incidence}.}
\end{table}

%% file: tables/table_attenuation.tex
\begin{table}
\caption{Response of the drift and volatility measurements, measured by injection--recovery with the nominal classifier. With the ridge measured on the detrended light curve, both quantities are recovered with little attenuation.}
\label{tab:attenuation}
\centering
\small
\setlength{\tabcolsep}{4pt}
\begin{tabular}{lcccc}
\hline\hline
\noalign{\smallskip}
Injected value & Median measured & Ratio & Above cut & $N$ \\
\noalign{\smallskip}
\hline
\noalign{\smallskip}
\multicolumn{5}{l}{Drift $|\varepsilon|$ [yr$^{-1}$], stationarity cut $|\varepsilon|_{\rm meas}\ge0.02$:} \\
\quad $0.01$ & $0.0084$ & $1.2\times$ & $0\%$ & 469 \\
\quad $0.02$ & $0.0173$ & $1.2\times$ & $8\%$ & 448 \\
\quad $0.04$ & $0.0345$ & $1.2\times$ & $99\%$ & 398 \\
\quad $0.06$ & $0.0503$ & $1.2\times$ & $100\%$ & 311 \\
\quad $0.08$ & $0.0625$ & $1.3\times$ & $100\%$ & 207 \\
\quad $0.12$ & $0.0846$ & $1.4\times$ & $100\%$ & 65 \\
\noalign{\smallskip}
\hline
\noalign{\smallskip}
\multicolumn{5}{l}{Volatility $\sigma_{\ln P}$:} \\
\quad $0.15$ & $0.0358$ & $4\times$ & \dots & 1456 \\
\noalign{\smallskip}
\hline
\end{tabular}
\tablefoot{Measured over injections with $Q>3$; drifting injections span $P_{\mathrm{inj}}=1.5$, $2.1$ and $2.5$~yr and SNR $1$, $2$ and $4$, and volatility injections the grid of Sect.~\ref{subsec:robustness} with SNR $\ge1$. The drift ratio is close to $1.2$ at every period and amplitude tested and rises only slowly for the fastest drifts; half of the injections with $\varepsilon_{\mathrm{inj}}\simeq0.03$~yr$^{-1}$ exceed the stationarity cut. Stationary injections recover $|\varepsilon|=0.0012$~yr$^{-1}$ and $\sigma_{\ln P}=0.0052$ in the median, the measurement floor of each parameter. $\varepsilon_{\mathrm{inj}}$ imposes a linear fractional period drift, whereas $\varepsilon$ is a power-weighted average of the local gradient of a median-smoothed ridge; $\sigma_{\ln P,\mathrm{inj}}$ sets the step of a cycle-to-cycle random walk in $\ln P$, whereas $\sigma_{\ln P}$ is the cycle-weighted rms of $\ln P$ about a fitted log-linear trend.}
\end{table}

%% file: tables/table_class.tex
\begin{table*}
\caption{Morphological classification of blazar QPO candidates, in order of decreasing compatibility with a stable clock. The rank names are observational labels describing the recovered ridge; the scenario column gives the physical interpretation available for each morphology, not an outcome established by the classification (Sect.~\ref{sec:discussion}). The ranks are mutually exclusive and exhaustive.}
\label{tab:ranking_hierarchy}
\centering
\small
\setlength{\tabcolsep}{2.5pt}
\begin{tabular}{llccccl}
\hline\hline
\noalign{\smallskip}
Class & Scenario & Coherence & Cycles$^{\,a}$ & Rigidity tests & Persistence & Morphology \\
      &          & $Q-\Delta Q$ & $N_{\mathrm{active}}$ & $|\varepsilon|+\Delta\varepsilon$ $[\mathrm{yr}^{-1}]$ \ \& \ $\sigma_{\ln P}+\Delta\sigma_{\ln P}$ & $\Pi-\Delta\Pi$ & \\
\noalign{\smallskip}
\hline
\noalign{\smallskip}
Rank I   & A & $>3$ & $>3$ & $<0.02$ \ and \ $<0.06$ & $>0.90$ & Stable clock \\
Rank II  & B & $>3$ & $>3$ & $\ge0.02$ \ or \ $\ge0.06$; \ and $\sigma_{\ln P}+\Delta\sigma_{\ln P}<0.15$ & $>0.60$ & Coherent, evolving period \\
\noalign{\smallskip}
\hline
\noalign{\smallskip}
\multicolumn{7}{l}{Rank III, residual; subclass of the first test failed, in the order coherence, cycles, volatility, persistence:} \\
\quad III$_{Q}$      & \dots & $\le3$ & \dots & \dots & \dots & Insufficient coherence \\
\quad III$_{P}$      & \dots & \dots & $\le3$, $P\ge P_{\mathrm{ceil}}$ & \dots & \dots & Period beyond the ceiling \\
\quad III$_{\sigma}$ & C$^{\,b}$ & \dots & \dots & $\sigma_{\ln P}+\Delta\sigma_{\ln P}\ge0.15$ & \dots & Plasma-level volatility \\
\quad III$_{\Pi}$    & \dots & \dots & $\le3$, $P<P_{\mathrm{ceil}}$ & \dots & otherwise & Insufficient persistence \\
\noalign{\smallskip}
\hline
\end{tabular}
\tablefoot{Assignments use the one-sided worst-case $1\sigma$ bounds of Sect.~\ref{subsec:uncertainties}, i.e. $\Delta X = \max(\Delta X_{\mathrm{lo}},\Delta X_{\mathrm{hi}})$. $\Pi$ is the cone-of-influence-corrected persistence of Eq.~(\ref{eq:Pi}); at $P\simeq2$~yr the criterion $\Pi-\Delta\Pi>0.90$ is numerically equivalent to $p>0.61$ in the uncorrected statistic. Rank~II is written as the logical complement of the Rank~I rigidity test, so that no coherent, long-lived candidate falls through to Rank~III merely because its drift or volatility is poorly constrained. III$_{\Pi}$ collects the coherent candidates that are not persistent enough for Rank~I or II, including any with a stable period but $\Pi-\Delta\Pi\le0.90$. $^{a}$Because $N_{\mathrm{active}}$ accumulates only outside the cone of influence, the cycle gate is attainable only for $P<P_{\mathrm{ceil}}=T_{\mathrm{obs}}/(3+2\lambda)\simeq3.1$~yr (Eq.~\ref{eq:pceil}); longer-period ridges are tracked and reported but cannot be classified on an $18.1$-yr baseline. $^{b}$Scenario~C is the interpretation available for plasma-level volatility, but in this sample the volatility of III$_{\sigma}$ is consistent with red noise (Sect.~\ref{subsec:volatility_regime}).}
\end{table*}

%% file: tables/table_ranki.tex
\begin{table*}
\caption{Rank~I: persistent, high-coherence candidates, ordered by decreasing cycle-weighted coherence $Q$. Both the raw persistence $p$ and the cone-of-influence-corrected persistence $\Pi$ (Eq.~\ref{eq:Pi}) are listed; the classification uses $\Pi$.}
\label{tab:rank1_sources}
\centering
\small
\setlength{\tabcolsep}{3.0pt}
\begin{tabular}{llcccccccc}
\hline\hline
\noalign{\smallskip}
4FGL name & Association & Class & $z$ & $P_{\mathrm{obs}}$ & $Q$ & $|\varepsilon|$ & $\sigma_{\ln P}$ & $p$ & $\Pi$ \\
 &  &  &  & [yr] &  & [yr$^{-1}$] &  &  &  \\
\noalign{\smallskip}
\hline
\noalign{\smallskip}
J1555.7+1111 & PG 1553+113 & BL Lac & $0.433^{\dagger}$ & $2.11 \pm 0.01$ & $10.2 \pm 0.6$ & $0.0057 \pm 0.0021$ & $0.0150 \pm 0.0057$ & $0.67$ & $1.00 \pm 0.01$ \\
J1217.9+3007 & B2 1215+30 & BL Lac & \dots & $2.89 \pm 0.02$ & $5.2 \pm 0.3$ & $0.0158 \pm 0.0028$ & $0.0200 \pm 0.0052$ & $0.55$ & $1.01 \pm 0.01$ \\
\noalign{\smallskip}
\hline
\end{tabular}
\tablefoot{Source classes follow 4FGL-DR3 \citep{Abdollahi2022}, identical for these sources to 4LAC-DR3 \citep{Ajello2022}; redshifts are from the compilation of Appendix~\ref{sec:appendix_z}, which supersedes the catalog values. The method behind each redshift is given in Table~\ref{tab:zcomp}. $^{\dagger}$PG~1553+113 has no intrinsic nuclear lines; the value quoted is that of the galaxy group in which the source appears to reside \citep{TorresZafra2018,Johnson2019}. Uncertainties, including those on $P_{\mathrm{obs}}$, are the worst-case (maximum) half-widths of the flux-randomization interval (Sect.~\ref{subsec:uncertainties}).}
\end{table*}

%% file: tables/table_stability.tex
\begin{table}
\caption{Period-stability limits for Rank~I. The intrinsic drift is bounded by the measured upper bound corrected for the measured attenuation of the drift measurement, $|\varepsilon|_{\rm int}\leq 1.22\,(|\varepsilon|+\Delta\varepsilon)$, and the period-stability timescale by $\tau_P=|P/\dot{P}|\geq1/|\varepsilon|_{\rm int}$.}
\label{tab:stability}
\centering
\small
\setlength{\tabcolsep}{3.5pt}
\begin{tabular}{lccccc}
\hline\hline
\noalign{\smallskip}
Association & $P_{\mathrm{obs}}$ & $|\varepsilon|$ & $\Delta\varepsilon$ & $|\varepsilon|_{\rm int}$ & $\tau^{\mathrm{min}}_{P}$ \\
 & [yr] & [yr$^{-1}$] & [yr$^{-1}$] & [yr$^{-1}$] & [yr] \\
\noalign{\smallskip}
\hline
\noalign{\smallskip}
PG 1553+113 & $2.11 \pm 0.01$ & 0.0057 & 0.0021 & 0.0095 & 105 \\
B2 1215+30 & $2.89 \pm 0.02$ & 0.0158 & 0.0028 & 0.0225 & 44 \\
\noalign{\smallskip}
\hline
\end{tabular}
\tablefoot{The attenuation factor, $1.22$, is the median ratio of injected to measured drift for drifting injections (Table~\ref{tab:attenuation}); $\Delta\varepsilon$ is the worst-case half-width of the flux-randomization interval. The gravitational-wave decay timescale of a compact SMBBH at the fiducial values of Sect.~\ref{sec:background} is $\tau_{\mathrm{GW}}\simeq7\times10^{3}$~yr (Eq.~\ref{eq:tau_gw}).}
\end{table}

%% file: tables/table_demographics.tex
\begin{table}
\caption{Blazar subclass demographics of the morphological ranks.}
\label{tab:demographics}
\centering
\small
\setlength{\tabcolsep}{6pt}
\begin{tabular}{lcccc}
\hline\hline
\noalign{\smallskip}
Rank & BL Lac & FSRQ & Total & $f_{\mathrm{BL\,Lac}}$ \\
\noalign{\smallskip}
\hline
\noalign{\smallskip}
Rank I & 2 & 0 & 2 & 1.00 \\
Rank II & 4 & 2  & 6 & 0.67 \\
Rank III & 64 & 25  & 89 & 0.72 \\
All candidates & 70 & 27  & 97 & 0.72 \\
\noalign{\smallskip}
\hline
\end{tabular}
\tablefoot{Subclasses from 4FGL-DR3 \citep{Abdollahi2022}, identical to 4LAC-DR3 \citep{Ajello2022} for these sources.}
\end{table}

%% file: tables/table_zcomp.tex
\begin{table*}
\caption{Spectroscopic redshift compilation for the 97 selected candidates. Values are assessed
against the primary spectroscopic literature and supersede the catalog entries: $>$ denotes a lower
limit set by an intervening absorption system, and \dots\ a featureless continuum for which no
systemic redshift is established, and $<$ an upper limit at $84\%$ confidence from the attenuation of the \textit{Fermi}-LAT spectrum by the extragalactic background light \citep{Dominguez2024}.}
\label{tab:zcomp}
\centering
\footnotesize
\setlength{\tabcolsep}{3pt}
\renewcommand{\arraystretch}{0.95}
\begin{tabular}{lllll l}
\hline\hline
\noalign{\smallskip}
4FGL name & Association & Rank & $z$ & Method & Reference \\
\noalign{\smallskip}
\hline
\noalign{\smallskip}
J1217.9+3007 & B2 1215+30 & Rank~I & \dots & featureless continuum & \citet{Paiano2017} \\
J1555.7+1111 & PG 1553+113 & Rank~I & $0.433$ & group/cluster association & \citet{TorresZafra2018}, \citet{Johnson2019} \\
J0144.6+2705 & TXS 0141+268 & Rank~II & $<2.38^{+0.53}_{-0.39}$ & EBL limit & \citet{Shaw2013}, \citet{Massaro2015} \\
J0222.6+4302 & 3C 66A & Rank~II & $0.340$ & group/cluster association & \citet{Paiano2017}, \citet{TorresZafra2018} \\
J0538.8$-$4405 & PKS 0537$-$441 & Rank~II & $0.892$ & broad emission lines & \citet{Heidt2003} \\
J1104.4+3812 & Mkn 421 & Rank~II & $0.031$ & host stellar absorption & \citet{Ulrich1975} \\
J1504.4+1029 & PKS 1502+106 & Rank~II & $1.838$ & broad emission lines & \citet{AdelmanMcCarthy2008} \\
J1512.2+0202 & PKS 1509+022 & Rank~II & $0.220$ & broad emission lines & \citet{HewettWild2010} \\
J0022.5+0608 & PKS 0019+058 & Rank~III & $<1.33^{+0.53}_{-0.47}$ & EBL limit &  \\
J0051.2$-$6242 & 1RXS J005117.7$-$624154 & Rank~III & \dots & no published redshift &  \\
J0100.3+0745 & GB6 J0100+0745 & Rank~III & \dots & no published redshift &  \\
J0112.1+2245 & S2 0109+22 & Rank~III & $0.265$ & host stellar absorption & \citet{Paiano2017} \\
J0120.4$-$2701 & PKS 0118$-$272 & Rank~III & $<0.79^{+0.20}_{-0.19}$ & EBL limit &  \\
J0136.5+3906 & B3 0133+388 & Rank~III & $<0.73^{+0.13}_{-0.12}$ & EBL limit & \citet{Shaw2013}, \citet{Paiano2017} \\
J0141.4$-$0928 & PKS 0139$-$09 & Rank~III & $0.733$ & narrow emission lines & \citet{StockeRector1997} \\
J0210.7$-$5101 & PKS 0208$-$512 & Rank~III & $1.003$ & broad emission lines & \citet{Stickel1991} \\
J0211.2+1051 & MG1 J021114+1051 & Rank~III & $0.200$ & host stellar absorption & \citet{Shaw2013} \\
J0217.2+0837 & ZS 0214+083 & Rank~III & $0.085$ & host stellar absorption & \citet{Falomo2002} \\
J0221.1+3556 & B2 0218+357 & Rank~III & $0.944$ & broad emission lines & \citet{Cohen2003} \\
J0237.8+2848 & 4C +28.07 & Rank~III & $1.206$ & broad emission lines & \citet{Shaw2012} \\
J0303.4$-$2407 & PKS 0301$-$243 & Rank~III & $0.266$ & host stellar absorption & \citet{Sbarufatti2005} \\
J0303.6+4716 & 4C +47.08 & Rank~III & $<0.63^{+0.37}_{-0.40}$ & EBL limit &  \\
J0334.2$-$4008 & PKS 0332$-$403 & Rank~III & \dots & no published redshift &  \\
J0428.6$-$3756 & PKS 0426$-$380 & Rank~III & $1.111$ & broad emission lines & \citet{Heidt2004} \\
J0457.0$-$2324 & PKS 0454$-$234 & Rank~III & $1.003$ & broad emission lines & \citet{Stickel1989} \\
J0509.4+0542 & TXS 0506+056 & Rank~III & $0.3365$ & narrow emission lines & \citet{Paiano2018} \\
J0516.7$-$6207 & PKS 0516$-$621 & Rank~III & $1.300$ & broad emission lines & \citet{Shaw2013} \\
J0521.7+2112 & TXS 0518+211 & Rank~III & $0.108$ & host stellar absorption & \citet{Archambault2013} \\
J0532.6+0732 & OG 050 & Rank~III & $1.254$ & broad emission lines & \citet{Healey2008} \\
J0607.4+4739 & TXS 0603+476 & Rank~III & $<1.96^{+0.34}_{-0.34}$ & EBL limit &  \\
J0612.8+4122 & B3 0609+413 & Rank~III & $>1.108$ & intervening absorption & \citet{Goldoni2021} \\
J0630.9$-$2406 & TXS 0628$-$240 & Rank~III & $>1.238$ & intervening absorption & \citet{Landt2012} \\
J0650.7+2503 & 1ES 0647+250 & Rank~III & \dots & featureless continuum & \citet{Paiano2017b} \\
J0700.5$-$6610 & PKS 0700$-$661 & Rank~III & $<1.53^{+0.33}_{-0.27}$ & EBL limit &  \\
J0721.9+7120 & S5 0716+71 & Rank~III & $>0.2315$ & intervening absorption & \citet{Danforth2013} \\
J0730.3$-$1141 & PKS 0727$-$11 & Rank~III & $1.591$ & broad emission lines & \citet{Zensus2002} \\
J0738.1+1742 & PKS 0735+17 & Rank~III & $>0.424$ & intervening absorption & \citet{Carswell1974} \\
J0739.2+0137 & PKS 0736+01 & Rank~III & $0.189$ & broad emission lines & \citet{Ho2009} \\
J0757.1+0956 & PKS 0754+100 & Rank~III & $0.266$ & narrow emission lines & \citet{Carangelo2003} \\
J0811.4+0146 & OJ 014 & Rank~III & $>1.148$ & intervening absorption & \citet{Sbarufatti2005}, \citet{Shaw2013} \\
J0825.9$-$2230 & PKS 0823$-$223 & Rank~III & $>0.911$ & intervening absorption & \citet{Sbarufatti2009} \\
J0904.9$-$5734 & PKS 0903$-$57 & Rank~III & $0.2621$ & narrow emission lines & \citet{Goldoni2024} \\
J0909.7$-$0230 & PKS 0907$-$023 & Rank~III & $0.957$ & emission lines, uncertain & \citet{Hewitt1989} \\
J0953.0$-$0840 & PMN J0953$-$0840 & Rank~III & \dots & no published redshift &  \\
J0957.6+5523 & 4C +55.17 & Rank~III & $0.8955$ & broad emission lines & \citet{AdelmanMcCarthy2008} \\
J1015.0+4926 & 1H 1013+498 & Rank~III & $0.212$ & host stellar absorption & \citet{Sbarufatti2005} \\
J1037.7+5711 & GB6 J1037+5711 & Rank~III & $>0.830$ & intervening absorption & \citet{Shaw2013} \\
J1048.4+7143 & S5 1044+71 & Rank~III & $1.150$ & broad emission lines & \citet{Polatidis1995} \\
J1058.4+0133 & 4C +01.28 & Rank~III & $0.890$ & narrow emission lines & \citet{Shaw2012} \\
\noalign{\smallskip}
\hline
\end{tabular}
\end{table*}

\begin{table*}
\caption{Continued.}
\label{tab:zcomp_b}
\centering
\footnotesize
\setlength{\tabcolsep}{3pt}
\renewcommand{\arraystretch}{0.95}
\begin{tabular}{lllll l}
\hline\hline
\noalign{\smallskip}
4FGL name & Association & Rank & $z$ & Method & Reference \\
\noalign{\smallskip}
\hline
\noalign{\smallskip}
J1058.6+5627 & TXS 1055+567 & Rank~III & \dots & featureless continuum & \citet{Finke2008} \\
J1059.2$-$1134 & PKS B1056$-$113 & Rank~III & \dots & no published redshift &  \\
J1103.9$-$5357 & PKS 1101$-$536 & Rank~III & \dots & no published redshift &  \\
J1132.7+0034 & PKS B1130+008 & Rank~III & $1.223$ & broad emission lines & \citet{Abazajian2009}, \citet{Bergeron2011} \\
J1136.4+7009 & Mkn 180 & Rank~III & $0.045$ & host stellar absorption & \citet{Ulrich1975} \\
J1146.9+3958 & S4 1144+40 & Rank~III & $1.090$ & broad emission lines & \citet{HewettWild2010} \\
J1147.0$-$3812 & PKS 1144$-$379 & Rank~III & $1.048$ & broad emission lines & \citet{Stickel1989} \\
J1159.5+2914 & Ton 599 & Rank~III & $0.729$ & broad emission lines & \citet{Schneider2010} \\
J1221.5+2814 & W Comae & Rank~III & $0.102$ & narrow emission lines & \citet{Paiano2017b} \\
J1229.0+0202 & 3C 273 & Rank~III & $0.158$ & broad emission lines & \citet{Schmidt1963} \\
J1239.5+0443 & MG1 J123931+0443 & Rank~III & $1.762$ & broad emission lines & \citet{Halpern2003} \\
J1248.3+5820 & PG 1246+586 & Rank~III & $<2.68^{+0.18}_{-0.18}$ & EBL limit & \citet{Paiano2017b} \\
J1253.2+5301 & S4 1250+53 & Rank~III & $<0.81^{+0.32}_{-0.32}$ & EBL limit & \citet{Shaw2013}, \citet{Paiano2017} \\
J1256.1$-$0547 & 3C 279 & Rank~III & $0.536$ & broad emission lines & \citet{Marziani1996} \\
J1304.3$-$4353 & 1RXS J130421.2$-$435308 & Rank~III & \dots & featureless continuum & \citet{Kasai2023} \\
J1309.4+4305 & B3 1307+433 & Rank~III & \dots & no published redshift &  \\
J1418.4$-$0233 & NVSS J141826$-$023336 & Rank~III & \dots & no published redshift &  \\
J1427.0+2348 & PKS 1424+240 & Rank~III & $0.6047$ & narrow emission lines & \citet{Paiano2017b}, \citet{DAmmando2024} \\
J1427.9$-$4206 & PKS 1424$-$41 & Rank~III & $1.522$ & broad emission lines & \citet{White1988} \\
J1443.9+2501 & PKS 1441+25 & Rank~III & $0.939$ & broad emission lines & \citet{Shaw2012} \\
J1443.9$-$3908 & PKS 1440$-$389 & Rank~III & \dots & featureless continuum & \citet{Shaw2013} \\
J1448.0+3608 & RBS 1432 & Rank~III & $<1.76^{+0.23}_{-0.20}$ & EBL limit &  \\
J1454.4+5124 & TXS 1452+516 & Rank~III & $<1.29^{+0.26}_{-0.25}$ & EBL limit &  \\
J1512.8$-$0906 & PKS 1510$-$089 & Rank~III & $0.360$ & broad emission lines & \citet{Thompson1990} \\
J1517.7$-$2422 & AP Librae & Rank~III & $0.048$ & host stellar absorption & \citet{Woo2005} \\
J1543.0+6130 & GB6 J1542+6129 & Rank~III & $0.117$ & host stellar absorption & \citet{Bade1998} \\
J1630.7+5221 & TXS 1629+524 & Rank~III & $>1.545$ & intervening absorption & \citet{Shaw2013} \\
J1647.5+4950 & SBS 1646+499 & Rank~III & $0.045$ & broad emission lines & \citet{Marcha1996} \\
J1653.8+3945 & Mkn 501 & Rank~III & $0.0336$ & host stellar absorption & \citet{Stickel1993} \\
J1748.6+7005 & S4 1749+70 & Rank~III & $0.770$ & narrow emission lines & \citet{RectorStocke2001} \\
J1749.0+4321 & B3 1747+433 & Rank~III & \dots & no published redshift &  \\
J1751.5+0938 & OT 081 & Rank~III & $0.322$ & narrow emission lines & \citet{Stickel1988} \\
J1806.8+6949 & 3C 371 & Rank~III & $0.051$ & narrow emission lines & \citet{Miller1975} \\
J1824.1+5651 & 4C +56.27 & Rank~III & $0.663$ & narrow emission lines & \citet{Marcha1996} \\
J1833.6$-$2103 & PKS 1830$-$211 & Rank~III & $2.507$ & broad emission lines & \citet{Lidman1999} \\
J1838.8+4802 & GB6 J1838+4802 & Rank~III & \dots & no published redshift &  \\
J1849.2+6705 & S4 1849+67 & Rank~III & $0.657$ & broad emission lines & \citet{Healey2008} \\
J1903.2+5540 & TXS 1902+556 & Rank~III & $>0.727$ & intervening absorption & \citet{Goldoni2021} \\
J2000.0+6508 & 1ES 1959+650 & Rank~III & $0.047$ & host stellar absorption & \citet{Perlman1996} \\
J2005.5+7752 & S5 2007+77 & Rank~III & $0.342$ & host stellar absorption & \citet{Paiano2017} \\
J2121.0+1901 & OX 131 & Rank~III & $2.180$ & broad emission lines & \citet{Sahakyan2024} \\
J2139.4$-$4235 & MH 2136$-$428 & Rank~III & $<1.96^{+0.22}_{-0.23}$ & EBL limit & \citet{Landoni2014} \\
J2158.8$-$3013 & PKS 2155$-$304 & Rank~III & $0.116$ & host stellar absorption & \citet{Falomo1993} \\
J2202.7+4216 & BL Lac & Rank~III & $0.0686$ & broad emission lines & \citet{Vermeulen1995} \\
J2232.6+1143 & CTA 102 & Rank~III & $1.037$ & broad emission lines & \citet{Schmidt1965} \\
J2243.9+2021 & RGB J2243+203 & Rank~III & \dots & featureless continuum & \citet{RosaGonzalez2019} \\
J2253.9+1609 & 3C 454.3 & Rank~III & $0.859$ & broad emission lines & \citet{JacksonBrowne1991} \\
J2323.5$-$0317 & PKS 2320$-$035 & Rank~III & $1.393$ & broad emission lines & \citet{Pacciani2014} \\
\noalign{\smallskip}
\hline
\end{tabular}
\tablefoot{Redshifts obtained from host-galaxy stellar absorption, from broad or narrow emission
lines, or from a spectroscopically confirmed group or cluster association are treated as systemic.
Values derived from intervening Mg\,\textsc{ii} or Fe\,\textsc{ii} doublets constrain only the
foreground absorber and are quoted as lower limits on the source redshift.}
\end{table*}

%% file: tables/table_evolving.tex
\begin{table*}
\caption{Rank~II candidates: coherent, long-lived modulation that fails the rigidity test through a drift bound above the stationarity threshold ($|\varepsilon|+\Delta\varepsilon\geq0.02$~yr$^{-1}$) or intermediate volatility ($0.06\leq\sigma_{\ln P}+\Delta\sigma_{\ln P}<0.15$), with $\Pi-\Delta\Pi>0.60$, ordered by decreasing $Q$.}
\label{tab:evolving}
\centering
\small
\setlength{\tabcolsep}{3.5pt}
\begin{tabular}{llcccccccc}
\hline\hline
\noalign{\smallskip}
4FGL name & Association & Class & $z$ & $P_{\mathrm{obs}}$ & $Q$ & $|\varepsilon|$ & $\sigma_{\ln P}$ & $p$ & $\Pi$ \\
 &  &  &  & [yr] &  & [yr$^{-1}$] &  &  &  \\
\noalign{\smallskip}
\hline
\noalign{\smallskip}
J0144.6+2705 & TXS 0141+268 & BL Lac & $<2.38$ & $2.40 \pm 0.18$ & $6.5 \pm 0.7$ & $0.0445 \pm 0.0101$ & $0.0634 \pm 0.0253$ & $0.64$ & $1.02 \pm 0.23$ \\
J0222.6+4302 & 3C 66A & BL Lac & $0.340$ & $2.34 \pm 0.07$ & $6.0 \pm 1.1$ & $0.0174 \pm 0.0105$ & $0.0124 \pm 0.0528$ & $0.52$ & $0.83 \pm 0.15$ \\
J1512.2+0202 & PKS 1509+022 & FSRQ & $0.220$ & $2.39 \pm 0.13$ & $6.0 \pm 1.8$ & $0.0050 \pm 0.0208$ & $0.0269 \pm 0.0534$ & $0.57$ & $0.92 \pm 0.24$ \\
J0538.8$-$4405 & PKS 0537$-$441 & BL Lac & $0.892$ & $2.68 \pm 0.01$ & $5.6 \pm 0.2$ & $0.0314 \pm 0.0009$ & $0.0340 \pm 0.0041$ & $0.60$ & $1.04 \pm 0.02$ \\
J1504.4+1029 & PKS 1502+106 & FSRQ & $1.838$ & $2.46 \pm 0.01$ & $5.5 \pm 0.2$ & $0.0227 \pm 0.0011$ & $0.0147 \pm 0.0014$ & $0.54$ & $0.88 \pm 0.02$ \\
J1104.4+3812 & Mkn 421 & BL Lac & $0.031$ & $2.63 \pm 0.02$ & $5.2 \pm 0.2$ & $0.0183 \pm 0.0109$ & $0.0595 \pm 0.0174$ & $0.54$ & $0.92 \pm 0.01$ \\
\noalign{\smallskip}
\hline
\end{tabular}
\tablefoot{Columns as in Table~\ref{tab:rank1_sources}. Redshifts follow the compilation of Appendix~\ref{sec:appendix_z}.}
\end{table*}

%% file: tables/table_rankiii.tex
\begin{table*}
\caption{Rank~III candidates, ordered by decreasing cycle-weighted coherence $Q$.}
\label{tab:rank3_sources}
\centering
\scriptsize
\setlength{\tabcolsep}{2pt}
\begin{tabular}{llcccccccccl}
\hline\hline
\noalign{\smallskip}
4FGL name & Association & Class & $z$ & $P_{\mathrm{obs}}$ & $Q$ & $|\varepsilon|$ & $\sigma_{\ln P}$ & $N_{\mathrm{active}}$ & $p$ & $\Pi$ & Subclass \\
 &  &  &  & [yr] &  & [yr$^{-1}$] &  & &  &  & \\
\noalign{\smallskip}
\hline
\noalign{\smallskip}
J2158.8$-$3013 & PKS 2155$-$304 & BL Lac & $0.116$ & $1.41 \pm 0.02$ & $15.6 \pm 0.7$ & $0.0294 \pm 0.0083$ & $0.1666 \pm 0.0118$ & $9.86 \pm 0.48$ & $0.77$ & $0.99 \pm 0.04$ & III$_{\sigma}$ \\
J1304.3$-$4353 & 1RXS J130421.2$-$435308 & BL Lac & \dots & $1.79 \pm 0.21$ & $12.3 \pm 6.6$ & $0.0014 \pm 0.0291$ & $0.2812 \pm 0.2425$ & $8.08 \pm 3.85$ & $0.80$ & $1.12 \pm 0.48$ & III$_{\sigma}$ \\
J1427.0+2348 & PKS 1424+240 & BL Lac & $0.6047$ & $1.66 \pm 0.15$ & $12.1 \pm 2.8$ & $0.0148 \pm 0.0125$ & $0.0577 \pm 0.0963$ & $8.03 \pm 1.40$ & $0.74$ & $1.00 \pm 0.20$ & III$_{\sigma}$ \\
J1136.4+7009 & Mkn 180 & BL Lac & $0.045$ & $1.45 \pm 0.18$ & $12.0 \pm 7.9$ & $0.0063 \pm 0.0400$ & $0.0696 \pm 0.0820$ & $9.57 \pm 6.32$ & $0.77$ & $1.00 \pm 0.67$ & III$_{\sigma}$ \\
J0700.5$-$6610 & PKS 0700$-$661 & BL Lac & $<1.53$ & $1.58 \pm 0.33$ & $9.6 \pm 5.0$ & $0.0071 \pm 0.0418$ & $0.1795 \pm 0.1508$ & $7.90 \pm 4.30$ & $0.69$ & $0.92 \pm 0.54$ & III$_{\sigma}$ \\
J0303.6+4716 & 4C +47.08 & BL Lac & $<0.63$ & $1.06 \pm 0.22$ & $9.0 \pm 5.1$ & $0.0020 \pm 0.0277$ & $0.0274 \pm 0.0662$ & $5.76 \pm 2.71$ & $0.34$ & $0.40 \pm 0.21$ & III$_{\Pi}$ \\
J2139.4$-$4235 & MH 2136$-$428 & BL Lac & $<1.96$ & $1.66 \pm 0.19$ & $8.5 \pm 3.2$ & $0.0283 \pm 0.0290$ & $0.1802 \pm 0.1376$ & $6.67 \pm 2.52$ & $0.61$ & $0.83 \pm 0.23$ & III$_{\sigma}$ \\
J0825.9$-$2230 & PKS 0823$-$223 & BL Lac & $>0.911$ & $1.54 \pm 0.15$ & $8.5 \pm 4.4$ & $0.0209 \pm 0.0342$ & $0.0888 \pm 0.0510$ & $6.88 \pm 3.63$ & $0.59$ & $0.77 \pm 0.38$ & III$_{\Pi}$ \\
J1749.0+4321 & B3 1747+433 & BL Lac & \dots & $1.99 \pm 0.57$ & $8.5 \pm 5.9$ & $0.0350 \pm 0.0651$ & $0.1787 \pm 0.1719$ & $6.50 \pm 4.44$ & $0.72$ & $1.06 \pm 0.68$ & III$_{Q}$ \\
J1132.7+0034 & PKS B1130+008 & BL Lac & $1.223$ & $1.10 \pm 0.07$ & $8.0 \pm 3.1$ & $0.0003 \pm 0.0299$ & $0.0567 \pm 0.0317$ & $5.03 \pm 1.28$ & $0.31$ & $0.37 \pm 0.09$ & III$_{\Pi}$ \\
J1103.9$-$5357 & PKS 1101$-$536 & BL Lac & \dots & $2.16 \pm 0.23$ & $7.4 \pm 1.1$ & $0.0077 \pm 0.0194$ & $0.1390 \pm 0.0919$ & $5.18 \pm 0.57$ & $0.62$ & $0.94 \pm 0.21$ & III$_{\sigma}$ \\
J0211.2+1051 & MG1 J021114+1051 & BL Lac & $0.200$ & $2.39 \pm 0.52$ & $7.3 \pm 4.4$ & $0.0226 \pm 0.0313$ & $0.1952 \pm 0.1821$ & $5.07 \pm 2.94$ & $0.67$ & $1.08 \pm 0.46$ & III$_{Q}$ \\
J1037.7+5711 & GB6 J1037+5711 & BL Lac & $>0.830$ & $1.32 \pm 0.03$ & $7.2 \pm 2.4$ & $0.0112 \pm 0.0153$ & $0.0186 \pm 0.0376$ & $5.14 \pm 1.81$ & $0.38$ & $0.48 \pm 0.18$ & III$_{\Pi}$ \\
J1630.7+5221 & TXS 1629+524 & BL Lac & $>1.545$ & $1.89 \pm 0.25$ & $7.0 \pm 2.7$ & $0.0196 \pm 0.0258$ & $0.0573 \pm 0.1013$ & $5.49 \pm 2.14$ & $0.57$ & $0.82 \pm 0.33$ & III$_{\sigma}$ \\
J0953.0$-$0840 & PMN J0953$-$0840 & BL Lac & \dots & $1.08 \pm 0.14$ & $6.9 \pm 3.5$ & $0.0301 \pm 0.0449$ & $0.0327 \pm 0.0435$ & $4.78 \pm 2.09$ & $0.29$ & $0.35 \pm 0.14$ & III$_{\Pi}$ \\
J1253.2+5301 & S4 1250+53 & BL Lac & $<0.81$ & $2.13 \pm 0.30$ & $6.9 \pm 2.5$ & $0.0056 \pm 0.0186$ & $0.2370 \pm 0.2144$ & $5.43 \pm 1.98$ & $0.64$ & $0.96 \pm 0.22$ & III$_{\sigma}$ \\
J1443.9$-$3908 & PKS 1440$-$389 & BL Lac & \dots & $1.10 \pm 0.19$ & $6.9 \pm 3.3$ & $0.0263 \pm 0.0380$ & $0.0309 \pm 0.0306$ & $5.06 \pm 2.14$ & $0.31$ & $0.37 \pm 0.11$ & III$_{\Pi}$ \\
J0721.9+7120 & S5 0716+71 & BL Lac & $>0.2315$ & $2.43 \pm 0.03$ & $6.7 \pm 0.2$ & $0.0202 \pm 0.0037$ & $0.1769 \pm 0.0177$ & $5.08 \pm 0.10$ & $0.68$ & $1.11 \pm 0.01$ & III$_{\sigma}$ \\
J0904.9$-$5734 & PKS 0903$-$57 & FSRQ & $0.2621$ & $1.90 \pm 0.13$ & $6.6 \pm 1.0$ & $0.0174 \pm 0.0457$ & $0.1822 \pm 0.0658$ & $5.08 \pm 0.93$ & $0.53$ & $0.76 \pm 0.14$ & III$_{\sigma}$ \\
J1838.8+4802 & GB6 J1838+4802 & BL Lac & \dots & $2.43 \pm 0.26$ & $6.5 \pm 1.4$ & $0.0306 \pm 0.0332$ & $0.1178 \pm 0.0858$ & $4.86 \pm 1.00$ & $0.65$ & $1.06 \pm 0.17$ & III$_{\sigma}$ \\
J0303.4$-$2407 & PKS 0301$-$243 & BL Lac & $0.266$ & $2.18 \pm 0.27$ & $6.5 \pm 2.5$ & $0.0185 \pm 0.0281$ & $0.0327 \pm 0.2022$ & $4.56 \pm 2.06$ & $0.55$ & $0.84 \pm 0.18$ & III$_{\sigma}$ \\
J0650.7+2503 & 1ES 0647+250 & BL Lac & \dots & $1.45 \pm 0.06$ & $6.5 \pm 2.8$ & $0.0035 \pm 0.0218$ & $0.0073 \pm 0.0661$ & $4.62 \pm 2.11$ & $0.37$ & $0.48 \pm 0.20$ & III$_{\Pi}$ \\
J1806.8+6949 & 3C 371 & BL Lac & $0.051$ & $1.62 \pm 0.15$ & $6.4 \pm 1.2$ & $0.0556 \pm 0.0455$ & $0.0602 \pm 0.1119$ & $4.90 \pm 0.93$ & $0.44$ & $0.59 \pm 0.18$ & III$_{\sigma}$ \\
J1309.4+4305 & B3 1307+433 & BL Lac & \dots & $1.18 \pm 0.19$ & $6.3 \pm 3.8$ & $0.0026 \pm 0.0413$ & $0.0939 \pm 0.0718$ & $5.11 \pm 2.31$ & $0.33$ & $0.41 \pm 0.22$ & III$_{Q}$ \\
J2323.5$-$0317 & PKS 2320$-$035 & FSRQ & $1.393$ & $1.04 \pm 0.23$ & $6.3 \pm 3.7$ & $0.0129 \pm 0.0273$ & $0.0336 \pm 0.0269$ & $3.80 \pm 1.55$ & $0.22$ & $0.27 \pm 0.07$ & III$_{Q}$ \\
J0112.1+2245 & S2 0109+22 & BL Lac & $0.265$ & $1.79 \pm 0.02$ & $6.3 \pm 0.4$ & $0.0297 \pm 0.0032$ & $0.0250 \pm 0.0041$ & $4.38 \pm 0.28$ & $0.43$ & $0.60 \pm 0.03$ & III$_{\Pi}$ \\
J0221.1+3556 & B2 0218+357 & FSRQ & $0.944$ & $1.67 \pm 0.33$ & $5.7 \pm 3.3$ & $0.0520 \pm 0.0753$ & $0.0256 \pm 0.0101$ & $4.22 \pm 2.00$ & $0.39$ & $0.53 \pm 0.26$ & III$_{Q}$ \\
J1059.2$-$1134 & PKS B1056$-$113 & BL Lac & \dots & $1.29 \pm 0.14$ & $5.5 \pm 1.8$ & $0.0430 \pm 0.0561$ & $0.0358 \pm 0.0280$ & $4.31 \pm 1.41$ & $0.31$ & $0.39 \pm 0.07$ & III$_{\Pi}$ \\
J1147.0$-$3812 & PKS 1144$-$379 & BL Lac & $1.048$ & $1.82 \pm 0.33$ & $5.3 \pm 3.2$ & $0.0119 \pm 0.0141$ & $0.0092 \pm 0.0884$ & $3.75 \pm 2.23$ & $0.38$ & $0.53 \pm 0.49$ & III$_{Q}$ \\
J1454.4+5124 & TXS 1452+516 & BL Lac & $<1.29$ & $2.01 \pm 0.13$ & $5.2 \pm 1.7$ & $0.0340 \pm 0.0247$ & $0.0274 \pm 0.1165$ & $4.10 \pm 1.28$ & $0.46$ & $0.67 \pm 0.28$ & III$_{\Pi}$ \\
J0521.7+2112 & TXS 0518+211 & BL Lac & $0.108$ & $2.95 \pm 0.21$ & $4.9 \pm 1.1$ & $0.0120 \pm 0.0199$ & $0.1588 \pm 0.1379$ & $3.72 \pm 0.93$ & $0.61$ & $1.14 \pm 0.16$ & III$_{\sigma}$ \\
J1647.5+4950 & SBS 1646+499 & BL Lac & $0.045$ & $2.43 \pm 0.87$ & $4.8 \pm 2.5$ & $0.0285 \pm 0.0758$ & $0.0472 \pm 0.2058$ & $3.89 \pm 1.81$ & $0.52$ & $0.85 \pm 0.49$ & III$_{Q}$ \\
J1248.3+5820 & PG 1246+586 & BL Lac & $<2.68$ & $2.16 \pm 0.37$ & $4.8 \pm 4.6$ & $0.0071 \pm 0.0298$ & $0.0104 \pm 0.2276$ & $3.65 \pm 3.15$ & $0.44$ & $0.66 \pm 0.33$ & III$_{Q}$ \\
J0730.3$-$1141 & PKS 0727$-$11 & FSRQ & $1.591$ & $1.84 \pm 0.20$ & $4.6 \pm 2.6$ & $0.0115 \pm 0.0431$ & $0.0511 \pm 0.0860$ & $3.56 \pm 2.05$ & $0.36$ & $0.51 \pm 0.21$ & III$_{Q}$ \\
J1146.9+3958 & S4 1144+40 & FSRQ & $1.090$ & $3.24 \pm 0.01$ & $4.6 \pm 0.1$ & $0.0019 \pm 0.0014$ & $0.0119 \pm 0.0017$ & $2.79 \pm 0.04$ & $0.50$ & $1.03 \pm 0.01$ & III$_{P}$ \\
J0532.6+0732 & OG 050 & FSRQ & $1.254$ & $2.86 \pm 0.21$ & $4.5 \pm 2.6$ & $0.0192 \pm 0.0096$ & $0.0376 \pm 0.0291$ & $3.45 \pm 1.88$ & $0.55$ & $0.99 \pm 0.49$ & III$_{Q}$ \\
J1748.6+7005 & S4 1749+70 & BL Lac & $0.770$ & $1.24 \pm 0.40$ & $4.5 \pm 2.8$ & $0.0641 \pm 0.0624$ & $0.0644 \pm 0.0610$ & $3.95 \pm 2.49$ & $0.27$ & $0.34 \pm 0.17$ & III$_{Q}$ \\
J2121.0+1901 & OX 131 & FSRQ & $2.180$ & $2.98 \pm 0.33$ & $4.4 \pm 2.1$ & $0.0012 \pm 0.0260$ & $0.0115 \pm 0.1091$ & $3.26 \pm 1.34$ & $0.54$ & $1.01 \pm 0.36$ & III$_{Q}$ \\
J1448.0+3608 & RBS 1432 & BL Lac & $<1.76$ & $2.66 \pm 0.65$ & $4.4 \pm 2.1$ & $0.0004 \pm 0.0292$ & $0.2208 \pm 0.2128$ & $3.56 \pm 1.80$ & $0.52$ & $0.90 \pm 0.37$ & III$_{Q}$ \\
J1048.4+7143 & S5 1044+71 & FSRQ & $1.150$ & $3.12 \pm 0.01$ & $4.2 \pm 0.0$ & $0.0015 \pm 0.0010$ & $0.0097 \pm 0.0006$ & $2.84 \pm 0.03$ & $0.49$ & $0.97 \pm 0.01$ & III$_{P}$ \\
J1512.8$-$0906 & PKS 1510$-$089 & FSRQ & $0.360$ & $1.68 \pm 0.19$ & $4.2 \pm 0.3$ & $0.0672 \pm 0.0575$ & $0.0511 \pm 0.0060$ & $3.28 \pm 0.28$ & $0.30$ & $0.41 \pm 0.10$ & III$_{\Pi}$ \\
J1903.2+5540 & TXS 1902+556 & BL Lac & $>0.727$ & $3.09 \pm 0.06$ & $4.1 \pm 0.3$ & $0.0060 \pm 0.0062$ & $0.0332 \pm 0.0103$ & $2.85 \pm 0.10$ & $0.49$ & $0.95 \pm 0.02$ & III$_{P}$ \\
J1221.5+2814 & W Comae & BL Lac & $0.102$ & $2.85 \pm 0.22$ & $3.8 \pm 1.2$ & $0.0168 \pm 0.0445$ & $0.0091 \pm 0.0585$ & $3.06 \pm 0.92$ & $0.48$ & $0.88 \pm 0.26$ & III$_{Q}$ \\
J0022.5+0608 & PKS 0019+058 & BL Lac & $<1.33$ & $2.95 \pm 0.41$ & $3.8 \pm 1.7$ & $0.0145 \pm 0.0293$ & $0.0877 \pm 0.0669$ & $3.12 \pm 1.15$ & $0.51$ & $0.95 \pm 0.08$ & III$_{Q}$ \\
J0217.2+0837 & ZS 0214+083 & BL Lac & $0.085$ & $3.55 \pm 0.05$ & $3.6 \pm 0.3$ & $0.0044 \pm 0.0058$ & $0.0057 \pm 0.0048$ & $2.23 \pm 0.06$ & $0.44$ & $0.99 \pm 0.01$ & III$_{P}$ \\
J0607.4+4739 & TXS 0603+476 & BL Lac & $<1.96$ & $3.16 \pm 0.27$ & $3.3 \pm 1.4$ & $0.0136 \pm 0.0437$ & $0.0116 \pm 0.0955$ & $2.79 \pm 1.13$ & $0.49$ & $0.97 \pm 0.42$ & III$_{Q}$ \\
\noalign{\smallskip}
\hline
\end{tabular}
\end{table*}

\begin{table*}
\caption{Continued.}
\label{tab:rank3_sources_b}
\centering
\scriptsize
\setlength{\tabcolsep}{2pt}
\begin{tabular}{llcccccccccl}
\hline\hline
\noalign{\smallskip}
4FGL name & Association & Class & $z$ & $P_{\mathrm{obs}}$ & $Q$ & $|\varepsilon|$ & $\sigma_{\ln P}$ & $N_{\mathrm{active}}$ & $p$ & $\Pi$ & Subclass \\
 &  &  &  & [yr] &  & [yr$^{-1}$] &  & &  &  & \\
\noalign{\smallskip}
\hline
\noalign{\smallskip}
J0428.6$-$3756 & PKS 0426$-$380 & BL Lac & $1.111$ & $3.59 \pm 0.02$ & $3.3 \pm 0.1$ & $0.0025 \pm 0.0014$ & $0.0033 \pm 0.0013$ & $2.17 \pm 0.03$ & $0.43$ & $1.00 \pm 0.01$ & III$_{P}$ \\
J0957.6+5523 & 4C +55.17 & FSRQ & $0.8955$ & $3.27 \pm 1.09$ & $3.2 \pm 2.3$ & $0.0417 \pm 0.0535$ & $0.0328 \pm 0.0310$ & $2.59 \pm 1.75$ & $0.47$ & $0.97 \pm 0.52$ & III$_{Q}$ \\
J0457.0$-$2324 & PKS 0454$-$234 & FSRQ & $1.003$ & $3.47 \pm 0.01$ & $3.2 \pm 0.1$ & $0.0033 \pm 0.0009$ & $0.0038 \pm 0.0008$ & $2.35 \pm 0.03$ & $0.45$ & $1.00 \pm 0.01$ & III$_{P}$ \\
J0100.3+0745 & GB6 J0100+0745 & BL Lac & \dots & $2.78 \pm 0.67$ & $3.2 \pm 2.3$ & $0.0336 \pm 0.0628$ & $0.0619 \pm 0.0596$ & $2.75 \pm 1.55$ & $0.42$ & $0.75 \pm 0.27$ & III$_{Q}$ \\
J1058.4+0133 & 4C +01.28 & BL Lac & $0.890$ & $3.64 \pm 0.04$ & $3.1 \pm 0.2$ & $0.0115 \pm 0.0055$ & $0.0133 \pm 0.0072$ & $2.19 \pm 0.07$ & $0.44$ & $1.04 \pm 0.02$ & III$_{Q}$ \\
J0612.8+4122 & B3 0609+413 & BL Lac & $>1.108$ & $3.21 \pm 0.65$ & $3.1 \pm 1.4$ & $0.0197 \pm 0.0316$ & $0.1066 \pm 0.0949$ & $2.46 \pm 0.98$ & $0.44$ & $0.88 \pm 0.33$ & III$_{Q}$ \\
J1443.9+2501 & PKS 1441+25 & FSRQ & $0.939$ & $3.31 \pm 0.02$ & $3.1 \pm 0.1$ & $0.0063 \pm 0.0017$ & $0.0294 \pm 0.0045$ & $2.47 \pm 0.04$ & $0.46$ & $0.97 \pm 0.01$ & III$_{Q}$ \\
J0237.8+2848 & 4C +28.07 & FSRQ & $1.206$ & $3.54 \pm 0.01$ & $3.0 \pm 0.1$ & $0.0132 \pm 0.0011$ & $0.0096 \pm 0.0011$ & $2.23 \pm 0.06$ & $0.44$ & $0.99 \pm 0.02$ & III$_{Q}$ \\
J1418.4$-$0233 & NVSS J141826$-$023336 & BL Lac & \dots & $2.59 \pm 0.25$ & $3.0 \pm 1.2$ & $0.0019 \pm 0.0132$ & $0.0107 \pm 0.0586$ & $2.10 \pm 1.06$ & $0.30$ & $0.51 \pm 0.36$ & III$_{Q}$ \\
J1824.1+5651 & 4C +56.27 & BL Lac & $0.663$ & $3.77 \pm 0.35$ & $2.9 \pm 0.9$ & $0.0016 \pm 0.0359$ & $0.0087 \pm 0.0719$ & $1.90 \pm 0.81$ & $0.40$ & $0.98 \pm 0.11$ & III$_{Q}$ \\
J2000.0+6508 & 1ES 1959+650 & BL Lac & $0.047$ & $3.82 \pm 0.03$ & $2.8 \pm 0.2$ & $0.0171 \pm 0.0025$ & $0.0035 \pm 0.0025$ & $1.86 \pm 0.04$ & $0.39$ & $0.99 \pm 0.01$ & III$_{Q}$ \\
J2253.9+1609 & 3C 454.3 & FSRQ & $0.859$ & $1.18 \pm 0.00$ & $2.6 \pm 0.4$ & $0.0040 \pm 0.0028$ & $0.0076 \pm 0.0018$ & $2.30 \pm 0.31$ & $0.15$ & $0.19 \pm 0.02$ & III$_{Q}$ \\
J0516.7$-$6207 & PKS 0516$-$621 & BL Lac & $1.300$ & $3.86 \pm 0.11$ & $2.6 \pm 0.4$ & $0.0129 \pm 0.0115$ & $0.0047 \pm 0.0069$ & $1.83 \pm 0.20$ & $0.39$ & $1.00 \pm 0.04$ & III$_{Q}$ \\
J0909.7$-$0230 & PKS 0907$-$023 & FSRQ & \dots & $2.31 \pm 1.20$ & $2.5 \pm 1.2$ & $0.0064 \pm 0.0143$ & $0.0062 \pm 0.0124$ & $2.09 \pm 0.95$ & $0.27$ & $0.43 \pm 0.15$ & III$_{Q}$ \\
J1543.0+6130 & GB6 J1542+6129 & BL Lac & $0.117$ & $3.99 \pm 0.04$ & $2.4 \pm 0.1$ & $0.0071 \pm 0.0028$ & $0.0080 \pm 0.0016$ & $1.73 \pm 0.04$ & $0.38$ & $1.03 \pm 0.01$ & III$_{Q}$ \\
J1256.1$-$0547 & 3C 279 & FSRQ & $0.536$ & $3.50 \pm 0.01$ & $2.4 \pm 0.0$ & $0.0197 \pm 0.0011$ & $0.0265 \pm 0.0015$ & $2.13 \pm 0.03$ & $0.41$ & $0.92 \pm 0.01$ & III$_{Q}$ \\
J0509.4+0542 & TXS 0506+056 & BL Lac & $0.3365$ & $3.87 \pm 0.04$ & $2.3 \pm 0.1$ & $0.0260 \pm 0.0027$ & $0.0054 \pm 0.0034$ & $1.81 \pm 0.06$ & $0.39$ & $0.99 \pm 0.02$ & III$_{Q}$ \\
J1653.8+3945 & Mkn 501 & BL Lac & $0.0336$ & $4.01 \pm 0.06$ & $2.3 \pm 0.2$ & $0.0027 \pm 0.0043$ & $0.0052 \pm 0.0017$ & $1.70 \pm 0.14$ & $0.38$ & $1.02 \pm 0.07$ & III$_{Q}$ \\
J0051.2$-$6242 & 1RXS J005117.7$-$624154 & BL Lac & \dots & $3.08 \pm 1.59$ & $2.3 \pm 1.5$ & $0.0094 \pm 0.0471$ & $0.1224 \pm 0.1204$ & $2.04 \pm 1.37$ & $0.35$ & $0.68 \pm 0.32$ & III$_{Q}$ \\
J1833.6$-$2103 & PKS 1830$-$211 & FSRQ & $2.507$ & $3.73 \pm 0.21$ & $2.3 \pm 0.3$ & $0.0239 \pm 0.0012$ & $0.0044 \pm 0.0136$ & $2.00 \pm 0.22$ & $0.41$ & $1.00 \pm 0.03$ & III$_{Q}$ \\
J1239.5+0443 & MG1 J123931+0443 & FSRQ & $1.762$ & $1.39 \pm 0.16$ & $2.2 \pm 0.8$ & $0.0141 \pm 0.0303$ & $0.0080 \pm 0.0510$ & $1.83 \pm 0.76$ & $0.14$ & $0.18 \pm 0.11$ & III$_{Q}$ \\
J0811.4+0146 & OJ 014 & BL Lac & $>1.148$ & $4.13 \pm 0.05$ & $2.2 \pm 0.1$ & $0.0043 \pm 0.0028$ & $0.0021 \pm 0.0013$ & $1.52 \pm 0.06$ & $0.35$ & $0.99 \pm 0.01$ & III$_{Q}$ \\
J1058.6+5627 & TXS 1055+567 & BL Lac & \dots & $3.94 \pm 0.40$ & $2.0 \pm 0.9$ & $0.0380 \pm 0.0285$ & $0.0345 \pm 0.0284$ & $1.70 \pm 0.76$ & $0.37$ & $0.98 \pm 0.48$ & III$_{Q}$ \\
J0757.1+0956 & PKS 0754+100 & BL Lac & $0.266$ & $4.02 \pm 0.92$ & $1.9 \pm 1.1$ & $0.0276 \pm 0.0527$ & $0.0086 \pm 0.0544$ & $1.67 \pm 1.01$ & $0.37$ & $1.02 \pm 0.56$ & III$_{Q}$ \\
J0739.2+0137 & PKS 0736+01 & FSRQ & $0.189$ & $4.18 \pm 0.04$ & $1.8 \pm 0.0$ & $0.0129 \pm 0.0013$ & $0.0026 \pm 0.0008$ & $1.46 \pm 0.03$ & $0.34$ & $0.99 \pm 0.01$ & III$_{Q}$ \\
J0136.5+3906 & B3 0133+388 & BL Lac & $<0.73$ & $4.36 \pm 0.45$ & $1.8 \pm 0.5$ & $0.0070 \pm 0.0131$ & $0.0017 \pm 0.0168$ & $1.27 \pm 0.33$ & $0.31$ & $0.98 \pm 0.27$ & III$_{Q}$ \\
J2005.5+7752 & S5 2007+77 & BL Lac & $0.342$ & $4.00 \pm 0.21$ & $1.8 \pm 0.5$ & $0.0075 \pm 0.0349$ & $0.0119 \pm 0.0220$ & $1.54 \pm 0.38$ & $0.34$ & $0.92 \pm 0.13$ & III$_{Q}$ \\
J2232.6+1143 & CTA 102 & FSRQ & $1.037$ & $3.99 \pm 0.01$ & $1.7 \pm 0.0$ & $0.0089 \pm 0.0008$ & $0.0209 \pm 0.0011$ & $1.57 \pm 0.00$ & $0.35$ & $0.94 \pm 0.00$ & III$_{Q}$ \\
J1849.2+6705 & S4 1849+67 & FSRQ & $0.657$ & $3.70 \pm 0.25$ & $1.7 \pm 0.7$ & $0.0738 \pm 0.0738$ & $0.0922 \pm 0.0865$ & $1.46 \pm 0.44$ & $0.30$ & $0.72 \pm 0.30$ & III$_{Q}$ \\
J0210.7$-$5101 & PKS 0208$-$512 & FSRQ & $1.003$ & $4.18 \pm 0.22$ & $1.6 \pm 0.5$ & $0.0211 \pm 0.0021$ & $0.0144 \pm 0.0861$ & $1.41 \pm 0.41$ & $0.33$ & $0.96 \pm 0.10$ & III$_{Q}$ \\
J1517.7$-$2422 & AP Librae & BL Lac & $0.048$ & $4.49 \pm 0.52$ & $1.4 \pm 0.7$ & $0.0112 \pm 0.0282$ & $0.0071 \pm 0.1324$ & $1.15 \pm 0.57$ & $0.29$ & $0.98 \pm 0.03$ & III$_{Q}$ \\
J1751.5+0938 & OT 081 & BL Lac & $0.322$ & $4.68 \pm 0.05$ & $1.3 \pm 0.0$ & $0.0134 \pm 0.0032$ & $0.0032 \pm 0.0010$ & $1.03 \pm 0.03$ & $0.27$ & $1.01 \pm 0.01$ & III$_{Q}$ \\
J0120.4$-$2701 & PKS 0118$-$272 & BL Lac & $<0.79$ & $4.88 \pm 0.45$ & $1.1 \pm 0.4$ & $0.0121 \pm 0.0180$ & $0.0019 \pm 0.0134$ & $0.86 \pm 0.36$ & $0.23$ & $1.00 \pm 0.03$ & III$_{Q}$ \\
J1159.5+2914 & Ton 599 & FSRQ & $0.729$ & $4.01 \pm 0.04$ & $1.1 \pm 0.2$ & $0.0307 \pm 0.0038$ & $0.0066 \pm 0.0011$ & $1.04 \pm 0.14$ & $0.23$ & $0.63 \pm 0.06$ & III$_{Q}$ \\
J0738.1+1742 & PKS 0735+17 & BL Lac & $>0.424$ & $4.90 \pm 0.07$ & $1.1 \pm 0.1$ & $0.0020 \pm 0.0031$ & $0.0014 \pm 0.0006$ & $0.83 \pm 0.05$ & $0.23$ & $0.99 \pm 0.01$ & III$_{Q}$ \\
J1427.9$-$4206 & PKS 1424$-$41 & FSRQ & $1.522$ & $5.16 \pm 0.01$ & $0.9 \pm 0.0$ & $0.0033 \pm 0.0010$ & $0.0012 \pm 0.0001$ & $0.65 \pm 0.00$ & $0.19$ & $1.02 \pm 0.01$ & III$_{Q}$ \\
J1229.0+0202 & 3C 273 & FSRQ & $0.158$ & $4.56 \pm 0.13$ & $0.8 \pm 0.1$ & $0.0514 \pm 0.0052$ & $0.0344 \pm 0.0144$ & $0.70 \pm 0.07$ & $0.18$ & $0.63 \pm 0.09$ & III$_{Q}$ \\
J0141.4$-$0928 & PKS 0139$-$09 & BL Lac & $0.733$ & $5.13 \pm 0.09$ & $0.8 \pm 0.1$ & $0.0169 \pm 0.0042$ & $0.0015 \pm 0.0011$ & $0.67 \pm 0.07$ & $0.19$ & $0.99 \pm 0.02$ & III$_{Q}$ \\
J0334.2$-$4008 & PKS 0332$-$403 & BL Lac & \dots & $5.35 \pm 0.05$ & $0.7 \pm 0.0$ & $0.0127 \pm 0.0042$ & $0.0016 \pm 0.0005$ & $0.53 \pm 0.04$ & $0.16$ & $0.99 \pm 0.02$ & III$_{Q}$ \\
J0630.9$-$2406 & TXS 0628$-$240 & BL Lac & $>1.238$ & $5.44 \pm 0.25$ & $0.6 \pm 0.2$ & $0.0099 \pm 0.0099$ & $0.0021 \pm 0.0011$ & $0.47 \pm 0.15$ & $0.14$ & $0.99 \pm 0.02$ & III$_{Q}$ \\
J2202.7+4216 & BL Lac & BL Lac & $0.0686$ & $5.46 \pm 0.03$ & $0.5 \pm 0.0$ & $0.0269 \pm 0.0026$ & $0.0015 \pm 0.0004$ & $0.47 \pm 0.02$ & $0.14$ & $1.01 \pm 0.03$ & III$_{Q}$ \\
J1015.0+4926 & 1H 1013+498 & BL Lac & $0.212$ & $5.61 \pm 0.08$ & $0.5 \pm 0.1$ & $0.0146 \pm 0.0040$ & $0.0011 \pm 0.0004$ & $0.37 \pm 0.05$ & $0.12$ & $0.99 \pm 0.02$ & III$_{Q}$ \\
J2243.9+2021 & RGB J2243+203 & BL Lac & \dots & $5.73 \pm 0.10$ & $0.4 \pm 0.1$ & $0.0088 \pm 0.0052$ & $0.0011 \pm 0.0002$ & $0.28 \pm 0.07$ & $0.09$ & $0.97 \pm 0.05$ & III$_{Q}$ \\
\noalign{\smallskip}
\hline
\end{tabular}
\tablefoot{Columns as in Table~\ref{tab:rank1_sources}, with the addition of $N_{\mathrm{active}}$ and the Rank~III subclass. Redshifts follow the compilation of Appendix~\ref{sec:appendix_z}. Subclasses (Sect.~\ref{subsec:rank3}): III$_{Q}$, insufficient coherence ($Q-\Delta Q\le3$); III$_{\sigma}$, plasma-level volatility ($\sigma_{\ln P}+\Delta\sigma_{\ln P}\ge0.15$); III$_{\Pi}$, coherent but insufficiently persistent or long-lived; III$_{P}$, period beyond the ceiling $P_{\mathrm{ceil}}\simeq3.09$~yr (Eq.~\ref{eq:pceil}), for which the cone-of-influence-free baseline does not contain three cycles at any amplitude.}
\end{table*}